\documentclass[a4paper,11pt]{article}
\pdfoutput=1
\usepackage{jheppub}
\usepackage{hyperref}
\usepackage{amsmath,amssymb,latexsym}
\usepackage{braket}
\usepackage{bm}
\usepackage{graphicx}
\usepackage{hepunits}
\usepackage{slashed}
\usepackage[svgnames]{xcolor}
\usepackage{multirow}

\allowdisplaybreaks

\DeclareMathOperator{\Tr}{Tr}
\DeclareMathOperator{\Imag}{Im}

\newcommand{\ff}{f\!\!f}

\preprint{IPPP/20/10, MITP/20-022}

\title{Top quark pair production near threshold: single/double distributions and mass determination}

\author[c]{Wan-Li Ju,}
\author[a]{Guoxing Wang,}
\author[d]{Xing Wang,}
\author[e]{Xiaofeng Xu,}
\author[a]{Yongqi Xu,}
\author[b]{Li Lin Yang}

\affiliation[a]{School of Physics and State Key Laboratory of Nuclear Physics and Technology,\\
Peking University, Beijing 100871, China}
\affiliation[b]{Zhejiang Institute of Modern Physics, Department of Physics, Zhejiang University, Hangzhou 310027, China}
\affiliation[c]{Institute for Particle Physics Phenomenology, Durham University, Durham DH1 3LE, UK}
\affiliation[d]{PRISMA+ Cluster of Excellence \& Mainz Institute for Theoretical Physics, Johannes Gutenberg University, 55099 Mainz, Germany}
\affiliation[e]{Institut f\"ur Theoretische Physik, Universit\"at Bern, Sidlerstrasse 5, CH-3012 Bern, Switzerland}

\emailAdd{wan.l.ju@durham.ac.uk}
\emailAdd{wangguoxing2015@pku.edu.cn}
\emailAdd{x.wang@uni-mainz.de}
\emailAdd{pkuxxf@gmail.com}
\emailAdd{xuyongqi@pku.edu.cn}
\emailAdd{yanglilin@zju.edu.cn}

\abstract{
We investigate top quark pair production near the threshold where the pair invariant mass $M_{t\bar{t}}$ approaches $2m_t$, which provides sensitive observables to extract the top quark mass $m_t$. Using the effective field theory methods, we derive a factorization and resummation formula for kinematic distributions in the threshold limit up to the next-to-leading power, which resums higher order Coulomb corrections to all orders in the strong coupling constant. Our formula is similar to those in the literature but differs in several important aspects. We apply our formula to the $M_{t\bar{t}}$ distribution, as well as to the double differential cross section with respect to $M_{t\bar{t}}$ and the rapidity of the $t\bar{t}$ pair. We find that the resummation effects significantly increase the cross sections near the threshold, and lead to predictions better compatible with experimental data than the fixed-order ones. We demonstrate that incorporating resummation effects in the top quark mass determination can shift the extracted value of $m_t$ by as large as \unit{1.4}{\GeV}. The shift is much larger than the estimated uncertainties in previous experimental studies, and leads to a value of the top quark pole mass more consistent with the current world average.
}

\begin{document}

\maketitle

\section{Introduction}

The top quark is the heaviest elementary particle in the Standard Model (SM). Its large mass plays important roles in many frontiers of particle physics. In the SM, the top quark mass $m_t$ comes exclusively from the $O(1)$ Yukawa coupling between the top quark and the Higgs field. Therefore, the top quark is believed to be crucial to understand the electroweak symmetry breaking and properties of the Higgs sector. For example, the stability of the electroweak vacuum is quite sensitive to the top quark mass. The same is true for the fine-tuning of the Higgs boson mass and the indirect constraints on new physics beyond the SM. Consequently, precise measurement of the top quark mass is a highly important quest of the Large Hadron Collider (LHC) and future high energy colliders.

Traditionally, the top quark mass is measured by reconstructing the top quark from its decay products, and fitting the resulting invariant mass distribution against that generated by Monte Carlo (MC) event generators. Such a mass is often referred to as the ``MC mass''. Thanks to the large amount of data collected by the ATLAS and CMS detectors at the LHC, the precision for the measured MC mass has been greatly improved in recent years. The current world average for the MC mass is given by $m_t^{\text{MC}} = \unit{$172.9 \pm 0.4$}{\GeV}$~\cite{Tanabashi:2018oca}. Despite the high precision of the experimental result, it turns out to be difficult to relate the MC mass to a well-defined mass parameter in the Lagrangian of the associated quantum field theory with a certain renormalization scheme (see, e.g., Refs.~\cite{Nason:2017cxd, Corcella:2019tgt}). The difficulties are mostly related to the fact that top quarks (and their decay products) are strongly-interacting particles who may radiate additional gluons and quarks which end up as hadrons in the detectors. These effects are described approximately by parton shower algorithms and hadronization models in MC event generators. Both the perturbative and non-perturbative aspects of the generators need to be carefully studied in order to relate the MC mass to a field-theoretic mass. There have been ongoing researches on these issues~\cite{Kieseler:2015jzh, Butenschoen:2016lpz, Hoang:2018zrp, Boronat:2019cgt}, but no final quantitative conclusion has been reached.

Instead of measuring the MC mass from the decay products of the top quark, it is possible to directly extract a Lagrangian mass by comparing experimental measurements and theoretical predictions for certain observables (e.g., total or differential cross sections of scattering processes involving the top quark). For that purpose, not only the experimental measurements, but also the theoretical predictions for these observables have to achieve rather high accuracies in order to extract a relatively precise value of the top quark mass. Such theoretical predictions necessarily involve higher order perturbative corrections. In their calculations ultraviolet (UV) divergences appear at intermediate steps and one has to adopt a renormalization scheme to arrive at finite predictions. The definition of the Lagrangian mass therefore depends on the renormalization scheme. In practice, one often employs the on-shell scheme or the modified minimal subtraction ($\overline{\text{MS}}$) scheme. In the on-shell scheme, one defines the so-called ``pole mass'' of the top quark in perturbation theory.\footnote{Note however that due to the strongly-interacting nature of the top quark, it actually has no pole mass non-perturbatively. A related issue is the renormalon ambiguity of the perturbatively-defined top quark pole mass \cite{Beneke:1994sw, Beneke:1994rs, Beneke:2016cbu, Hoang:2017btd, FerrarioRavasio:2018ubr, Kataev:2018gle, Kataev:2018mob}.} This is the most widely used mass scheme in perturbative calculations for top quark related scattering processes, and we will only discuss this mass definition in the current work. The current world average for the top quark pole mass, extracted from cross section measurements, is given by $m_t^{\text{pole}} = \unit{$173.1 \pm 0.9$}{\GeV}$~\cite{Tanabashi:2018oca}. The value of the extracted pole mass is rather close to the MC mass, and their exact relationship is an important question to be addressed \cite{Kieseler:2015jzh, Butenschoen:2016lpz, Hoang:2018zrp, Boronat:2019cgt}.

Following the above discussions, it is clear that to extract the top quark mass, one needs to use observables that are strongly dependent on $m_t$, and in the same time can be experimentally measured and theoretically calculated with high precisions. An often used observable is the $t\bar{t}$ pair invariant-mass distribution and related multi-differential cross sections in the top quark pair production process \cite{Aaboud:2017ujq, Sirunyan:2019zvx}. It can be easily anticipated that the kinematic region most sensitive to $m_t$ is where the pair invariant mass $M_{t\bar{t}}$ is near the $2m_t$ threshold. Precision theoretical predictions for this observable, especially in the threshold region, are therefore highly demanded to achieve the goal of extracting the top quark mass. A closely related observable $\rho_s$ (and similar ones) in $t\bar{t} + \text{jet}$ production was employed in \cite{Alioli:2013mxa, Aad:2015waa, Fuster:2017rev, Bevilacqua:2017ipv, Aad:2019mkw}, where $\rho_s$ is defined as
\begin{equation}
\rho_s = \frac{2m_0}{\sqrt{s_{t\bar{t}j}}} \, ,
\end{equation}
where $m_0$ is an arbitrarily chosen scale of the order of $m_t$, and $s_{t\bar{t}j}$ is the invariant mass of the top quark,  the anti-top quark and the additional jet. It was shown in \cite{Alioli:2013mxa} that the region most sensitive to $m_t$ is where $\rho_s$ is near its maximal value. In that region, the $t\bar{t}$ invariant mass $M_{t\bar{t}}$ is pushed to the $2m_t$ threshold. Consequently, understanding the threshold behavior of $M_{t\bar{t}}$ is crucial also when using the $\rho_s$ variable to extract the top quark mass.

In this work, we will investigate the $M_{t\bar{t}}$ distribution in top quark pair production, especially its behavior in the threshold region. The $t\bar{t} + \text{jet}$ production process will be studied in a forthcoming article. In perturbation theory, the differential cross section receives corrections from both strong and electroweak (EW) interactions. One can therefore organize the theoretical result as a double series in the strong coupling constant $\alpha_s$ and the fine-structure constant $\alpha$. We will mainly be concerned with strong-interaction contributions described by quantum chromodynamics (QCD). It is possible to incorporate EW effects in the future in a similar way as in \cite{Pagani:2016caq, Czakon:2017wor, Czakon:2019txp}. In QCD, the current benchmark of fixed-order calculations is at the level of next-to-next-to-leading order (NNLO)~\cite{Baernreuther:2012ws, Czakon:2012zr, Czakon:2012pz, Czakon:2013goa, Czakon:2014xsa, Czakon:2015owf, Czakon:2016dgf, Catani:2019iny, Catani:2019hip}. Upon the NNLO result, all-order resummation of soft logarithms \cite{Kidonakis:1996aq, Kidonakis:1997gm, Ahrens:2010zv} combined with resummation of small-mass logarithms \cite{Ferroglia:2012ku, Pecjak:2016nee, Pecjak:2018lif} up to the NNLL$'$ accuracy can be added which improves the theoretical precision, particularly in the high $M_{t\bar{t}}$ (a.k.a. boosted) region. This results in the state-of-the-art QCD prediction at NNLO+NNLL$'$~\cite{Czakon:2018nun}.

\begin{figure}[t!]
\centering
\includegraphics[width=0.7\textwidth]{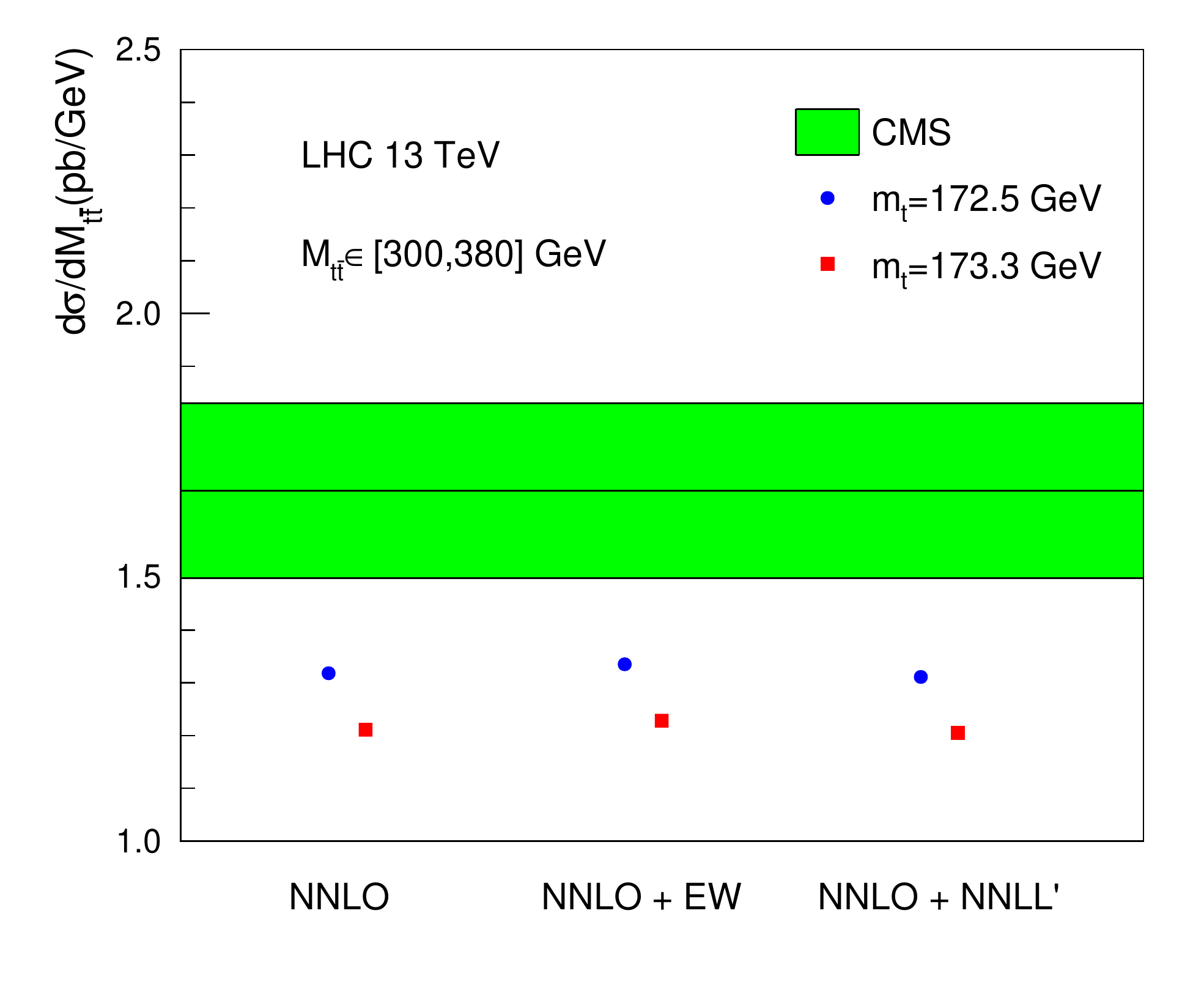}
\vspace{-3ex}
\caption{\label{fig:CMS_vs_old}The averaged $M_{t\bar{t}}$ distribution in the range \unit{$[300,380]$}{\GeV}. The CMS result in the di-lepton channel \cite{Sirunyan:2018ucr} is shown as the green band. The central-values of various theoretical predictions are shown in comparison.}
\end{figure}

The high precision theoretical predictions are compared to the experimental measurements by the ATLAS and CMS collaborations at the \unit{13}{\TeV} LHC in, e.g., Refs.~\cite{Aaboud:2018eqg, Sirunyan:2018wem, Sirunyan:2018ucr, CMS:2018jcg, Aad:2019ntk}. Overall excellent agreement between theory and data is found in almost all phase space regions. However, there exists an interesting discrepancy in the threshold region of the $M_{t\bar{t}}$ distribution found in both the lepton+jets and di-lepton data of the CMS experiment \cite{Sirunyan:2018wem, Sirunyan:2018ucr}. To see that more clearly, we show in Fig.~\ref{fig:CMS_vs_old} the CMS result in the di-lepton channel \cite{Sirunyan:2018ucr} for the averaged $M_{t\bar{t}}$ distribution in the \unit{$[300,380]$}{\GeV} range, where the green band reflects the combined statistical and systematical uncertainty of the experimental measurement. The central values of various theoretical predictions (NNLO from \cite{Czakon:2016dgf, Catani:2019hip}, NNLO+EW from \cite{Czakon:2019txp}, and NNLO+NNLL$'$ from \cite{Czakon:2018nun}) are shown in comparison. It can be seen that there exists a clear gap between the experimental and theoretical results. While this is just a small discrepancy in a vast collection of observables which is normally not very important, the threshold region of the $M_{t\bar{t}}$ distribution is somewhat special since it is strongly sensitive to the top quark mass. This can be easily observed from Fig.~\ref{fig:CMS_vs_old}, where we have shown theoretical predictions using two values of $m_t$: \unit{172.5}{\GeV} (blue points) and \unit{173.3}{\GeV} (red points). Therefore, this small discrepancy has profound implications on the top quark mass measurement. As a matter of fact, such a measurement using the data of \cite{Sirunyan:2018ucr} has already been performed in \cite{Sirunyan:2019zvx}. It is found that the extracted top quark pole mass is around \unit{171}{\GeV} (with an uncertainty of about \unit{0.7}{\GeV}), which is significantly smaller than the current world average $m_t^{\text{pole}} = \unit{$173.1 \pm 0.9$}{\GeV}$ and $m_t^{\text{MC}} = \unit{$172.9 \pm 0.4$}{\GeV}$. The main driving force towards the lower value is exactly the mismatch between theory and data in the threshold region $M_{t\bar{t}} \sim 2m_t$. Note that Ref.~\cite{Sirunyan:2019zvx} has only used an integrated luminosity of \unit{35.9}{\invfb} compared to the full LHC Run 2 dataset of \unit{150}{\invfb}. The Run 3 of the LHC will further collect much more data in the near future. With the large amount of $t\bar{t}$ events, future extractions of the top quark mass will have much smaller experimental uncertainties. One should therefore take this discrepancy seriously if it persists in the future.

It is known that in the threshold region $M_{t\bar{t}} \sim 2m_t$, there is a class of higher-order contributions not included in the current state-of-the-art QCD predictions of Refs.~\cite{Czakon:2016dgf, Pecjak:2016nee, Czakon:2018nun}. They are of the form $\alpha_s^n/\beta^m$ where $\beta \equiv \sqrt{1-4m^2_t/M^2_{t\bar{t}}}$ is the speed of the top quark in the $t\bar{t}$ rest frame. In the threshold region where the top and anti-top quarks are slowly moving with respect to each other, one has $\beta \sim 0$, and the $\alpha_s^n/\beta^m$ contributions are enhanced. These corrections arise from exchanges of Coulomb-like gluons, and can be systematically resummed to all orders in $\alpha_s$ \cite{Fadin:1990wx, Bodwin:1994jh, Petrelli:1997ge, Hagiwara:2008df, Kiyo:2008bv}. A physical effect of this resummation is that the value of $M_{t\bar{t}}$ can be lower than the $2m_t$ threshold, due to bound-state effects caused by the virtual gluon exchanges. In Ref.~\cite{Sirunyan:2019zvx}, the authors use the result of \cite{Kiyo:2008bv} to estimate that these higher-order corrections will lead to a shift of \unit{+0.7}{\GeV} to the extracted $m_t$, which is of the similar size of the total experimental uncertainty. However, there are a few concerns which may invalidate the direct application of the result of \cite{Kiyo:2008bv}. First of all, Ref.~\cite{Kiyo:2008bv} only gives numeric results for $M_{t\bar{t}} \geq \unit{335}{\GeV}$ which does not fully cover the range $M_{t\bar{t}} \geq \unit{300}{\GeV}$ used in the experimental analysis. While the contributions below \unit{335}{\GeV} may not be very important, it is best to be clarified quantitatively.\footnote{Note that the shape of the distribution below $2m_t$ threshold strongly depends on the decay width $\Gamma_t$ of the top quark.} Secondly, the prediction of Ref.~\cite{Kiyo:2008bv} (as well as the first bin of the experimental data) extends to $M_{t\bar{t}} = \unit{380}{\GeV}$, where $\beta \approx 0.4$ is not so small. One should therefore carefully treat the subleading-power contributions in $\beta$ in order not to introduce unrealistic corrections into the theoretical prediction. Last but not least, on top of the small-$\beta$ threshold limit, Ref.~\cite{Kiyo:2008bv} also considers the ``soft'' limit $z \equiv M_{t\bar{t}}^2/\hat{s} \to 1$, where $\sqrt{\hat{s}}$ is the center-of-mass energy of initial-state partons in the hard scattering.\footnote{\label{fn:zsoft}Later we will also study the behaviors of soft gluons in the $\beta \to 0$ limit. To avoid confusion, we will refer to the $z \to 1$ soft limit as the $z$-soft limit, and refer to the $\beta \to 0$ soft limit as the $\beta$-soft limit, respectively.} Given the high energy (\unit{13}{\TeV}) of the LHC compared to $2m_t \approx \unit{345}{\GeV}$, it is necessary to assess the validity of the $z$-soft limit in the current context.

The goal of this paper is two-fold. Firstly, we reexamine the three points raised above. Our main findings can be summarized as following: 1) The contribution from the region $M_{t\bar{t}} \in \unit{[300,335]}{\GeV}$ is about 4\% of the integrated cross section in the bin \unit{[300,380]}{\GeV}, which is non-negligible for current and future high precision measurements;\footnote{However, we note that one should take this result with a grain of salt, since in this deeply-bounded region the on-shell top quark approximation is no longer valid. It is therefore necessary to properly define the variable $M_{t\bar{t}}$ in terms of the decay products, consistently in both experimental and theoretical analyzes. This requires the incorporation of electroweak effects into the framework \cite{Hoang:1999zc, Beneke:2003xh, Beneke:2004km, Hoang:2004tg, Beneke:2010mp, Beneke:2017rdn}.} 2) It is necessary to modify the resummation formula to take into account the subleading power corrections such that the formula is valid up to $M_{t\bar{t}} \sim \unit{380}{\GeV}$; 3) The soft limit $z \to 1$ does not provide a reasonable approximation for the kinematic region of interest, therefore soft resummation should either not be performed, or be applied very carefully. The second goal of this paper is to combine the Coulomb resummation with the NNLO results of \cite{Czakon:2016dgf, Catani:2019hip}, to achieve the best prediction in the threshold region, and to extend the prediction to higher $M_{t\bar{t}}$ values. For that purpose, we need to modify the factorization formula of \cite{Bodwin:1994jh, Petrelli:1997ge, Kiyo:2008bv} to deal with the dynamic renormalization and factorization scales used in the NNLO calculation. We also need to calculate a new hard function with kinematic dependence which is an essential ingredient in our factorization formula. Note that some of the results in this work have already been presented in \cite{Ju:2019mqc}. This paper aims at a more thorough analysis with more technique details and more phenomenological results and discussions.


This paper is organized as follows. In Section~\ref{sec:fac} we discuss the fixed-order QCD corrections for the $M_{t\bar{t}}$ distribution and derive the factorization and resummation formula relevant in the threshold region. In Section~\ref{sec:hard} we calculate the hard function which is an essential ingredient in the factorization formula. We then use these analytic results to perform numeric calculations and present the phenomenological results in Section~\ref{sec:num}. We summarize in Section~\ref{sec:summary} and give additional details in the Appendices.

\section{Fixed-order results and factorization}
\label{sec:fac}

\subsection{Fixed-order results}

In this work we consider the hadronic process
\begin{align}
h_1(P_1) + h_2(P_2) \to t(p_t) + \bar{t}(p_{\bar{t}}) + X_h \, ,
\label{ep:process}
\end{align}
where $h_1$ and $h_2$ are two incoming hadrons, while $X_h$ denotes all final-state particles except the top quark and the anti-top quark. We are mainly interested in the invariant mass of the $t\bar{t}$ pair, which is defined as
\begin{equation}
M_{t\bar{t}}^2 \equiv (p_t + p_{\bar{t}})^2 \, .
\end{equation}
In QCD factorization \cite{Collins:1989gx}, the invariant-mass distribution can be written as a convolution of partonic differential cross sections and non-perturbative parton luminosity functions:
\begin{equation}
\frac{d\sigma}{dM_{t\bar{t}}} = \sum_{i,j} \int^1_{\tau} \frac{dz}{z} \, \frac{\tau}{z} \int d\Theta \, \frac{d\hat{\sigma}_{ij}(z,\mu_f)}{dM_{t\bar{t}} \, d\Theta} \, \ff_{ij}(\tau/z,\mu_f) \, ,
\label{eq:fac1}
\end{equation}
where $i,j \in \{q,\bar{q},g\}$ denote partons within the colliding hadrons; $z \equiv M_{t\bar{t}}^2/\hat{s}$, $\tau \equiv M_{t\bar{t}}^2/s$, with $\sqrt{s}$ and $\sqrt{\hat{s}}$ being the hadronic and partonic center-of-mass energies, respectively; and $\mu_f$ is the factorization scale. The symbol $\Theta$ denotes a collection of extra kinematic variables (other than $m_t$ and $M_{t\bar{t}}$) upon which $\mu_f$ may depend.
The functions $\ff_{ij}(y,\mu_f)$ are the parton luminosity functions defined by
\begin{equation}\label{eq:lumi}
\ff_{ij}(y,\mu_f) \equiv \int^1_{y} \frac{d\xi}{\xi} \, f_{i/h_1}(\xi,\mu_f) \, f_{j/h_2}(y/\xi,\mu_f) \, ,
\end{equation}
where $f_{i/h}$ is the parton distribution function (PDF) of the parton $i$ in the hadron $h$. They are non-perturbative objects which can be extracted from experimental data, and can be obtained using, e.g., the program package \texttt{LHAPDF} \cite{Buckley:2014ana}.

The partonic differential cross sections can be calculated in perturbation theory. In this work, we are concerned with QCD corrections to this quantity. At the leading order (LO) in the strong coupling constant $\alpha_s$, only the $q\bar{q}$ and $gg$ channels give non-vanishing contributions
\begin{align}
\frac{d^2\hat{\sigma}_{q\bar{q}}^{(0)}}{dM_{t\bar{t}} \, d\cos\theta_t} &= \frac{2\pi\beta\alpha_s^2(\mu_r)}{M_{t\bar{t}}^3} \frac{C_FC_A}{9} \, c_{q\bar{q},8}(\cos\theta_t) \, \delta(1-z) \, , \nonumber
\\
\frac{d^2\hat{\sigma}_{gg}^{(0)}}{dM_{t\bar{t}} \, d\cos\theta_t} &= \frac{2\pi\beta\alpha_s^2(\mu_r)}{M_{t\bar{t}}^3} \bigg[ \frac{C_F}{32} \, c_{gg,1}(\cos\theta_t) + \frac{(C_A^2-4)C_F}{64} \, c_{gg,8}(\cos\theta_t) \bigg] \, \delta(1-z) \, ,
\label{eq:sigma0_hat}
\end{align}
where $\mu_r$ is the renormalization scale, $\theta_t$ is the scattering angle of the top quark in the $t\bar{t}$ rest frame (which coincides with the partonic center-of-mass frame at LO). The coefficient functions $c_{ij,\alpha}$, with $\alpha=1,8$ labelling the color configuration of the $t\bar{t}$ system, are given by
\begin{align}
c_{q\bar{q},8}(\cos\theta_t) &= \frac{1}{4} \big[ 2 - \beta^2(1-\cos^2\theta_t) \big] \, , \nonumber
\\
c_{gg,1}(\cos\theta_t) &= \frac{1}{2(1-\beta^2\cos^2\theta_t)^2} \Big[ 4 - 2(1-\beta^2)^2 - 2\beta^2(1-\beta^2\cos^2\theta_t) - (1+\beta^2\cos^2\theta_t)^2 \Big] \, , \nonumber
\\
c_{gg,8}(\cos\theta_t) &= 2c_{gg,1}(\cos\theta_t) \, \bigg[ \frac{16}{5} - \frac{9}{10} (3-\beta^2\cos^2\theta_t) \bigg] \, ,
\label{eq:cijalpha}
\end{align}
with
\begin{equation}
\beta \equiv \sqrt{1-\frac{4m_t^2}{M_{t\bar{t}}^2}} \, .
\end{equation}
Plugging Eq.~\eqref{eq:sigma0_hat} into Eq.~\eqref{eq:fac1}, we obtain the LO hadronic differential cross sections
\begin{align}
\frac{d\sigma^{(0)}_{q\bar{q}}}{dM_{t\bar{t}}} &= \int_{-1}^1 d\cos\theta_t \, \frac{2\pi\beta\alpha_s^2(\mu_r)}{s M_{t\bar{t}}} \frac{C_FC_A}{9} \, c_{q\bar{q},8}(\cos\theta_t) \, \ff_{q\bar{q}}(\tau,\mu_f) \, , \nonumber
\\
\frac{d\sigma^{(0)}_{gg}}{dM_{t\bar{t}}} &= \int_{-1}^1 d\cos\theta_t \,
\frac{2\pi\beta\alpha_s^2(\mu_r)}{s M_{t\bar{t}}} \bigg[ \frac{C_F}{32} \, c_{gg,1}(\cos\theta_t) + \frac{(C_A^2-4)C_F}{64} \, c_{gg,8}(\cos\theta_t) \bigg] \, \ff_{gg}(\tau,\mu_f) \, .
\label{eq:sigma0}
\end{align}

At the next-to-leading order (NLO) and the next-to-next-to-leading order (NNLO) in QCD, there are no analytic formulas for the partonic differential cross sections, and one relies on numeric methods to perform the phase-space integrals as well as loop integrals (at NNLO). The NLO results were calculated in \cite{Nason:1989zy, Mangano:1991jk, Frixione:1995fj}, and can be obtained using the program package \texttt{MCFM} \cite{Campbell:2010ff}. The NNLO results were calculated in \cite{Baernreuther:2012ws, Czakon:2012zr, Czakon:2012pz, Czakon:2013goa, Czakon:2014xsa, Czakon:2015owf, Czakon:2016dgf, Catani:2019iny, Catani:2019hip}, and we obtain the invariant-mass distribution from \cite{Czakon:2016dgf, Catani:2019hip, Britzger:2012bs, Czakon:2017dip}.

Besides the above fixed-order QCD calculations, there are also results implementing all-order resummation of certain classes of large logarithms \cite{Kidonakis:1996aq, Kidonakis:1997gm, Ahrens:2010zv, Ferroglia:2012ku, Pecjak:2016nee, Pecjak:2018lif} as well as electroweak corrections \cite{Beenakker:1993yr, Bernreuther:2005is, Kuhn:2005it,  Bernreuther:2006vg, Kuhn:2006vh, Hollik:2007sw,  Bernreuther:2008md, Bernreuther:2010ny, Hollik:2011ps, Kuhn:2011ri, Manohar:2012rs, Bernreuther:2012sx, Kuhn:2013zoa, Campbell:2015vua, Pagani:2016caq, Czakon:2017wor, Czakon:2017lgo, Gutschow:2018tuk}. We however do not incorporate them into our final predictions in the current work. This can be done in the future straightforwardly using combination methods similar as \cite{Czakon:2019txp}.

\subsection{Factorization near threshold}

In the threshold region $M_{t\bar{t}} \sim 2m_t$, higher order QCD corrections are enhanced by contributions of the form $(\alpha_s/\beta)^n$ as well as $\alpha_s^n\ln^m\beta$, which arise from exchanges of Coulomb-type gluons and soft gluons between the top and anti-top quarks. Using the method of regions, we identify the following relevant momentum regions in the $t\bar{t}$ rest frame:
\begin{align}
\text{hard: } \; &k^\mu \sim M_{t\bar{t}} \, , \nonumber
\\
\text{potential: } \; &k^0 \sim M_{t\bar{t}}\beta^2 \, , \; \vec{k} \sim M_{t\bar{t}}\beta \, , \nonumber
\\
\text{soft: } \; &k^\mu \sim M_{t\bar{t}}\beta \, , \nonumber
\\
\text{ultrasoft: } \; &k^\mu \sim M_{t\bar{t}}\beta^2 \, , \nonumber
\\
\text{collinear: } \; &k^\mu = (\bar{n}_i \cdot k , n_i \cdot k , k_\perp) \sim M_{t\bar{t}}(1, \beta^2, \beta) \, .
\label{eq:modes}
\end{align}
Note that later we will also consider the ultrasoft region in the $z \to 1$ limit, i.e., the $z$-soft limit introduced in footnote~\ref{fn:zsoft} on page~\pageref{fn:zsoft}. That should not be confused with the $\beta$-soft limit here. In the last equation above, the light-like 4-vector $n^\mu_i$ is along the momentum of each massless energetic parton in the initial and final states. The light-like 4-vectors $\bar{n}^\mu_i$ satisfy $n_i \cdot \bar{n}_i = 2$. Later we will show that the collinear modes are irrelevant at the order considered in this work. We nevertheless list them here for completeness.

The momentum modes in Eq.~\eqref{eq:modes} can be described in the language of effective field theories (EFTs). The relevant EFT is potential non-relativistic QCD (pNRQCD) \cite{Pineda:1997bj, Brambilla:1999xf, Beneke:1999zr, Beneke:1999qg}, possibly supplemented by soft-collinear effective theory (SCET) \cite{Bauer:2000ew, Bauer:2000yr, Bauer:2001yt, Beneke:2002ph, Beneke:2002ni}. The EFT of pNRQCD describes interactions among potential, soft and ultrasoft fields, while SCET describes interactions among ultrasoft and collinear fields. Both theories admit a power expansion in the small parameter $\beta \ll 1$. In this work, we will consider the power expansion up to the next-to-leading power (NLP). In order to resum the $(\alpha_s/\beta)^n$ terms up to all orders in $\alpha_s$, pNRQCD adopts an additional power counting $\alpha_s \sim \beta$, such that all $(\alpha_s/\beta)^n$ terms are $\mathcal{O}(1)$ and are incorporated already at the leading power (LP).

We begin with the partonic differential cross section with respect to $M_{t\bar{t}}^2$
\begin{multline}
\frac{d\hat{\sigma}_{ij}}{dM_{t\bar{t}}^2} = \frac{1}{2\hat{s}} \sum_X \int \frac{d^4P_{t\bar{t}}}{(2\pi)^4} \, \delta(P_{t\bar{t}}^2-M_{t\bar{t}}^2) \, (2\pi)^4 \delta^{(4)}(p_1+p_2-P_{t\bar{t}}-P_X)
\\
\times \int \frac{d^4p_t}{(2\pi)^4} \frac{d^4p_{\bar{t}}}{(2\pi)^4} \, (2\pi) \delta(p_t^2-m_t^2) \, (2\pi) \delta(p_{\bar{t}}^2-m_t^2)  \, (2\pi)^4 \delta^{(4)}(P_{t\bar{t}}-p_t-p_{\bar{t}})
\\
\times \big| \mathcal{M}(i+j \to t\bar{t} + X) \big|^2 \, ,
\label{eq:dsigma_dM}
\end{multline}
where the summation over final-state polarization and color indices and the average over initial-state ones are understood. In the $t\bar{t}$ rest frame, the momenta of the top and anti-top quarks are given by
\begin{equation}
p_t^\mu = \frac{M_{t\bar{t}}}{2} v^\mu + \frac{q^\mu}{2} \, , \quad p_{\bar{t}}^\mu = \frac{M_{t\bar{t}}}{2} v^\mu - \frac{q^\mu}{2} \, ,
\label{eq:momenta}
\end{equation}
where $v^\mu=(1,0,0,0)$ and the relative momentum $q^\mu$ behaves as the potential mode in Eq.~\eqref{eq:modes}. The extra radiations $X$ are generically counted as the hard mode in our setup, since we count $1-z = 1-M_{t\bar{t}}^2/\hat{s}$ as an $\mathcal{O}(1)$ quantity. In other words, we do not consider the limit $z \to 1$ besides the threshold limit $\beta \to 0$. The reason will be clear later.

In the $\beta \to 0$ limit, the scattering amplitude in Eq.~\eqref{eq:dsigma_dM} can be described in pNRQCD up to the NLP as
\begin{equation}
\mathcal{M}(i+j \to t\bar{t} + X) = C^{a_1a_2}_{ij,X}(p_1,p_2,P_{t\bar{t}},P_X) \braket{t_{a_1}\bar{t}_{a_2} | \psi^\dagger \chi | 0} \, ,
\label{eq:amp}
\end{equation}
where the fields $\psi$ and $\chi$ are heavy quark fields in pNRQCD describing the top and anti-top quarks, respectively; and $C^{a_1a_2}_{ij,X}$ are Wilson coefficients which encode fluctuations at the hard scale $M_{t\bar{t}}$. They receive contributions from both virtual exchanges and real emissions of hard gluons. They depend on total momentum of the $t\bar{t}$ pair as well as the momenta of other external particles. They also depend on the color indices of the external particles, in particular, the color indices $a_1$ and $a_2$ of the top and anti-top quarks, which are contracted with the corresponding indices of the operator matrix elements in Eq.~\eqref{eq:amp}. The squared amplitude in Eq.~\eqref{eq:dsigma_dM} can then be expressed as
\begin{multline}
\big| \mathcal{M}(i+j \to t\bar{t} + X) \big|^2 = \frac{1}{N_{ij}} \, C^{a_1a_2}_{ij,X}(p_1,p_2,P_{t\bar{t}},P_X) \, C^{a_3a_4\dagger}_{ij,X}(p_1,p_2,P_{t\bar{t}},P_X)
\\
\times \braket{0 | \chi^\dagger \psi | t_{a_3}\bar{t}_{a_4}} \braket{t_{a_1}\bar{t}_{a_2} | \psi^\dagger \chi | 0} \, ,
\label{eq:sqramp}
\end{multline}
where the summation over polarization and color indices are understood, and the $1/N_{ij}$ factor takes into account the average over initial states.

The contraction of color indices in Eq.~\eqref{eq:sqramp} can be simplified by inserting a complete set of orthonormal color projectors $P^\alpha_{\{a\}}$ given by
\begin{equation}
P^1_{a_1a_2a_3a_4} = \frac{1}{3} \delta_{a_1a_2} \delta_{a_3a_4} \, , \quad
P^8_{a_1a_2a_3a_4} = 2 T^c_{a_1a_2} T^c_{a_4a_3} \, ,
\end{equation}
where $\alpha=1,8$ denote the singlet and octet color configurations of the $t\bar{t}$ pair.
We can now define the hard functions as
\begin{align}
H_{ij,\alpha}(z,M_{t\bar{t}},Q_T,Y,\mu_r,\mu_f) &= \frac{z M_{t\bar{t}}^2}{32 \pi^3 \alpha_s^2} \sum_X \int \frac{d^4P_{t\bar{t}}}{(2\pi)^4} \, (2\pi) \,\delta(P_{t\bar{t}}^2-M_{t\bar{t}}^2) \, \delta(P^2_{T,t\bar{t}}-Q^2_T)  \nonumber
\\
&\times \delta(Y_{t\bar{t}}-Y) \, (2\pi)^4 \delta^{(4)}(p_1+p_2-P_{t\bar{t}}-P_X) \nonumber
\\
&\times \frac{1}{N_{ij}} \, P^\alpha_{\{a\}} \, C^{a_1a_2}_{ij,X}(p_1,p_2,P_{t\bar{t}},P_X) \, C^{a_3a_4\dagger}_{ij,X}(p_1,p_2,P_{t\bar{t}},P_X) \, ,
\label{eq:hard}
\end{align}
where $Q_T$ and $Y$ are the transverse momentum and the rapidity of the $t\bar{t}$ pair in the initial-state center-of-mass frame, respectively. The reason for keeping their dependence in the hard functions will be clear later. The hard functions can be calculated in perturbation theory, where both ultraviolet (UV) and infrared (IR) divergences appear. The UV divergences are removed via renormalization. Part of the IR divergences cancels when adding virtual and real contributions, while the remaining collinear divergences are absorbed into the PDFs. After these procedures, the hard functions develop dependencies on the renormalization scale $\mu_r$ and the factorization scale $\mu_f$.

Plugging Eqs.~\eqref{eq:sqramp} and \eqref{eq:hard} into Eq.~\eqref{eq:dsigma_dM}, we find that the remaining integrals are over $p_t$ and $p_{\bar{t}}$, or equivalently, over the potential-scaling relative momentum $q^\mu$ as given in Eq.~\eqref{eq:momenta}. We can then define a potential function describing fluctuations of the potential, soft and ultrasoft modes as
\begin{multline}
J^\alpha(E) = M_{t\bar{t}}^2 \int \frac{d^4p_t}{(2\pi)^4} \frac{d^4p_{\bar{t}}}{(2\pi)^4} \, (2\pi) \delta(p_t^2-m_t^2) \, (2\pi) \delta(p_{\bar{t}}^2-m_t^2)  \, (2\pi)^4 \delta^{(4)}(P_{t\bar{t}}-p_t-p_{\bar{t}})
\\
\times P^\alpha_{\{a\}} \braket{0 | \chi^\dagger \psi | t_{a_3}\bar{t}_{a_4}} \braket{t_{a_1}\bar{t}_{a_2} | \psi^\dagger \chi | 0} \, ,
\label{eq:Coulombfunction}
\end{multline}
where $E \equiv M_{t\bar{t}} - 2m_t$ represents the residue kinetic energy of the top and anti-top quarks in the $t\bar{t}$ rest frame. The partonic differential cross section can then be written in the factorized form up to the NLP:
\begin{multline}
\frac{d\hat{\sigma}_{ij}}{dM_{t\bar{t}} \, d\Theta} = \frac{16\pi^2\alpha_s^2(\mu_r)}{M_{t\bar{t}}^5} \sqrt{\frac{M_{t\bar{t}}+2m_t}{2M_{t\bar{t}}}} \sum_{\alpha} c_{ij,\alpha}(\cos\theta_t)
\\
\times H_{ij,\alpha}(z,M_{t\bar{t}},Q_T,Y, \mu_r,\mu_f) \, J^{\alpha}(E)  + \mathcal{O}(\beta^3) \, ,
\label{eq:fac2}
\end{multline}
where the coefficient functions $c_{ij,\alpha}$ are included such that the leading order expansion of the factorization formula coincides with the exact results in Eq.~\eqref{eq:sigma0_hat}. They are given in Eq.~\eqref{eq:cijalpha} for $(ij,\alpha)=(q\bar{q},8), (gg,1), (gg,8)$, and we choose $c_{ij,\alpha}=1$ for all other cases.
The kinematic variables contained in $\Theta$ include $Q_T^2$, $Y$, as well as $\theta_t$ and $\phi_t$ being the scattering angle and the azimuthal angle of the top quark in the $t\bar{t}$ rest frame.

The formula \eqref{eq:fac2} holds for rather generic choices of $\mu_r$ and $\mu_f$. Near the threshold $M_{t\bar{t}} \sim 2m_t$, it is reasonable to associate the scales to either $m_t$ or $M_{t\bar{t}}$. On the other hand, we have in mind that our results can be extended to a much larger range of $M_{t\bar{t}}$ through a combination with fixed-order calculations \cite{Baernreuther:2012ws, Czakon:2012zr, Czakon:2012pz, Czakon:2013goa, Czakon:2014xsa, Czakon:2015owf, Czakon:2016dgf, Catani:2019iny, Catani:2019hip} and with soft-gluon resummation calculations \cite{Ahrens:2010zv, Ferroglia:2012ku, Pecjak:2016nee, Pecjak:2018lif}. We will therefore also consider the scale choices adopted by those calculations, where the scales are correlated with the variable
\begin{equation}
H_T \equiv \sqrt{p_{T,t}^2+m_t^2} + \sqrt{p_{T,\bar{t}}^2+m_t^2} \, ,
\label{eq:HT}
\end{equation}
where $p_{T,t}$ and $p_{T,\bar{t}}$ are the transverse momenta of the top and anti-top quarks in the initial-state center-of-mass frame. The variable $H_T$ is a (complicated) function of $M_{t\bar{t}}$, $\theta_t$, $\phi_t$, $Q_T$ and $Y$. This is essentially the reason why we need to keep these variables unintegrated in Eq.~\eqref{eq:fac2}, as collected in the symbol $\Theta$.

\subsection{Absence of additional structures up to NLP}

At this point, it is worthwhile to briefly discuss the derivation of the factorization formula \eqref{eq:amp}. Such a factorization is straightforward if one could count all parton exchanges and radiations (except those within the $t\bar{t}$ system) as hard. In this case the only EFT required to describe the process is pNRQCD, and hence the standard matching formula \eqref{eq:amp}. On the other hand, IR divergences appearing at higher orders in perturbation theory may spoil this simple assumption. If that happens, one will need to utilize other EFTs such as the SCET to describe, e.g., the collinear modes, and introduce new structures into the factorization formula. In the following, we will show that such new structures are not required at LP and NLP.

Besides the dynamics described by pNRQCD, the remaining IR divergences arise from soft and/or collinear interactions. The strategy we are going to take is then to use SCET (combined with pNRQCD) to analyze the behavior of the differential cross section in those limits. At LP in $\beta$, the interactions of ultrasoft gluons with initial-state and final-state partons are both governed by the eikonal approximation. The interactions among collinear fields are the same as in the full QCD. The cancellation of soft divergences and final-state collinear divergences therefore follows similarly as the KLN theorem~\cite{Kinoshita:1962ur, Lee:1964is}. The remaining initial-state collinear divergences can be absorbed into the PDFs through factorization~\cite{Collins:1989gx}. Note that the above discussions apply to arbitrary orders in $\alpha_s$ at LP in $\beta$. We will explicitly demonstrate these cancellations through the calculation of the NLO hard functions in the next section.

Using the EFT language, the ultrasoft and collinear interactions are described by the LP Lagrangians of SCET and pNRQCD, written as
\begin{align}
\mathcal{L}_{\text{SCET}}^{0}(x) &= \sum_{n \in \{n_i\}} \bigg[ \bar{\xi}_n \Big( i n \cdot D_n + g_s n \cdot A_{\text{us}} + i \slashed{D}_{n\perp} \frac{1}{i \bar{n} \cdot D_n} i \slashed{D}_{n\perp} \Big) \frac{\slashed{\bar{n}}}{2} \xi_n - \frac{1}{2} \Tr F^{\mu\nu}_n F_{\mu\nu}^n \bigg] \nonumber
\\
&- \frac{1}{2} \Tr F^{\mu\nu}_{\text{us}} F_{\mu\nu}^{\text{us}} \, ,
\label{eq:L0SCET}
\\
\mathcal{L}_{\text{pNRQCD}}^{0}(x) &=\psi^{\dag} \left( i \partial^0 + g_s n \cdot A_{\text{us}}^0 +\frac{\vec{\partial}^2}{2m_t} \right) \psi + \chi^{\dag} \left( i \partial^0 + g_s n \cdot A_{\text{us}}^0 - \frac{\vec{\partial}^2}{2m_t} \right) \chi \nonumber
\\
&- \int \mathrm{d}^3\vec{r} \, \psi^{\dag} T^{a} \psi \big( x^0,\vec{x}+\vec{r} \big) \left( \frac{\alpha_s}{r} \right) \chi^{\dag} T^{a}\chi \big(x^0,\vec{x}\big) \, ,
\label{eq:L0pNRQCD}
\end{align}
where $n^\mu$ takes each of the light-like 4-vectors $n_i^\mu$ along initial-state and final-state massless energetic partons; $\xi_n$ is the collinear quark field along the $n$ direction; $\psi$ and $\chi$ are Pauli spinor fields annihilating the top quark and creating the anti-top quark, respectively; $A_n$ (in the covariant derivative $D_n$) and $A_{\text{us}}$ represent the collinear and ultrasoft gluon fields, with $F^{\mu\nu}_{n(\text{us})}$ their field strength tensors. The ultrasoft eikonal interactions are manifest in the $n \cdot A_{\text{us}}$ terms in the above Lagrangians. One can perform the field redefinitions \cite{Bauer:2001yt, Beneke:2010da}
\begin{gather}
\xi_{n}(x) \to S^q_{n}(x) \xi_{n}(x) \, , \quad A_{n}(x) \to S^g_{n}(x) A_{n}(x) \, , \nonumber
\\
\psi(x) \to S_v(x) \psi(x) \, , \quad \chi(x) \to S_v(x) \chi(x) \, ,
\label{eq:decoupling}
\end{gather}
such that these interactions do not appear explicitly in the LP Lagrangians, where $S_v(x)$ and $S^{q}_{n}(x)$ are ultrasoft Wilson lines in the fundamental representation along the directions implied by the subscripts, while $S^{g}_{n}(x)$ are ultrasoft Wilson lines in the adjoint representation.
These interactions reappear in the effective operators describing the $t\bar{t}$ production process.

The partonic differential cross sections can then be decomposed into a hard sector (containing Wilson coefficients from matching the full QCD to the EFT), a potential sector (containing top and anti-top quarks as well as potential and soft modes), an ultrasoft sector (containing the ultrasoft Wilson lines), and several collinear sectors (containing the collinear fields along each of the incoming and outgoing energetic partons). Within each sector, one needs to perform the well-known multipole expansion \cite{Beneke:2002ph, Beneke:2002ni} to have a uniform power counting in $\beta$. However, the only physical scale which may enter the ultrasoft sector and the collinear sectors is given by the residue momentum $p_1+p_2-P_{t\bar{t}}$, which is counted as hard in our approach. As a result, the loop and phase-space integrals in the ultrasoft sector and the collinear sectors become scaleless and vanish in dimensional regularization. This effectively means that we do not need to consider them at LP in $\beta$ to start with, and hence the differential cross sections are factorized as in Eq.~\eqref{eq:fac2}.

At NLP in $\beta$, we need to consider the subleading Lagrangians of pNRQCD and SCET, as well as the subleading effective operators relevant for the process. The NLP pNRQCD Lagrangians are given by~\cite{Beneke:1999zr, Beneke:1999qg, Kniehl:2002br}
\begin{align}
\mathcal{L}_{\text{pNRQCD}}^{1a}(x) &= -\psi^{\dag}(x) \, g_s \, \vec{x} \cdot \vec{E}_{\text{us}}(x^0,\vec{0}) \, \psi(x) - \chi^{\dag}(x) \, g_s \, \vec{x} \cdot \vec{E}_{\text{us}}(x^0,\vec{0}) \, \chi(x) \, ,
\label{eq:L1apNRQCD}
\\
\mathcal{L}_{\text{pNRQCD}}^{1b}(x) &= - \int \mathrm{d}^3\vec{r} \, \psi^{\dag}T^{a} \psi \big( x^0,\vec{x}+\vec{r} \big) \frac{\alpha^2_s}{4\pi r} \Big[ a_1 + 2\beta_0 \, \ln \big( e^{\gamma_E} \mu r \big) \Big] \chi^{\dag} T^{a}\chi \big( x^0,\vec{x} \big) \, ,
\label{eq:L1bpNRQCD}
\end{align}
where $\vec{E}^{i}_{\text{us}}=F^{i0}_{\text{us}}$ are the chromoelectric components of the ultrasoft field strength tensor. The coefficient $a_1$ was calculated in \cite{Fischler:1977yf, Billoire:1979ih} and is given by $a_1 = 31 C_A/9 - 10 N_l/9$ with $N_l$ light quark flavors. The one-loop coefficient $\beta_0$ of the QCD $\beta$-function is $\beta_0=(11C_A-4T_FN_l)/3$. The NLP SCET Lagrangians are~\cite{Bauer:2000ew, Bauer:2000yr, Bauer:2001yt, Beneke:2002ph, Beneke:2002ni}
\begin{align}
\mathcal{L}_{\text{SCET}}^{1a} &= \bar{\xi}_n \Big(   x_{\perp}^{\mu} n^{\nu} W_n g_s F^{\text{us}}_{\mu\nu} W_{n}^{\dag} \Big)\frac{\slashed{\bar{n}}}{2} \xi_n + (n \leftrightarrow \bar{n}) \, ,
\\
\mathcal{L}_{\text{SCET}}^{1b} &= \Tr \bigg\lbrace n^{\mu} F_{\mu\nu}^n W_n i\left[ x^{\rho}_{\perp} \bar{n}^{\rho} F^{\text{us}}_{\rho\sigma}, W_n^{\dag} \left( iD^{\nu}_{n\perp} W_n \right) \right] W_n^{\dag} \bigg\rbrace - \Tr \Big\lbrace n^{\mu} F_{n}^{\mu\nu_{\perp}} W_n \bar{n}^{\rho} F^{\text{us}}_{\rho\nu_{\perp}} W_n^{\dag} \Big\rbrace \nonumber
\\
&+ (n \leftrightarrow \bar{n}) \, ,
\\
\mathcal{L}_{\text{SCET}}^{1c} &= \bar{\xi}_n i \slashed{D}_{n\perp} W_n q_{\text{us}} + \text{h.c.} + (n \leftrightarrow \bar{n}) \, ,
\end{align}
where $W_n$ is the collinear Wilson line and $q_{\text{us}}$ is the ultrasoft quark field.
It can be shown that single insertions of $\mathcal{L}_{\text{pNRQCD}}^{1a}$ give rise to vanishing results due to angular momentum conservation \cite{Beneke:2009ye, Beneke:2010da, Ju:2019lwp}, while $\mathcal{L}_{\text{SCET}}^{1c}$ does not contribute due to baryon number conservation. The terms in $\mathcal{L}_{\text{pNRQCD}}^{1b}$ involve subleading potentials between the top and anti-top quarks. These contributions can be incorporated by upgrading the potential function $J^\alpha(E)$ to the NLO, which we will discuss in the next subsection.

Besides the NLP Lagrangians which describe the low-energy interactions in the EFTs, we also need to consider the NLP effective operators describing the hard scattering processes. These are constructed out of gauge-invariant building blocks of pNRQCD and SCET fields, with the overall power counting of order $\beta^1$ (the LP operators are of order $\beta^0$). This extra power of $\beta$ comes either from the collinear fields or from the fields in the potential sector. Note that the ultrasoft mode scales as $\beta^2$ and therefore cannot provide a single power of $\beta$. The new operators from the potential sector may lead to new potential functions in addition to the LP one in Eq.~\eqref{eq:Coulombfunction}. For example, there could be contributions from matrix elements of the form
\begin{align}
\braket{0 | \chi^\dagger \psi | t_{a_3}\bar{t}_{a_4}} \braket{t_{a_1}\bar{t}_{a_2} | \psi^\dagger \vec{\partial} \chi | 0} \, .
\end{align}
However, such terms have an odd parity and always lead to a vanishing result when integrating over the phase space as in Eq.~\eqref{eq:Coulombfunction}. For the NLP operators in the collinear sector, and for the single insertions of $\mathcal{L}_{\text{SCET}}^{1a,1b}$, the situation is quite similar. Only the transverse component of a collinear momentum or a collinear gluon field can give rise to an order $\beta^1$ contribution. In the NLP collinear functions (beam or jet functions), one therefore generically encounters integrals similar as
\begin{equation}
\int d^4x \, e^{i q \cdot x} \sum_{X} \braket{ i | \Xi_n^\dagger(x) | X } \braket{ X | \partial_\perp^\mu \Xi_{n}(0) | i }  \propto q_\perp^\mu \, ,
\end{equation}
where $\Xi_n$ represents gauge-invariant building blocks of collinear fields, and $\partial_\perp^\mu$ might be replaced by $x^\mu_\perp$ or $\mathcal{A}^\mu_\perp$. Note that at NLP, such dependence on the transverse component can only appear once. This kind of contributions either vanish trivially, or vanish after phase-space integration. We therefore conclude that the factorization formula \eqref{eq:fac2} is not modified by NLP contributions, except that the potential function $J^\alpha(E)$ should be calculated up to order $\beta$.

\subsection{The perturbative ingredients and resummation at NLP}

The hard functions $H_{ij,\alpha}$ can be expanded in powers of the strong coupling $\alpha_s$:
\begin{equation}
H_{ij,\alpha}(z,M_{t\bar{t}},Q_T,Y,\mu_r,\mu_f) = H^{(0)}_{ij,\alpha} + \frac{\alpha_s(\mu_r)}{4\pi} H^{(1)}_{ij,\alpha} + \cdots \, .
\end{equation}
Due to soft and collinear divergences, $H_{ij,\alpha}$ are singular (in terms of distributions) in both the limits $z \to 1$ and $Q_T \to 0$. We work in dimensional regularization with the spacetime dimension $d=4-2\epsilon$.
The LO hard functions are simply given by
\begin{equation}
\begin{split}
H^{(0)}_{q\bar{q},1} &= h_{q\bar{q},1} = 0 \, ,
\\
H^{(0)}_{q\bar{q},8} &= h_{q\bar{q},8} \, \delta(1-z) \, \delta(Q_T^2) \, \delta(Y)  = \frac{C_A C_F(1-\epsilon)}{9}  \, \delta(1-z) \, \delta(Q^2_T) \, \delta(Y) \, ,
\\
H^{(0)}_{gg,1} &= h_{gg,1} \, \delta(1-z) \, \delta(Q_T^2) \, \delta(Y)  = \frac{C_F(1-2\epsilon)}{32(1-\epsilon)} \, \delta(1-z) \, \delta(Q_T^2) \, \delta(Y) \, ,
\\
H^{(0)}_{gg,8} &= h_{gg,8} \, \delta(1-z) \, \delta(Q_T^2) \, \delta(Y)  = \frac{(C_A^2-4)C_F(1-2\epsilon)}{64(1-\epsilon)} \, \delta(1-z) \, \delta(Q_T^2) \, \delta(Y)  \, ,
\label{eq:H0}
\end{split}
\end{equation}
where we have kept the dependence on $\epsilon$ which is needed for renormalization.
The NLO hard functions are much more complicated, and serve as one of the major new ingredients of this work. We will discuss their calculation in the next section.

We now turn to the potential function $J^\alpha(E)$, which can be related to the imaginary part of the pNRQCD Green function $G^\alpha(\vec{r}_1,\vec{r}_2;E)$ of the $t\bar{t}$ pair at origin \cite{Beneke:2010da}:
\begin{equation}
J^\alpha(E) = 2 \Imag G^\alpha(\vec{0},\vec{0};E) \, .
\end{equation}
Up to the NLP, the potential function can be written as
\begin{equation}
J^\alpha(E) = J_0^\alpha(E) + J_1^\alpha(E) \equiv 2 \Imag G^\alpha_0(\vec{0},\vec{0};E) + 2 \Imag G^\alpha_1(\vec{0},\vec{0};E) \, .
\end{equation}
The Green function can be obtained by solving a differential equation \cite{Beneke:1999qg, Beneke:1999zr, Pineda:2006ri, Beneke:2011mq}. It depends on an additional (hard) scale other than $E$, which is usually chosen as $m_t$. It is equally well to write the Green function in terms of $M_{t\bar{t}}$ and $E$, which corresponds to a reorganization of the power expansion in $\beta$.
Since $M_{t\bar{t}} = 2m_t(1 + \mathcal{O}(\beta^2))$, at NLP it is sufficient to simply replace $m_t \to M_{t\bar{t}}/2$. We can then write the Green function as
\begin{align}
G^\alpha_0(\vec{0},\vec{0}; E) &= \frac{M_{t\bar{t}}^2}{16\pi} \bigg\{ - \sqrt{\frac{-2E}{M_{t\bar{t}}}} + \frac{\alpha_s(\mu_J) D_{\alpha}}{2} \Big[ -2L_J + 2\psi(\lambda) + 2\gamma_{E} - 1 \Big] \bigg\rbrace \, , \nonumber
\\
G^\alpha_1(\vec{0},\vec{0}; E) &= -\frac{M_{t\bar{t}}^2 D_{\alpha} \alpha^2_s(\mu_J)}{64\pi^2} \bigg\lbrace a_1 \Big[ L_J + (1-\lambda) \psi'(\lambda) - \psi(\lambda) - \gamma_E \Big] \nonumber
\\
&\hspace{-4em} + \beta_0 \Big[ L_J^2 + 2L_J \big( (1-\lambda) \psi'(\lambda) - \psi(\lambda) - \gamma_E \big) + 4 {}_4F_3(1,1,1,1;2,2,\lambda;1) \nonumber
\\
&\hspace{-4em} + (1-\lambda) \psi''(\lambda) - 2 (1-\lambda) \big( \psi(\lambda) + \gamma_E \big) \psi'(\lambda) - \frac{\pi^2}{6} - 3\psi'(\lambda) + \big( \psi(\lambda) + \gamma_E \big)^2 \Big] \bigg\rbrace \, .
\end{align}
Here $a_1 = 31 C_A/9 - 10 N_l/9$, $D_1 = -C_F$, $D_8 = 1/(2N_c)$, and
\begin{equation}
L_J = -\frac{1}{2} \ln \bigg( -\frac{2M_{t\bar{t}} E}{\mu_J^2} \bigg) \, , \quad \lambda = 1 + \frac{\alpha_s(\mu_J) D_\alpha}{2\sqrt{-2E/M_{t\bar{t}}}} \, .
\end{equation}
From the form of the logarithm, it appears that the natural choice of the potential scale $\mu_J$ is $\sqrt{2M_{t\bar{t}}E}$. However, as $E$ approaches zero, this scale enters the non-perturbative regime. We therefore follow the prescription in \cite{Beneke:2010da, Ju:2019lwp} to set a lower bound $\mu_J^{\text{cut}}$ for the potential scale. It is set to be the solution to the equation $\mu_J^{\text{cut}} = C_F m_t \alpha_s(\mu_J^{\text{cut}})$, with a numeric value $\mu_J^{\text{cut}} \approx \unit{32}{\GeV}$. Finally, when $E$ is small, the top quark width effect becomes important. To deal with that we replace $E \to E + i\Gamma_t$, where $\Gamma_t \approx \unit{1.4}{\GeV}$.

Combining the hard functions and the potential functions and convoluting with the parton luminosities, we define the NLP resummed hadronic differential cross section as
\begin{multline}
\frac{d\sigma^{\text{NLP}}}{dM_{t\bar{t}}} = \int_\tau^1 \frac{dz}{z} \int_{-1}^1 d\cos\theta_t \int_{0}^{2\pi} \frac{d\phi_t}{2\pi}
\int_0^{Q_{T,\text{max}}^2} dQ^2_T \int^{Y_{\text{max}}}_{-Y_{\text{max}}} dY
\frac{16\pi^2\alpha_s^2(\mu_r)}{s \, M_{t\bar{t}}^3} \, \sqrt{\frac{M_{t\bar{t}}+2m_t}{2M_{t\bar{t}}}}
\\
\times \sum_{ij,\alpha} c_{ij,\alpha}(\cos\theta_t) \, \ff_{ij}(\tau/z,\mu_f) \, \frac{1}{z} \, K^{\text{NLP}}_{ij,\alpha}(z,M_{t\bar{t}},m_t,Q_T,Y,\mu_r,\mu_f) + \mathcal{O}(\beta^3) \, ,
\label{eq:sigma_nlp}
\end{multline}
with the NLP kernel
\begin{equation}
K_{ij,\alpha}^{\text{NLP}}(z,M_{t\bar{t}},m_t,Q_T,Y ,\mu_r,\mu_f) = H^{(0)}_{ij,\alpha} \big( J_0^\alpha(E) + J_1^\alpha(E) \big) + \frac{\alpha_s(\mu_r)}{4\pi} \, H^{(1)}_{ij,\alpha} \, J_0^\alpha(E) \, .
\label{eq:K_nlp}
\end{equation}
In Eq.~\eqref{eq:sigma_nlp}, the integration domain of $Q_T$ and $Y$ is determined by
\begin{equation}
Q_{T,\text{max}} = \frac{M_{t\bar{t}}(1-z)}{2\sqrt{z}} \, , \quad \cosh(Y_{\text{max}}) = \frac{M_{t\bar{t}}(1+z)}{2\sqrt{z}\sqrt{M_{t\bar{t}}^2+Q_T^2}} \, .
\label{eq:max}
\end{equation}
It is evident that in the limit $z\to 1$, where $\hat{s} \to M_{t\bar{t}}^2$, both $Q_T$ and $Y$ must approach zero.

In practice, it is often useful to have the perturbative expansion of the NLP kernel for $E = M_{t\bar{t}}-2m_t > 0$:
\begin{equation}
K_{ij,\alpha}^{\text{NLP}}(z,M_{t\bar{t}},m_t,Q_T,Y ,\mu_r,\mu_f)= \frac{M_{t\bar{t}}^2}{8\pi} \sqrt{\frac{2E}{M_{t\bar{t}}}} \sum_{n=0}^\infty \bigg( \frac{\alpha_s(\mu_r)}{4\pi} \bigg)^n K_{ij,\alpha}^{(n)} \, ,
\label{eq:K_expansion}
\end{equation}
where the coefficients for the first few orders are given by
\begin{align}
K_{ij,\alpha}^{(0)} &= H_{ij,\alpha}^{(0)} \, , \nonumber
\\
K_{ij,\alpha}^{(1)} &= - 2\pi^2 D_\alpha \sqrt{\frac{M_{t\bar{t}}}{2E}} \, H_{ij,\alpha}^{(0)} + H_{ij,\alpha}^{(1)} \, , \nonumber
\\
K_{ij,\alpha}^{(2)} &= \frac{4\pi^4 D_\alpha^2}{3} \frac{M_{t\bar{t}}}{2E} H_{ij,\alpha}^{(0)} + 2\pi^2 D_\alpha \sqrt{\frac{M_{t\bar{t}}}{2E}} \Big[ \big( \beta_0 L_r - a_1 \big) H_{ij,\alpha}^{(0)} - H_{ij,\alpha}^{(1)} \Big] \, , \nonumber
\\
K_{ij,\alpha}^{(3)} &= \frac{4\pi^2 D_\alpha^2}{3} \frac{M_{t\bar{t}}}{2E}
\bigg\{
\Big[ 2\pi^2 a_1 - 2\beta_0 \big( \pi^2 L_r + 12\zeta_3 \big) \Big]  H_{ij,\alpha}^{(0)}
+ \pi^2 H_{ij,\alpha}^{(1)}
\bigg\} \nonumber
\\
&+ 2\pi^2 D_\alpha \sqrt{\frac{M_{t\bar{t}}}{2E}} \, L_{Jr}  \bigg\{ \Big[ \beta_1 + 2a_1\beta_0 + \beta_0^2 \big( L_{Jr} - 2L_r \big) \Big] H_{ij,\alpha}^{(0)} + \beta_0 H_{ij,\alpha}^{(1)} \bigg\} \, , \nonumber
\\
K_{ij,\alpha}^{(4)} &= -\frac{16\pi^8 D_\alpha^4}{45} \bigg( \frac{M_{t\bar{t}}}{2E} \bigg)^2 H_{ij,\alpha}^{(0)}
+ 48\pi^4\zeta_3 \beta_0 D_\alpha^3 \bigg( \frac{M_{t\bar{t}}}{2E} \bigg)^{3/2} H_{ij,\alpha}^{(0)} \nonumber
\\
&- \frac{4\pi^2 D_\alpha^2}{3} \frac{M_{t\bar{t}}}{2E} \, L_{Jr} \bigg\{ \Big[ 2\pi^2 \beta_1 + 6\pi^2 \beta_0 a_1 - 72\zeta_3 \beta_0^2 + 3\pi^2 \beta_0^2 \big( L_{Jr} - 2L_r \big) \Big] H_{ij,\alpha}^{(0)} + 2\pi^2\beta_0 H_{ij,\alpha}^{(1)} \bigg\} \nonumber
\\
&+ \pi^2 D_\alpha \sqrt{\frac{M_{t\bar{t}}}{2E}} \, L_{Jr} \bigg\{ \big( 2\beta_1 - 2\beta_0^2 L_{Jr} \big) H_{ij,\alpha}^{(1)} \nonumber
\\
&\hspace{3em} + \Big[ 2\beta_2 + 4\beta_1a_1 - 4\beta_0^3 L_{Jr}^2 - 4\beta_0\beta_1 L_r + \beta_0 L_{Jr} \big( 6\beta_0^2 L_r - 6\beta_0 a_1 - \beta_1 \big) \Big] H_{ij,\alpha}^{(0)} \bigg\} \, ,
\label{eq:K_i_expansion}
\end{align}
where
\begin{equation}
L_r = \ln \frac{2M_{t\bar{t}} E}{\mu_r^2} \, , \quad L_{Jr} = \ln \frac{\mu_J^2}{\mu_r^2} \, .
\end{equation}
We note that $\sqrt{2E/M_{t\bar{t}}} = \beta + \mathcal{O}(\beta^3)$, and the above expansion makes the $1/\beta$ corrections explicit.

We still need to specify how to perform the integrations in Eq.~\eqref{eq:sigma_nlp}, and how to compute the variable $H_T$ in Eq.~\eqref{eq:HT}. These are in general quite complicated, but are simplified at NLP, where the extra radiation $X$ satisfies $P_X^2=0$.
In this case the transverse momenta of the top and anti-top quarks can be written as
\begin{align}
p_{T,t}^2 &= \frac{1}{4} \, \bigg[ \Big( Q_T + \sqrt{M_{t\bar{t}}^2+Q_T^2} \, \beta \sin\theta_t \cos\phi_t \Big)^2 + \big( M_{t\bar{t}} \beta \sin\theta_t \sin\phi_t \big)^2 \bigg] \, , \nonumber
\\
p_{T,\bar{t}}^2 &= p_{T,t}^2 \bigg|_{\beta \to -\beta} \, .
\label{eq:PTt_initialCMS}
\end{align}
It is then straightforward to compute the variable $H_T$ which enters the scales $\mu_r$ and $\mu_f$. The integrals in Eq.~\eqref{eq:sigma_nlp} can now be performed numerically. The only subtlety is that the NLP kernel $K^{\text{NLP}}_{ij,\alpha}$ contains singular distributions involving $z$, $Q_T$ and $Y$, which arise from the NLO hard functions to be discussed in the next section.

\subsection{Matching with fixed-order results}

The resummed result of Eq.~\eqref{eq:sigma_nlp} contains contributions enhanced by $1/\beta$ or $\ln\beta$ to all orders in $\alpha_s$ at the NLP accuracy. It is possible to add back the $\beta$-power suppressed contributions at NLO and NNLO to achieve a more precise prediction through a matching procedure. This is straightforward given the fixed-order expansion Eq.~\eqref{eq:K_expansion} of the resummation formula. We define the n$^k$LO differential cross sections (with $k=0,1,2,\ldots$) as
\begin{multline}
\frac{d\sigma^{\text{n}^k\text{LO}}}{dM_{t\bar{t}}} = \int_\tau^1 \frac{dz}{z} \int_{-1}^1 d\cos\theta_t \int_{0}^{2\pi} \frac{d\phi_t}{2\pi}
\int_0^{Q_{T,\text{max}}^2} dQ^2_T \int^{Y_{\text{max}}}_{-Y_{\text{max}}} dY
\frac{16\pi^2\alpha_s^2(\mu_r)}{s \, M_{t\bar{t}}^3} \, \sqrt{\frac{M_{t\bar{t}}+2m_t}{2M_{t\bar{t}}}}
\\
\times \sum_{ij,\alpha} c_{ij,\alpha}(\cos\theta_t) \, \ff_{ij}(\tau/z,\mu_f) \, \frac{1}{z} \, \frac{M_{t\bar{t}}^2}{8\pi} \sqrt{\frac{2E}{M_{t\bar{t}}}} \sum_{n=0}^k \bigg( \frac{\alpha_s(\mu_r)}{4\pi} \bigg)^n K_{ij,\alpha}^{(n)} \, .
\label{eq:sigma_nkLO}
\end{multline}
Note that the n$^0$LO cross section is exactly the same as the LO cross section \eqref{eq:sigma0} with our choice of normalization in the resummation formula, while the n$^k$LO cross sections provide approximations to the exact N$^k$LO results (with $\text{N}^1\text{LO} \equiv \text{NLO}$ and $\text{N}^2\text{LO} \equiv \text{NNLO}$). The validity of these approximations is very important for applying the resummation, which we will study numerically in Section~\ref{sec:num}. At the moment, we just note that the difference
\begin{align}
\frac{d\sigma^{\text{N}^k\text{LO}}}{dM_{t\bar{t}}} - \frac{d\sigma^{\text{n}^k\text{LO}}}{dM_{t\bar{t}}}
\end{align}
contains $\beta$-power suppressed contributions beyond NLP at N$^k$LO, which are exactly what we would like to incorporate through the matching procedure. The matching formula is then simply given by
\begin{equation}
\label{eq:matching_formula}
\frac{d\sigma^{\text{(N)NLO+NLP}}}{dM_{t\bar{t}}} = \frac{d\sigma^{\text{NLP}}}{dM_{t\bar{t}}} - \frac{d\sigma^{\text{(n)nLO}}}{dM_{t\bar{t}}} + \frac{d\sigma^{\text{(N)NLO}}}{dM_{t\bar{t}}} \, ,
\end{equation}
where $\text{nLO} \equiv \text{n}^1\text{LO}$ and $\text{nnLO} \equiv \text{n}^2\text{LO}$ as defined in Eq.~\eqref{eq:sigma_nkLO}.
The matched results at NLO+NLP and NNLO+NLP precisions are then our main results in this paper, based on which we will present our best predictions in Section~\ref{sec:num}. Before going into that, we first perform the calculation of the hard functions at NLO in the next section.

\section{The hard functions at NLO}
\label{sec:hard}

In this section, we discuss the calculation of the NLO hard functions, which were not available in the literature. The hard functions receive contributions from both virtual gluon exchanges and real emission subprocesses.
We first consider one-loop virtual corrections where no extra radiation is present. As a result they must be proportional to the tree-level results in Eq.~\eqref{eq:H0}. We generate the one-loop amplitudes using \texttt{FeynArts}~\cite{Hahn:2000kx}, manipulate them with \texttt{FeynCalc}~\cite{Mertig:1990an, Shtabovenko:2016sxi, Shtabovenko:2020gxv}, and reduce the relevant integrals to a set of master integrals using \texttt{Reduze2}~\cite{vonManteuffel:2012np}. The calculation of the master integrals is straightforward and we collect the results in Appendix~\ref{sec:integrals}.
Supplemented with the trivial one-body phase space integral, the bare virtual contributions to the NLO hard functions can be written as
\begin{equation}
\begin{split}
H^{(1),V,\text{bare}}_{q\bar{q},1} &= 0 \, ,
\\
H^{(1),V,\text{bare}}_{q\bar{q},8} &= H^{(0)}_{q\bar{q},8} \bigg[ -\frac{16}{3 \epsilon^2} - \frac{16L_M-44+4N_l}{3\epsilon} - \frac{8L_M^2}{3} + \bigg( \frac{44}{3} - \frac{4N_l}{3} \bigg) L_M
\\
&\hspace{4em} + \frac{280}{9} - \frac{20N_l}{9} + \frac{64\ln(2)}{3} + \frac{\pi^2}{9} \bigg] \, ,
\\
H^{(1),V,\text{bare}}_{gg,1} &= H^{(0)}_{gg,1} \bigg( -\frac{12}{\epsilon^2} - \frac{12L_M}{\epsilon} - 6 L_M^2 - \frac{44}{3} + \frac{16\pi^2}{3} \bigg)  \, ,
\\
H^{(1),V,\text{bare}}_{gg,8} &= H^{(0)}_{gg,8} \bigg( -\frac{12}{\epsilon^2} - \frac{12L_M+6}{\epsilon} - 6 L_M^2 - 6 L_M - \frac{8}{3} + \frac{23\pi^2}{6} \bigg) \, ,
\end{split}
\end{equation}
where $L_M=\ln(\mu_r^2/M_{t\bar{t}}^2)$. Note that we have put in the numerical values of the color factors $C_F=4/3$ and $C_A=N_c=3$ here and below for simplicity. The above results contain both UV and IR divergences. The UV ones are removed by renormalization. We renormalize the fields and the top quark mass in the on-shell scheme, and renormalize the strong coupling in the $\overline{\text{MS}}$ scheme with the top quark integrated out and $N_l=5$ active flavors. We collect the relevant renormalization constants in Appendix~\ref{sec:integrals}.
After renormalization, we get the UV-finite virtual contributions as follows:
\begin{equation}
\begin{split}
H^{(1),V}_{q\bar{q},1} &= 0 \, ,
\\
H^{(1),V}_{q\bar{q},8} &= H^{(0)}_{q\bar{q},8} \bigg[ -\frac{16}{3 \epsilon^2} - \frac{16L_M+42}{3\epsilon} - \frac{8 L_M^2}{3} + \bigg( 8 - \frac{4N_l}{3} \bigg) L_M
\\
&\hspace{4em} + \frac{184}{9} - \frac{20N_l}{9} + 8\ln(2) + \frac{\pi^2}{9} \bigg] \, ,
\\
H^{(1),V}_{gg,1} &= H^{(0)}_{gg,1} \bigg( -\frac{12}{\epsilon^2} - \frac{36L_M+66-4N_l}{3\epsilon} - 6 L_M^2 - \frac{44}{3} + \frac{16\pi^2}{3} \bigg) \, ,
\\
H^{(1),V}_{gg,8} &= H^{(0)}_{gg,8} \bigg( -\frac{12}{\epsilon^2} - \frac{36L_M+84-4N_l}{3\epsilon} - 6 L_M^2 - 6 L_M - \frac{8}{3} + \frac{23\pi^2}{6} \bigg) \, .
\label{eq:virtual}
\end{split}
\end{equation}

We now turn to the real emission subprocesses
\begin{align}
  i(p_1) + j(p_2) \to t\bar{t}(P_{t\bar{t}}) + X(k) \, .
\end{align}
The sum over $X$ in the definition Eq.~\eqref{eq:hard} of the hard function now involves integrating over the momentum $k$. This leads to the two-body phase-space integral
\begin{align}
\Phi_2 &= \mu_r^{2\epsilon} \int \frac{d^d k}{(2\pi)^d} (2\pi) \delta^+(k^2)
\frac{d^d P_{t\bar{t}}}{(2\pi)^d} (2\pi)\delta^+(P_{t\bar{t}}^2-M^2_{t\bar{t}}) \nonumber
\\
&\hspace{4em} \times \delta(P_{T,t\bar{t}}^2-Q_T^2) \, \delta \bigg( Y - \frac{1}{2} \ln \frac{P_{t\bar{t}}^0+P_{t\bar{t}}^3}{P_{t\bar{t}}^0-P_{t\bar{t}}^3} \bigg) \, (2\pi)^d  \delta^{(d)}(p_1+p_2-P_{t\bar{t}}-k) \, .
\label{eq:phi2}
\end{align}
At NLO, the kinematic variables either do not appear in the Wilson coefficient in Eq.~\eqref{eq:hard}, or are fixed by the delta functions in Eq.~\eqref{eq:phi2}. Therefore the whole integral can be carried out which leads to
\begin{align}
\Phi_2 &= \frac{(4\pi)^{\epsilon}}{16\pi\Gamma(1-\epsilon)} \left( \frac{\mu_r^2}{Q_T^2} \right)^{\epsilon} \frac{1+z}{M_{t\bar{t}}^2+Q_T^2} \, \delta \bigg( \cosh^2(Y) - \frac{M_{t\bar{t}}^2(1+z)^2}{4z(M_{t\bar{t}}^2+Q_T^2)} \bigg) \nonumber
\\
&= \frac{(4\pi)^{\epsilon}}{16\pi\Gamma(1-\epsilon)} \left( \frac{\mu_r^2}{Q_T^2} \right)^{\epsilon} \frac{1+z}{M_{t\bar{t}}^2+Q_T^2} \, \frac{\delta(Y-Y_{\text{max}}) + \delta(Y+Y_{\text{max}}) }{\sinh(2Y_{\text{max}})} \, ,
\label{eq:phi2r}
\end{align}
where $Y_{\text{max}}$ is a function of $Q_T$ and $z$ defined in Eq.~\eqref{eq:max}, satisfying
\begin{equation}
\cosh(Y_{\text{max}}) = \frac{M_{t\bar{t}}(1+z)}{2\sqrt{z}\sqrt{M_{t\bar{t}}^2+Q_T^2}} \, , \quad
\sinh(Y_{\text{max}}) = \sqrt{\frac{Q_{T,\text{max}}^2 - Q_T^2}{M_{t\bar{t}}^2 +Q_T^2}} \, ,
\end{equation}
where again $Q_{T,\text{max}}$ is defined in Eq.~\eqref{eq:max} as a function of $z$. Later we will often invoke the value of $Y_{\text{max}}$ at $Q_T = 0$. It therefore deserves a separate symbol which we write as
\begin{equation}
Y_{\text{max},0} \equiv Y_{\text{max}}(Q_T = 0) = -\frac{1}{2} \ln z \, ,
\end{equation}
which satisfies
\begin{equation}
\sinh(Y_{\text{max},0}) = \frac{1-z}{2\sqrt{z}} \, , \quad
\cosh(Y_{\text{max},0}) = \frac{1+z}{2\sqrt{z}} \, , \quad
\sinh(2Y_{\text{max},0}) = \frac{(1+z)(1-z)}{2z} \, .
\label{eq:Ymax0}
\end{equation}
The Wilson coefficients in the definition \eqref{eq:hard} of the hard functions are divergent in the limits $z \to 1$ and $Q_T \to 0$ which correspond to soft and collinear singularities. These singularities are regularized in dimensional regularization by the factor of $Q_T^{-2\epsilon}$ appearing in Eq.~\eqref{eq:phi2r}. In practice, it is useful to write
\begin{equation}
\left( \frac{\mu_r^2}{Q_T^2} \right)^{\epsilon} = \left( \frac{\mu_r^2}{M_{t\bar{t}}^2} \right)^{\epsilon} \left( \frac{Q_{T,\text{max}}^2}{Q_T^2} \right)^{\epsilon} 4^\epsilon \, z^{\epsilon} \, (1-z)^{-2\epsilon} \, .
\end{equation}
One can then perform the expansion in $\epsilon$ using
\begin{align}
(1-z)^{-1-2\epsilon} &=  -\frac{1}{2\epsilon} \delta(1-z) + \left( \frac{1}{1-z} \right)_+ - 2\epsilon \left[ \frac{\ln(1-z)}{1-z} \right]_+ + \cdots \, , \nonumber
\\
\frac{1}{Q_T^2} \left( \frac{Q_{T}^2}{Q_{T,\text{max}}^2} \right)^{-\epsilon} &= -\frac{1}{\epsilon} \delta(Q_T^2) + \left( \frac{1}{Q_T^2} \right)_* - \epsilon \left( \frac{1}{Q_T^2} \ln\frac{Q_T^2}{Q_{T,\text{max}}^2} \right)_* + \cdots \, ,
\end{align}
where the plus-distributions and star-distributions satisfy
\begin{align}
\int_0^1 dz \left[ \frac{\ln^n(1-z)}{1-z} \right]_+ f(z) &= \int_0^1 dz \, \frac{\ln^n(1-z)}{1-z} \, \big[ f(z) - f(1) \big] \, , \nonumber
\\
\int_0^{Q_{T,\text{max}}^2} dQ_T^2 \left( \frac{1}{Q_T^2} \ln^n\frac{Q_T^2}{Q_{T,\text{max}}^2} \right)_* f(Q_T^2) &= \int_0^{Q_{T,\text{max}}^2} \frac{dQ_T^2}{Q_T^2} \ln^n\frac{Q_T^2}{Q_{T,\text{max}}^2} \, \big[ f(Q_T^2) - f(0) \big] \, ,
\end{align}
for some test functions $f(z)$ and $f(Q_T^2)$.

It will be convenient to introduce the scattering angle $\theta$ of the $t\bar{t}$ pair in the partonic center-of-mass frame. It satisfies the relations
\begin{align}
Q_T = \frac{M_{t\bar{t}}(1-z)}{2\sqrt{z}} \sin\theta \, , \quad Y = \frac{1}{2} \ln \frac{1+z+(1-z)\cos\theta}{1+z-(1-z)\cos\theta} \, .
\label{eq:QTY}
\end{align}
The inverse relation reads
\begin{equation}
y \equiv \cos\theta = \frac{1+z}{1-z} \tanh(Y) \, .
\label{eq:costh}
\end{equation}
Using the delta functions in Eq.~\eqref{eq:phi2r}, it can be further expressed as
\begin{equation}
y \equiv \cos\theta = \pm \frac{1+z}{1-z} \tanh(Y_{\text{max}}) \, .
\end{equation}
It should be stressed that while there is a factor of $1-z$ in the denominator above, the value of $y$ is well-defined in the limit $z \to 1$. In fact, it is easy to see from Eq.~\eqref{eq:Ymax0} that
\begin{equation}
y \equiv \cos\theta \xrightarrow{Q_T \to 0} \pm 1 \, ,
\end{equation}
where the sign depends on the sign of $Y = \pm Y_{\text{max}}$. We further introduce a few abbreviations to shorten the expressions:
\begin{align}
\delta^Y_{\text{max}} &= \frac{1}{M_{t\bar{t}}^2+Q_T^2} \, \frac{\delta(Y-Y_{\text{max}}) + \delta(Y+Y_{\text{max}}) }{\sinh(2Y_{\text{max}})} \nonumber \, ,
\\
\delta^Y_{\text{max},0} &= \delta(Y-Y_{\text{max},0}) + \delta(Y+Y_{\text{max},0}) \, , \nonumber
\\
d(Q_T,z) &= \delta^Y_{\text{max}} \,\frac{ M_{t\bar{t}}^2 (1+z) (1-z)}{2z} \left( \frac{1}{Q_T^2} \right)_* \left( \frac{1}{1-z} \right)_+ \, .
\label{eq:abbrev}
\end{align}
The reason to include a factor of $1-z$ in the last equation is that the combination of $\delta^Y_{\text{max}}$ and $(1/Q_T^2)_*$ will produce a singularity as $z \to 1$ upon integration over $Y$ and $Q_T^2$. This can be easily seen from the integral
\begin{equation}
\int_0^{Q_{T,\text{max}}^2}
\frac{dQ_T^2}{Q_T^2} \left( \frac{Q_T^2}{Q_{T,\text{max}}^2} \right)^{-\epsilon} \frac{1}{\sqrt{Q_{T,\text{max}}^2-Q_T^2}} = \frac{1}{Q_{T,\text{max}}} \frac{\Gamma(1-\epsilon) \, \Gamma(-\epsilon)}{4^\epsilon \, \Gamma(1-2\epsilon)} \, .
\end{equation}
This singularity has to be cancelled by a corresponding factor of $1-z$ in the numerator, and we therefore include that factor explicitly here. This will help to identify the leading singular terms in the $z \to 1$ limit later.

The results of the hard functions will also involve the one-loop splitting functions given by
\begin{equation}
\begin{split}
P_{qq}^{(0)}(z) &= 2C_F \bigg[ \frac{1+z^2}{(1-z)_+} + \frac{3}{2} \delta(1-z) \bigg] \, ,
\\
P_{gg}^{(0)}(z) &= 4C_A \bigg[ \frac{z}{(1-z)_+} + \frac{1-z}{z} + z(1-z) \bigg] + \bigg( \frac{11}{3} C_A - \frac{4}{3} T_F N_l \bigg) \delta(1-z) \, ,
\\
P_{qg}^{(0)}(z) &= 2T_F \big[ z^2 + (1-z)^2 \big] \, ,
\\
P_{gq}^{(0)}(z) &= 2C_F \frac{1+(1-z)^2}{z} \, .
\end{split}\label{eq:spfunc}
\end{equation}
We can now write the real emission contributions as
\begin{align}
H_{q\bar{q},1}^{(1),R} &= \frac{2}{27} \, (1+y^2) z^2 (1+z) \, \delta^Y_{\text{max}}  \, , \nonumber
\\
H_{q\bar{q},8}^{(1),R} &= \bigg[ \frac{16}{3\epsilon^2} + \frac{1}{\epsilon} \bigg( \frac{16}{3} L_M + 14 \bigg) + \frac{8}{3} L_M^2 + 14 L_M + 12 + \frac{4\pi^2}{9} \bigg] \, H_{q\bar{q},8}^{(0)} \nonumber
\\
&- \bigg( \frac{1}{\epsilon} + L_M \bigg) z P_{qq}^{(0)}(z) \, h_{q\bar{q},8} \,  \delta(Q_T^2) \, \delta^Y_{\text{max},0} \nonumber
\\
&+ \frac{16z}{3} \bigg[ \frac{1-z}{2} - (1+z) \ln \frac{1-z}{2\sqrt{z}} + 2 \left( \frac{1}{1-z} \ln\frac{1-z}{2\sqrt{z}} \right)_+ \bigg] h_{q\bar{q},8} \, \delta(Q_T^2) \, \delta^Y_{\text{max},0} \nonumber
\\
&+ \frac{z}{54} \Big[ 8(7+9y^2)z + (1+y^2)(19+13y^2)(1-z)^2 - 5(1-y^4)(1-z)^3 \Big] d(Q_T,z) \, ,  \nonumber
\\
H_{gg,1}^{(1),R} &= \bigg[ \frac{12}{\epsilon^2} + \frac{1}{\epsilon} \bigg( 12 L_M + 22-\frac{4 N_l}{3} \bigg) + 6 L_M^2 + \bigg(22-\frac{4 N_l}{3}\bigg) L_M +\pi^2 \bigg] \, H_{gg,1}^{(0)} \nonumber
\\
&- \bigg( \frac{1}{\epsilon} + L_M \bigg) z P_{gg}^{(0)}(z) \, h_{gg,1} \,  \delta(Q_T^2) \, \delta^Y_{\text{max},0} \nonumber
\\
&+ 24z \bigg[  \bigg( \frac{(1-z)^2}{z} - z^2 \bigg) \ln \frac{1-z}{2\sqrt{z}} +  \left( \frac{1}{1-z} \ln\frac{1-z}{2\sqrt{z}} \right)_+ \bigg] h_{gg,1} \, \delta(Q_T^2) \, \delta^Y_{\text{max},0} \nonumber
\\
&+ \frac{z^2\cosh^4(Y)}{288(1+z)^4} \bigg[ 9(1-y^2)^2(3+y^2)^2 (1-z)^6 - 16(1-y^2)(61+4y^2+7y^4) (1-z)^5 \nonumber
\\
&\quad + 8(529-343y^2+107y^4-5y^6) (1-z)^4 - 32 (281-146y^2+9y^4) (1-z)^3 \nonumber
\\
&\quad + 16 (615-182y^2-y^4) (1-z)^2 - 1152(5-y^2)(1-z) + 2304 \bigg] \, d(Q_T,z) \, ,  \nonumber
\\
H_{gg,8}^{(1),R} &= \bigg[ \frac{12}{\epsilon^2} + \frac{1}{\epsilon} \bigg( 12 L_M + 28-\frac{4 N_l}{3} \bigg) + 6 L_M^2 + \bigg(28-\frac{4 N_l}{3}\bigg) L_M + 12 +\pi^2 \bigg] \, H_{gg,8}^{(0)} \nonumber
\\
&- \bigg( \frac{1}{\epsilon} + L_M \bigg) z P_{gg}^{(0)}(z) \, h_{gg,8} \,  \delta(Q_T^2) \, \delta^Y_{\text{max},0} \nonumber
\\
&+ 24z \bigg[  \bigg( \frac{(1-z)^2}{z} - z^2 \bigg) \ln \frac{1-z}{2\sqrt{z}} +  \left( \frac{1}{1-z} \ln\frac{1-z}{2\sqrt{z}} \right)_+ \bigg] h_{gg,8} \, \delta(Q_T^2) \, \delta^Y_{\text{max},0} \nonumber
\\
&+ \frac{z\cosh^4(Y)}{192(1+z)^4} \Big[ (1-y^2)^2 (1-z)^2 - 4(1-y^2) (1-z) + 4(3+y^2) \Big] \nonumber
\\
&\quad \times \Big[ -15(3+y^2)^2 (1-z)^5 + (713+46y^2-39y^4) (1-z)^4 - 8(161-11y^2) (1-z)^3 \nonumber
\\
&\qquad + 8(169-19y^2) (1-z)^2 - 720(1-z) + 240 \Big] \, d(Q_T,z) \nonumber \, ,
\\
H_{qg,1}^{(1),R} &= \bigg[ \frac{z P_{gq}^{(0)}(z)}{24} \bigg( -\frac{1}{\epsilon} - L_M + 2 \ln\frac{1-z}{2\sqrt{z}} + 1 \bigg) + \frac{z^2}{9} \bigg] \, \delta(Q_T^2) \, \delta(Y+Y_{\text{max},0}) \nonumber
\\
&+ \frac{z(1-y)}{36(1+z)[1+\tanh(Y)]^2} \big[ 4 + (1-z)^2(1-y)^2 \big]  \, \delta^Y_{\text{max}} \, M_{t\bar{t}}^2 \, (1-z) \left( \frac{1}{Q_T^2} \right)_* \, , \nonumber
\\
H_{qg,8}^{(1),R} &= \bigg[ \frac{5z P_{gq}^{(0)}(z)}{48} \bigg( -\frac{1}{\epsilon} - L_M + 2 \ln\frac{1-z}{2\sqrt{z}} + 1 \bigg) + \frac{5z^2}{18} \bigg] \, \delta(Q_T^2) \, \delta(Y+Y_{\text{max},0}) \nonumber
\\
&+ \bigg[ \frac{4z P_{qg}^{(0)}(z)}{9} \bigg( -\frac{1}{\epsilon} - L_M + 2 \ln\frac{1-z}{2\sqrt{z}} \bigg) + \frac{4z}{9} \bigg] \, \delta(Q_T^2) \, \delta(Y-Y_{\text{max},0}) \nonumber
\\
&+ \frac{1}{72(1+z)[1+\tanh(Y)]^2} \Big[ 2(1-y)^2(1+y)(5+2y+y^2)(1-z)^4 \nonumber
\\
&\quad - (1-y)(8-9y-6y^2-9y^3)(1-z)^3 + (49+29y+35y^2+15y^3) (1-z)^2 \nonumber
\\
&\quad - 4(12+11y+9y^2)(1-z) + 52+12y \Big] \, \delta^Y_{\text{max}} \, M_{t\bar{t}}^2 \, (1-z) \left( \frac{1}{Q_T^2} \right)_* \, .
\label{eq:real}
\end{align}

Combining the virtual contributions in Eq.~\eqref{eq:virtual} and the real contributions in Eq.~\eqref{eq:real}, the soft divergences cancel according to the KLN theorem. However, there are still collinear divergences remaining. These divergences must be absorbed into the PDFs, which is equivalent to adding the following counter-terms
\begin{align}
H^{(1),C}_{ij,\alpha} &= z \, \delta(Q_T^2) \, \frac{1}{\epsilon} \left(\frac{\mu_r^2}{\mu_f^2} \right)^{\epsilon} \sum_{k} \Big[ \delta(Y-Y_{\text{max},0}) P_{kj}^{(0)}(z) \, h_{ik,\alpha} + \delta(Y+Y_{\text{max},0}) P_{ki}^{(0)}(z) \, h_{kj,\alpha} \Big] \, .
\end{align}
Finally, we obtain the UV and IR finite NLO hard functions:
\begin{align}
H_{q\bar{q},1}^{(1)} &= \frac{2}{27} \, (1+y^2) z^2 (1+z) \, \delta^Y_{\text{max}}  \, , \nonumber
\\
H_{q\bar{q},8}^{(1)} &=\bigg( 2\beta_0 \, L_M + \frac{292}{9} + 8\ln(2) - \frac{20N_l}{9} + \frac{5\pi^2}{9} \bigg) \, H_{q\bar{q},8}^{(0)} -  L_f z P_{qq}^{(0)}(z) \, h_{q\bar{q},8} \,  \delta(Q_T^2) \, \delta^Y_{\text{max},0} \nonumber
\\
&+ \frac{16z}{3} \bigg[ \frac{1-z}{2} - (1+z) \ln \frac{1-z}{2\sqrt{z}} + 2 \left( \frac{1}{1-z} \ln\frac{1-z}{2\sqrt{z}} \right)_+ \bigg] h_{q\bar{q},8} \, \delta(Q_T^2) \, \delta^Y_{\text{max},0} \nonumber
\\
&+ \frac{z}{54} \Big[ 8(7+9y^2)z + (1+y^2)(19+13y^2)(1-z)^2 - 5(1-y^4)(1-z)^3 \Big] d(Q_T,z) \, ,  \nonumber
\\
H_{gg,1}^{(1)} &=\bigg( 2\beta_0 \, L_M - \frac{44}{3} + \frac{19\pi^2}{3} \bigg) \, H_{gg,1}^{(0)} -  L_f z P_{gg}^{(0)}(z) \, h_{gg,1} \,  \delta(Q_T^2) \, \delta^Y_{\text{max},0} \nonumber
\\
&+ 24z \bigg[  \bigg( \frac{(1-z)^2}{z} - z^2 \bigg) \ln \frac{1-z}{2\sqrt{z}} + \left( \frac{1}{1-z} \ln\frac{1-z}{2\sqrt{z}} \right)_+ \bigg] h_{gg,1} \, \delta(Q_T^2) \, \delta^Y_{\text{max},0} \nonumber
\\
&+ \frac{z^2\cosh^4(Y)}{288(1+z)^4} \bigg[ 9(1-y^2)^2(3+y^2)^2 (1-z)^6 - 16(1-y^2)(61+4y^2+7y^4) (1-z)^5 \nonumber
\\
&\quad + 8(529-343y^2+107y^4-5y^6) (1-z)^4 - 32 (281-146y^2+9y^4) (1-z)^3 \nonumber
\\
&\quad + 16 (615-182y^2-y^4) (1-z)^2 - 1152(5-y^2)(1-z) + 2304 \bigg] \, d(Q_T,z) \, ,  \nonumber
\\
H_{gg,8}^{(1)}&= \bigg(2\beta_0 \, L_M  + \frac{28}{3} + \frac{29\pi^2}{6} \bigg) \, H_{gg,8}^{(0)} -  L_f z P_{gg}^{(0)}(z) \, h_{gg,8} \,  \delta(Q_T^2) \, \delta^Y_{\text{max},0} \nonumber
\\
&+ 24z \bigg[ \bigg( \frac{(1-z)^2}{z} - z^2 \bigg) \ln \frac{1-z}{2\sqrt{z}} +  \left( \frac{1}{1-z} \ln\frac{1-z}{2\sqrt{z}} \right)_+ \bigg] h_{gg,8} \, \delta(Q_T^2) \, \delta^Y_{\text{max},0} \nonumber
\\
&+ \frac{z\cosh^4(Y)}{192(1+z)^4} \Big[ (1-y^2)^2 (1-z)^2 - 4(1-y^2) (1-z) + 4(3+y^2) \Big] \nonumber
\\
&\quad \times \Big[ -15(3+y^2)^2 (1-z)^5 + (713+46y^2-39y^4) (1-z)^4 - 8(161-11y^2) (1-z)^3 \nonumber
\\
&\qquad + 8(169-19y^2) (1-z)^2 - 720(1-z) + 240 \Big] \, d(Q_T,z) \nonumber \, ,
\\
H_{qg,1}^{(1)} &= \bigg[ \frac{z P_{gq}^{(0)}(z)}{24} \bigg(-L_f + 2 \ln\frac{1-z}{2\sqrt{z}}  \bigg) + \frac{z^2}{9} \bigg] \, \delta(Q_T^2) \, \delta(Y+Y_{\text{max},0}) \nonumber
\\
&+ \frac{z(1-y)}{36(1+z)[1+\tanh(Y)]^2} \big[ 4 + (1-z)^2(1-y)^2 \big]  \, \delta^Y_{\text{max}} \, M_{t\bar{t}}^2 \, (1-z) \left( \frac{1}{Q_T^2} \right)_* \, , \nonumber
\\
H_{qg,8}^{(1)} &= \bigg[ \frac{5z P_{gq}^{(0)}(z)}{48} \bigg( - L_f + 2 \ln\frac{1-z}{2\sqrt{z}} \bigg) + \frac{5z^2}{18} \bigg] \, \delta(Q_T^2) \, \delta(Y+Y_{\text{max},0}) \nonumber
\\
&+ \bigg[ \frac{4z P_{qg}^{(0)}(z)}{9} \bigg( - L_f + 2 \ln\frac{1-z}{2\sqrt{z}} \bigg) + \frac{8z^2}{9}(1-z) \bigg] \, \delta(Q_T^2) \, \delta(Y-Y_{\text{max},0}) \nonumber
\\
&+ \frac{1}{72(1+z)[1+\tanh(Y)]^2} \Big[ 2(1-y)^2(1+y)(5+2y+y^2)(1-z)^4 \nonumber
\\
&\quad - (1-y)(8-9y-6y^2-9y^3)(1-z)^3 + (49+29y+35y^2+15y^3) (1-z)^2 \nonumber
\\
&\quad - 4(12+11y+9y^2)(1-z) + 52+12y \Big] \, \delta^Y_{\text{max}} \, M_{t\bar{t}}^2 \, (1-z) \left( \frac{1}{Q_T^2} \right)_* \, ,
\label{eq:HnloQTY}
\end{align}
where $L_f=\ln(\mu^2_f/M^2_{t\bar{t}})$. The above expressions, when integrated over $Q_T$ and $Y$ as in Eq.~\eqref{eq:sigma_nlp}, can be rewritten in terms of integrations over $y$. Namely, we may define $\tilde{H}_{ij,\alpha}$ as functions of $y$ which satisfy
\begin{multline}
\int_0^{Q_{T,\text{max}}^2} dQ^2_T \int^{Y_{\text{max}}}_{-Y_{\text{max}}} dY \, H_{ij,\alpha}(z,M_{t\bar{t}},Q_T,Y,\mu_r,\mu_f) \, f(Q_T^2,Y)
\\
= \int_{-1}^1 dy \, \tilde{H}_{ij,\alpha}(z,M_{t\bar{t}},y,\mu_r,\mu_f) \, f(Q_T^2,Y) \, ,
\end{multline}
for a test function $f(Q_T^2,Y)$, where on the right side one should understand that $Q_T$ and $Y$ are determined by $y$ and $z$ through Eq.~\eqref{eq:QTY}. It is straightforward to obtain $\tilde{H}_{ij,\alpha}$ from the expressions of $H_{ij,\alpha}$, Eq.~\eqref{eq:H0} and \eqref{eq:HnloQTY}, by the following replacements:
\begin{align}
\delta(1-z) \, \delta(Q_T^2) \, \delta(Y) &\to \delta(1-z) \, \frac{\delta(1-y) + \delta(1+y)}{2} \, , \nonumber
\\
\delta(Q_T^2) \, \delta(Y \pm Y_{\text{max},0}) &\to \delta(1 \pm y) \, , \nonumber
\\
\delta^Y_{\text{max}} \, M_{t\bar{t}}^2 \, (1-z) \left( \frac{1}{Q_T^2} \right)_* &\to \frac{2z}{1+z}  \bigg[ \theta(y) \, \bigg( \frac{1}{(1-y)_+} + \frac{1}{1+y} + \ln(2) \, \delta(1-y) \bigg) \nonumber
\\
&\hspace{3em} + \theta(-y) \, \bigg( \frac{1}{(1+y)_+} + \frac{1}{1-y} +  \ln(2) \, \delta(1+y) \bigg) \bigg] \, , \nonumber
\\
\delta^{Y}_{\text{max}} &\to \frac{1-z}{1+z} \,  .
\end{align}
To illustrate the idea, we give the results for the $q\bar{q}$ channel:
\begin{align}
\tilde{H}_{q\bar{q},1}^{(1)} &= \frac{2}{27} \, (1+y^2) z^2 (1-z) \, , \nonumber
\\
\tilde{H}_{q\bar{q},8}^{(0)} &= h_{q\bar{q},8} \, \delta(1-z) \, \frac{\delta(1-y) + \delta(1+y)}{2} = \frac{C_AC_F}{9} \, \delta(1-z) \, \frac{\delta(1-y) + \delta(1+y)}{2} \, , \nonumber
\\
\tilde{H}_{q\bar{q},8}^{(1)}&=\bigg( 2\beta_0 \, L_M + \frac{292}{9} + 8\ln(2) - \frac{20N_l}{9} + \frac{5\pi^2}{9} \bigg) \, \tilde{H}_{q\bar{q},8}^{(0)} \nonumber
\\
&-  L_f z P_{qq}^{(0)}(z) \, h_{q\bar{q},8} \, \big[ \delta(1-y) + \delta(1+y) \big] \nonumber
\\
&+ \frac{16z}{3} \bigg[ \frac{1-z}{2} - (1+z) \ln \frac{1-z}{\sqrt{2z}} + 2 \left( \frac{1}{1-z} \ln\frac{1-z}{\sqrt{2z}} \right)_+ \bigg] h_{q\bar{q},8} \, \big[ \delta(1-y) + \delta(1+y) \big] \nonumber
\\
&+ \frac{z}{54} \Big[ 8(7+9y^2)z + (1+y^2)(19+13y^2)(1-z)^2 - 5(1-y^4)(1-z)^3 \Big] \nonumber
\\
&\hspace{1em} \times \bigg[ \theta(y) \bigg( \frac{1}{(1-y)_+} + \frac{1}{1+y} \bigg) + \theta(-y) \, \bigg( \frac{1}{(1+y)_+} + \frac{1}{1-y} \bigg) \bigg] \left( \frac{1}{1-z} \right)_+ \, .
\label{eq:HnloCosth}
\end{align}

The results for the NLO hard functions serve as an important ingredient in the factorization formula at NLP. Combining them with the other ingredients, we are now ready to perform various numerical analyses, which is the main topic of the next section.

\section{Numerical results and discussions}
\label{sec:num}

In this section, we use our resummation formula to carry out several numerical studies and present phenomenologically relevant results. We will discuss in more detail the three points raised in the Introduction concerning the difference between our result and the result of Ref.~\cite{Kiyo:2008bv}. Throughout this section we take $\Gamma_t = \unit{1.4}{\GeV}$, use the NNPDF3.1 NNLO PDFs \cite{Ball:2017nwa} with $\alpha_s(m_Z) = 0.118$, and set the renormalization scale $\mu_r$ to be the same as the factorization scale $\mu_f$. The default scale is chosen to be $H_T/4$, if not otherwise stated. To estimate the scale uncertainties of the differential cross sections, the two scales are varied simultaneously up and down by a factor of 2.

\subsection{Validity of the threshold approximation}

Any factorization and resummation formula is only valid in kinematic regions where higher order power corrections are small compared to the required accuracy. It is therefore necessary to check the validity of the relevant approximation in the region of interest before performing the resummation. One way to do that is to compare the fixed-order expansion of the resummation formula against the exact perturbative results. In the region of validity, the expansion should provide reasonable approximations to the exact results order-by-order.

In this subsection we carry out the validity check of our resummation formula in the region $\unit{300}{\GeV} \leq M_{t\bar{t}} \leq \unit{380}{\GeV}$ at the \unit{13}{\TeV} LHC. This is straightforward since we already have the fixed-order expansion of the resummation formula in Eq.~\eqref{eq:sigma_nkLO}. We just need to check whether the n$^k$LO results are good approximations to the exact N$^k$LO ones. We first note that due to our normalization of the factorization formula Eq.~\eqref{eq:fac2}, the n$^0$LO result (i.e., the first term in the fixed-order expansion) is precisely the same as the exact LO one in Eq.~\eqref{eq:sigma0}. The factorization formula of Ref.~\cite{Kiyo:2008bv}, on the other hand, has a different normalization than ours. Consequently, the first term of their expansion would not be the same as the exact LO. The difference, of course, is formally power-suppressed in $\beta$, but it has significant impact on the validity of the formula when $\beta$ is not so small, e.g., when $M_{t\bar{t}} \sim \unit{380}{\GeV}$.

\begin{figure}[t!]
\centering
\includegraphics[width=0.49\textwidth]{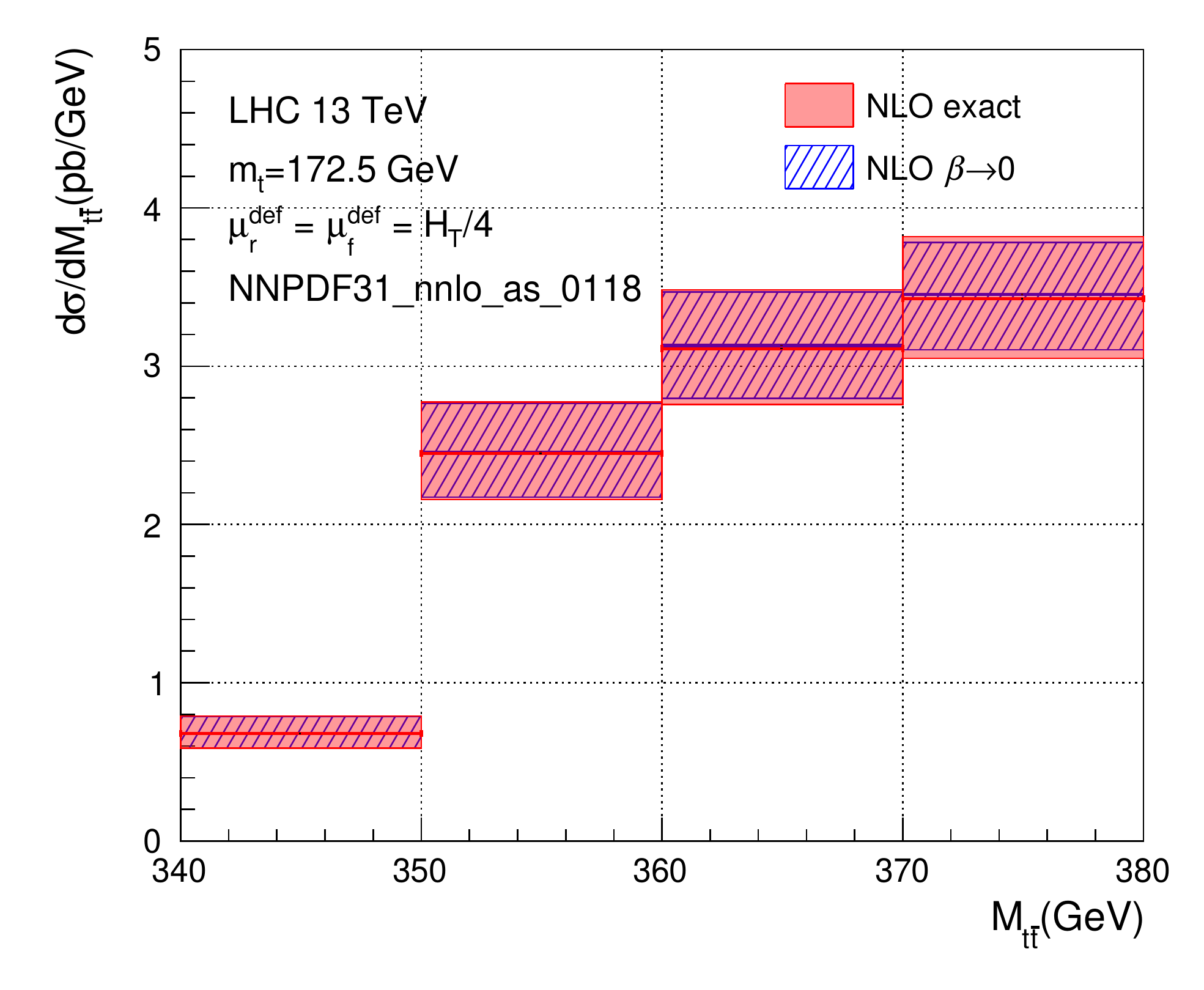}
\includegraphics[width=0.49\textwidth]{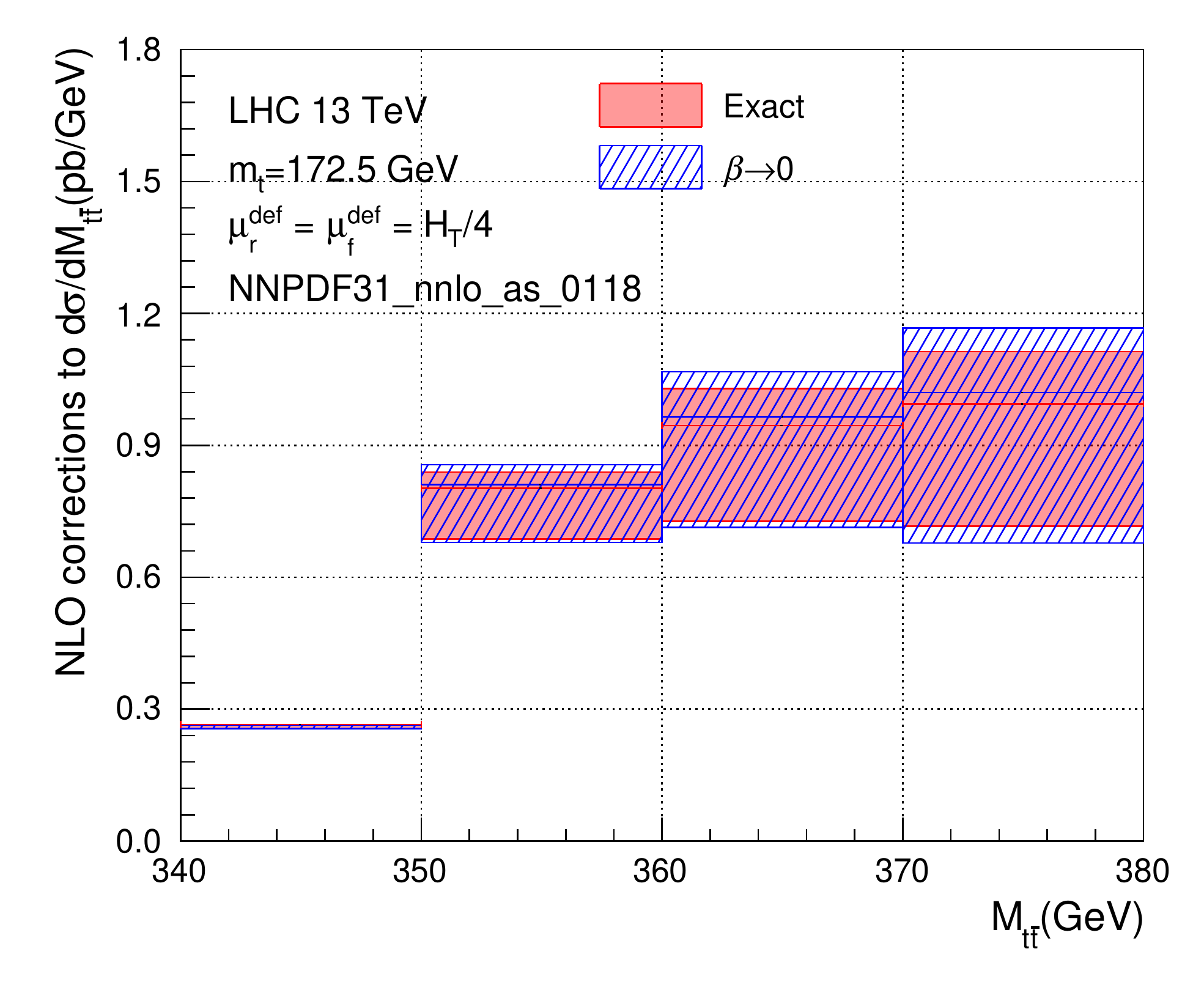}
\caption{Comparison between the exact NLO distribution (red band) and the NLO expansion of our resummation formula (blue shaded band) in the range $\unit{340}{\GeV} \leq M_{t\bar{t}} \leq \unit{380}{\GeV}$ at the \unit{13}{\TeV} LHC. The left plot shows the differential cross sections, while the right plot shows the NLO corrections only.}
\label{fig:NLOvsnLO}
\end{figure}

We now proceed to perform the comparison at NLO. We show the exact NLO $M_{t\bar{t}}$ distribution in the range \unit{[340-380]}{\GeV} in the left plot of Fig.~\ref{fig:NLOvsnLO} as the red band, while the nLO one from the expansion (labelled as ``NLO $\beta \to 0$'') is shown as the blue shaded band. It can be clearly seen that the nLO result provides an excellent approximation to the exact NLO one in the whole range, including scale variations. Since both the NLO and nLO results include the common LO term, it is interesting to compare just the corrections (i.e, the second term in the perturbative series). We show this comparison in the right plot of Fig.~\ref{fig:NLOvsnLO}. Again, the agreement is remarkable. The plot also shows clearly that the deviation between the two results gradually increases from small $\beta$ to larger $\beta$, but remains under-control even when $M_{t\bar{t}}$ is as large as \unit{380}{\GeV}. The agreement we just observed is a strong implication for the validity of the resummation formula Eq.~\eqref{eq:fac2} in the region of interest. We emphasize again that such an agreement is only possible due to the fact that we have correctly taken into account the subleading-power contributions in $\beta$ at LO in $\alpha_s$. If we had used a different normalization factor, the agreement at the upper edge of the region of interest would not be as good.

At this point, it is worthwhile to discuss the $z$-soft limit where $z \equiv M_{t\bar{t}}^2/\hat{s} \to 1$. Such a limit in the context of the $M_{t\bar{t}}$ distribution has been extensively studied in the literature~\cite{Kidonakis:1996aq, Kidonakis:1997gm, Ahrens:2010zv}. By taking this limit it is possible to resum logarithms of $1-z$ to all orders in $\alpha_s$, at the price that power corrections in $1-z$ are neglected. As such, it can be expected that this limit works better at larger values of $M_{t\bar{t}}$ than the threshold region. Furthermore, Ref.~\cite{Kiyo:2008bv} employed the double limit $\beta \to 0$ \emph{and} $z \to 1$ to perform a soft gluon resummation (on top of the Coulomb resummation), which neglects power corrections in both $\beta$ and $1-z$. Given the high collision energy of the LHC compared to the values of $M_{t\bar{t}}$ we are considering (hence $z$ is not necessarily close to 1), and the fact that $\beta$ is not so small at $M_{t\bar{t}} \sim \unit{380}{\GeV}$, one must carefully check the validity of such a double approximation in the region of interest.

\begin{figure}[t!]
\centering
\includegraphics[width=0.49\textwidth]{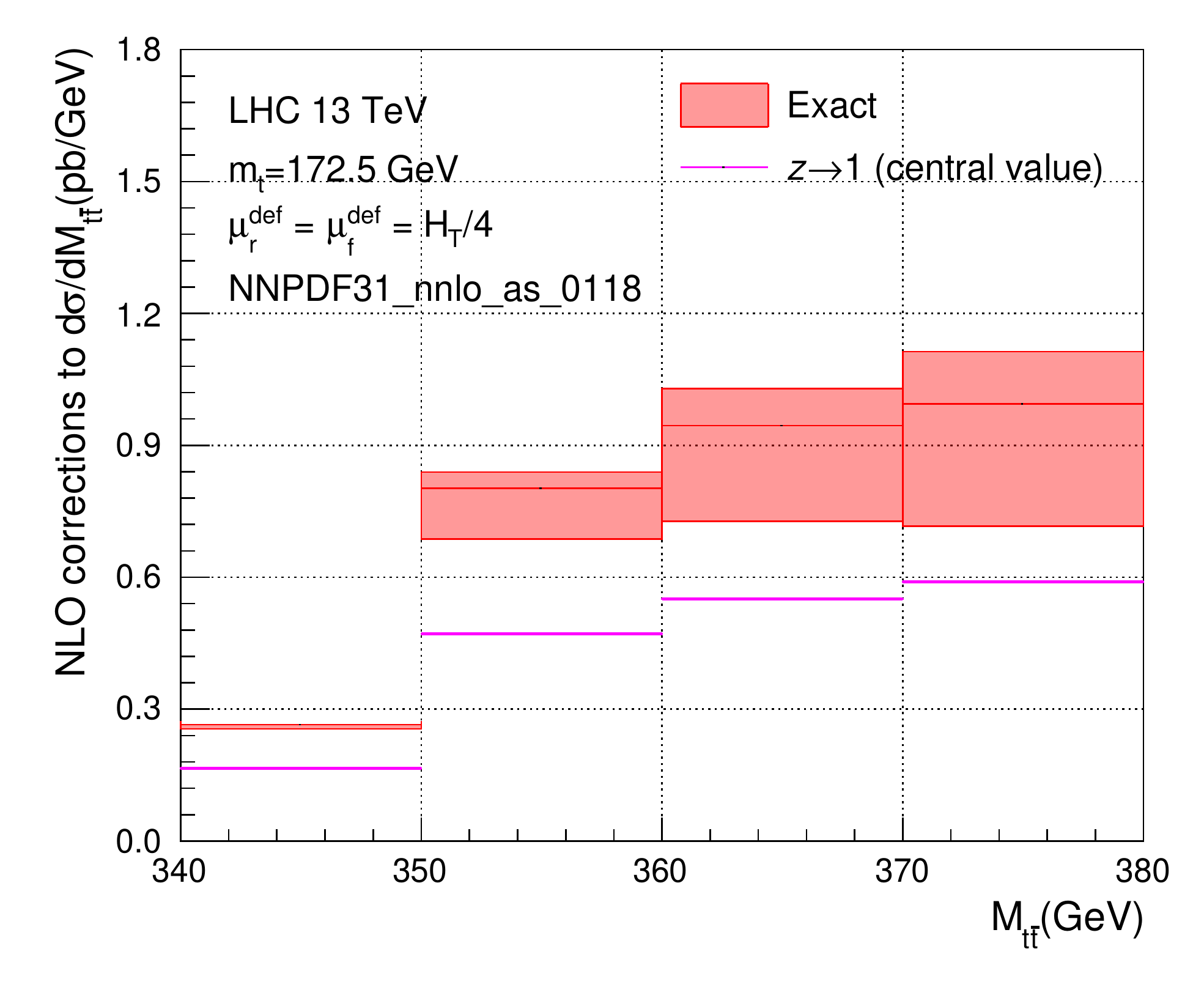}
\includegraphics[width=0.49\textwidth]{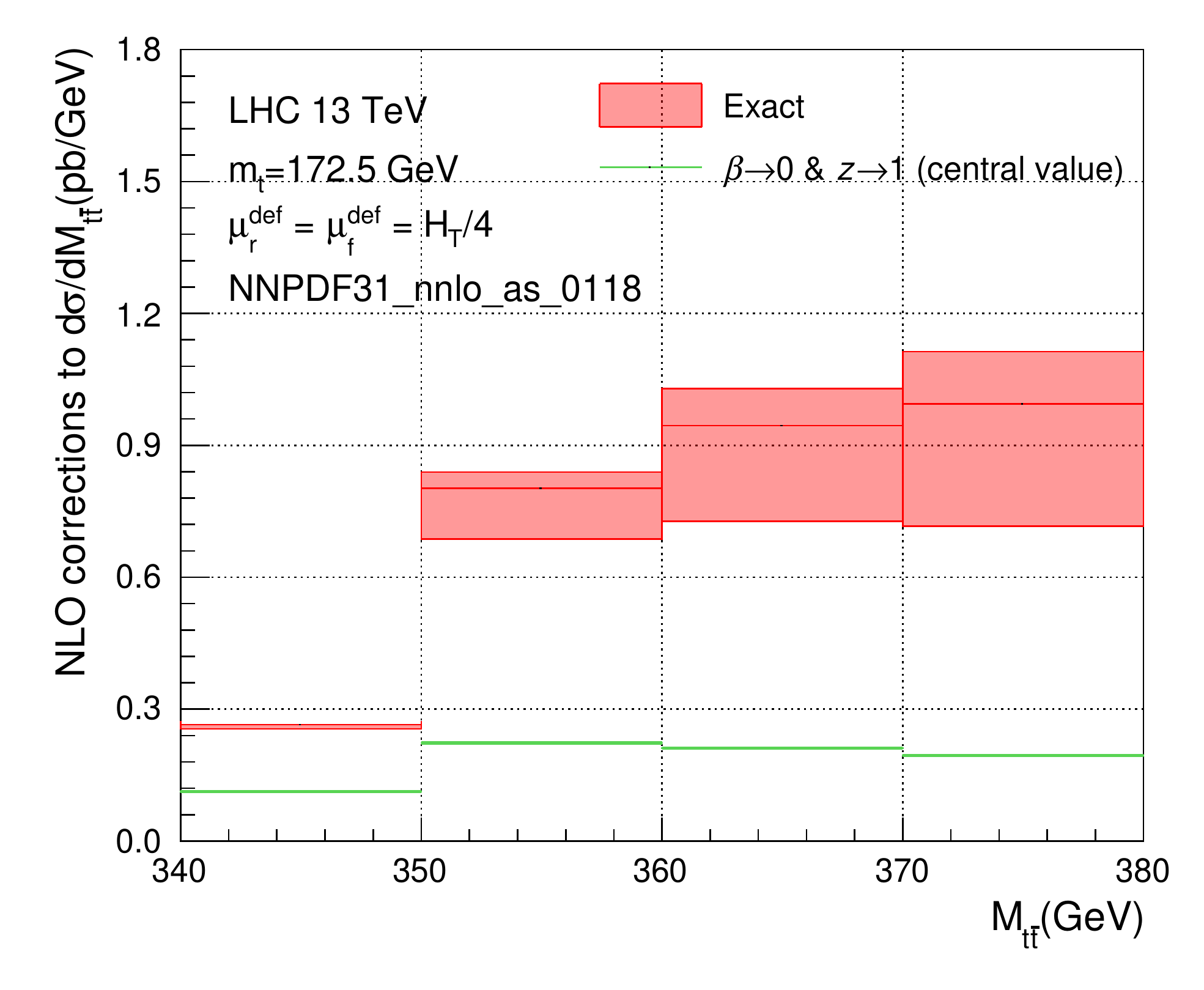}
\caption{\label{fig:zsoft}Comparison of the exact NLO corrections (red band) with the approximate results in the $z \to 1$ limit (pink line in the left plot) and in the double limit $\beta \to 0$ \& $z \to 1$ (green line in the right plot). For the approximate results, only the central values are shown.}
\end{figure}

The NLO result in the $z \to 1$ limit can be obtained from \cite{Ahrens:2010zv}. The result in the double limit $\beta \to 0$ and $z \to 1$ can be obtained from our formula Eq.~\eqref{eq:fac2} by further taking $z \to 1$. This amounts to keeping only the singular plus- and delta-distributions in the hard functions, which is straightforward given their expressions in Eq.~\eqref{eq:HnloQTY}. In this limit, only the flavor-diagonal channels (i.e., the $q\bar{q}$ and $gg$ channels) contribute. We collect the relevant analytic expressions in Appendix~\ref{sec:soft_limit}, and show the numeric results in Fig.~\ref{fig:zsoft}. Note that for the approximate results only the central values are shown. In the left plot, we compare the exact NLO corrections with that in the $z$-soft limit $z \to 1$. We see that although the agreement is not so good (as expected), the $z$-soft limit still captures a dominant portion of the NLO corrections. This is a justification for the application of the soft gluon resummation to this region as in \cite{Ahrens:2010zv, Pecjak:2016nee, Czakon:2018nun}. On the other hand, the NLO corrections in the double limit $\beta \to 0$ and $z \to 1$ are shown in the right plot of Fig.~\ref{fig:zsoft}. It is obvious that the double limit does not provide a reasonable approximation at all. Therefore, the factorization formula valid in the double limit cannot be applied to the region we are considering. Although such a factorization formula can be used to resum certain logarithmic terms to all orders in $\alpha_s$, they are not the dominant contributions and such a resummation may even lead to incorrect estimation of higher order corrections. In other words, the power corrections in $1-z$ are not under-control in this situation and consequently the results cannot be trusted. Based on the above observations, we do not perform the $z$-soft gluon resummation in the $\beta \to 0$ limit in our work, in contrast to \cite{Kiyo:2008bv}.\footnote{Note however that Ref.~\cite{Kiyo:2008bv} implemented the $z$-soft limit in the Mellin space (where the $z$-soft limit corresponds to $N \gg 1$ with $N$ the Mellin moment) instead of the momentum space. It is known that these two differ by terms suppressed by powers of $1-z$ (or equivalently powers of $1/N$). In the current context where $z$ is not so close to 1 (namely, $N$ is not so large), this difference could actually be rather big numerically. We discuss about this in Appendix~\ref{sec:soft_limit}.}

\subsection{NLP Resummation at \unit{13}{\TeV} LHC}
\label{sec:lhc13}

Given the perfect agreement between the approximate ($\beta \to 0$) and exact results at NLO, we will apply the small-$\beta$ resummation at NLP to the range $\unit{300}{\GeV} \leq M_{t\bar{t}} \leq \unit{380}{\GeV}$ at the \unit{13}{\TeV} LHC. Our starting point is the matching formula Eq.~\eqref{eq:matching_formula}, which combines the all-order resummation with the fixed-order results at NLO or NNLO. We will compare our numeric predictions with the experimental data \cite{Sirunyan:2018ucr}, and therefore we use $m_t=\unit{172.5}{\GeV}$ in accordance. In this subsection and the subsequent ones, whenever we present numeric results for a broader range of $M_{t\bar{t}}$, it should always be understood that the resummation is only applied to $M_{t\bar{t}} \leq \unit{380}{\GeV}$. We have checked that the results are insensitive to the exact point at which resummation is switched off. This should be clear from the analyses below.

\begin{figure}[t!]
\centering
\includegraphics[width=0.49\textwidth]{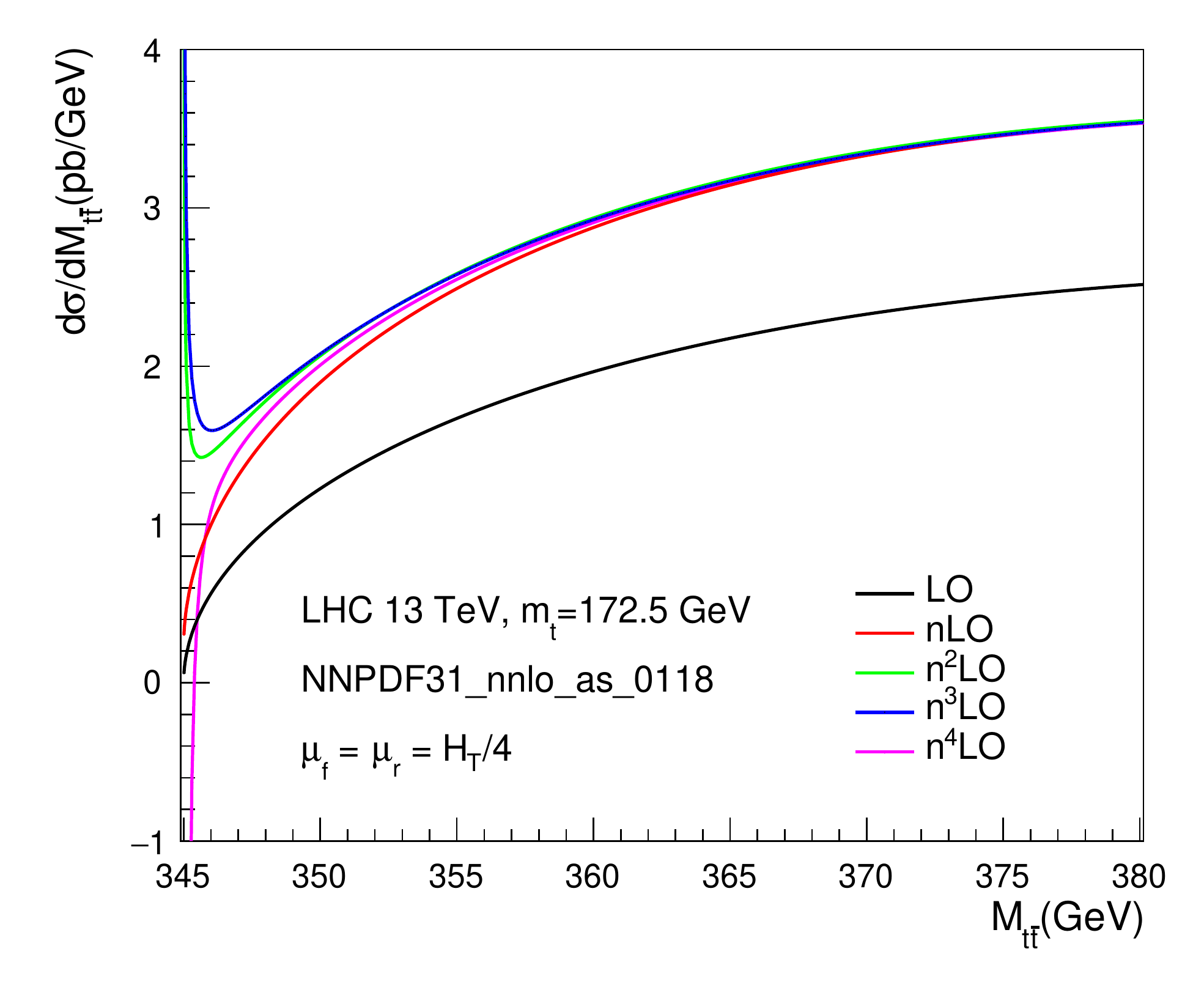}
\includegraphics[width=0.49\textwidth]{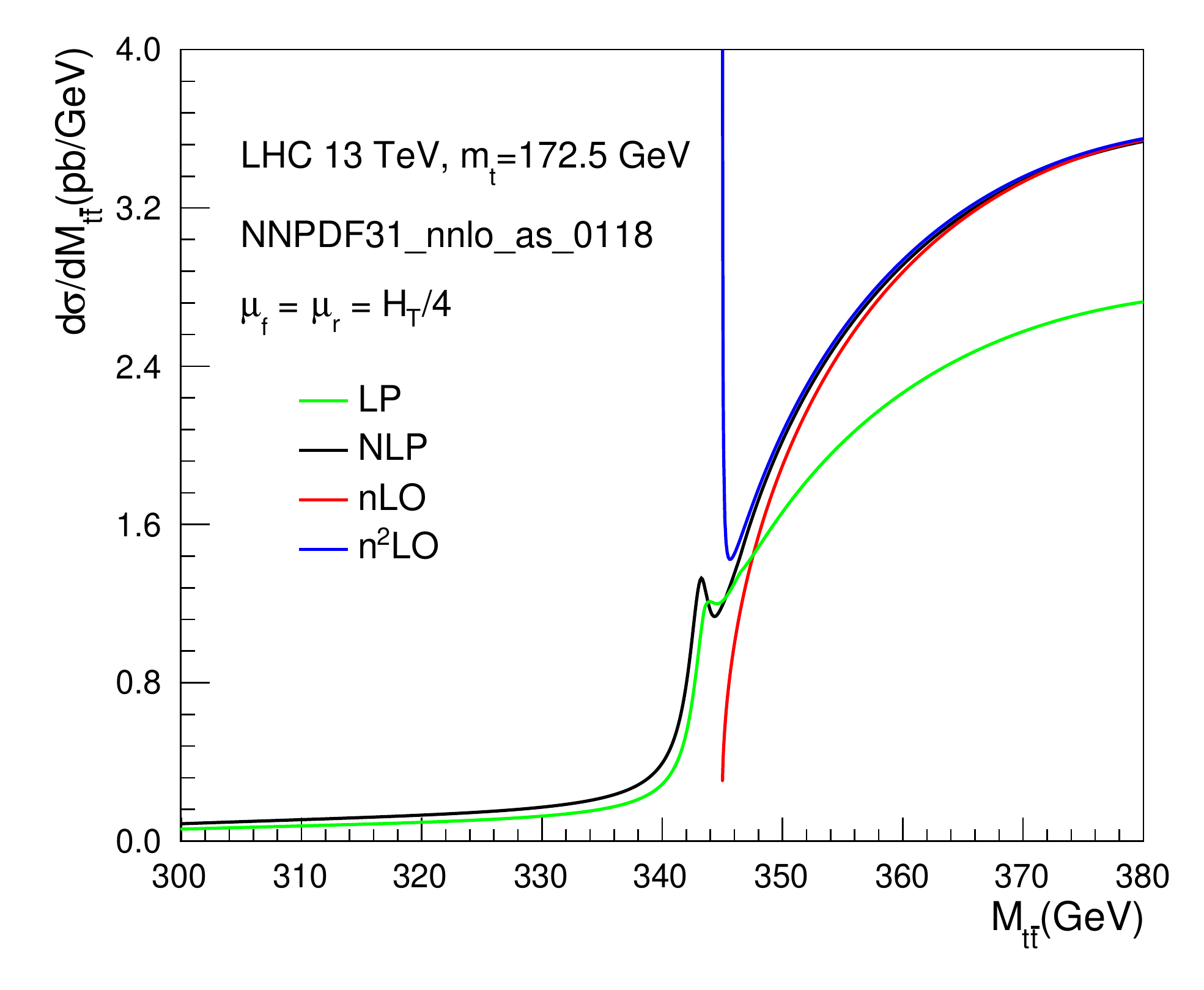}
\caption{\label{fig:NLP_expansion}
Left: the behaviors of the n$^k$LO expansion ($k=0,1,2,3,4$) of the NLP resummed result. Right: the comparison between the NLP resummed result and the LP resummed, nLO and nnLO ones.}
\end{figure}

First of all, given the matching formula \eqref{eq:matching_formula}, it is interesting to ask in which region the resummation effects (which are added onto the fixed-order results) are important. This information is encoded in the correction term
\begin{align}
\frac{d\sigma^{\text{NLP}}}{dM_{t\bar{t}}} - \frac{d\sigma^{\text{(n)nLO}}}{dM_{t\bar{t}}}
\label{eq:NLPcorrection}
\end{align}
of Eq.~\eqref{eq:matching_formula}. The first term in the above difference contains all-order information in the strong coupling. It is instructive to see its perturbative behavior order-by-order. This is shown in the left plot of Fig.~\ref{fig:NLP_expansion}, up to the 5th order in $\alpha_s$. We see that the perturbative expansion converges rather quickly for values of $M_{t\bar{t}}$ not too close to the $2m_t$ threshold. However, in the threshold region, the perturbative behavior goes wild. While the LO vanishes and the nLO approaches a constant value in the threshold limit $M_{t\bar{t}} \to 2m_t$, the differential cross section becomes divergent starting from nnLO. The nnLO and n$^3$LO distributions are still integrable, but the n$^4$LO one will give rise to infinite total cross section if one integrates down to the threshold.\footnote{Note that the integrability of the n$^3$LO distribution stems from the absence of the $\alpha_s^3/\beta^3$ term \cite{Beneke:2011mq}. It was shown that an additional contribution proportional to $\delta(\beta)$ needs to be added at this order to satisfy the dispersion relation \cite{Beneke:2016jpx}. Such a contribution is automatically contained in the NLP resummed results.} Such a breakdown of the perturbation theory in the threshold region is a natural reflection of the $(\alpha_s/\beta)^n$ terms from Coulomb gluon exchange.

The divergent behavior observed above is cured by the resummation. We show a comparison between the NLP resummed result and its perturbative expansion in the right plot of Fig.~\ref{fig:NLP_expansion}. We also show the LP resummed result for reference. The divergence in the threshold region is replaced by a small peak in the NLP resummed distribution. One can also observe that the NLP distribution extends below the $2m_t$ threshold, where the difference $2m_t - M_{t\bar{t}}$ can be viewed as the binding energy of the $t\bar{t}$ ``bound-state''. The low-energy tail of the distribution is rather long, all the way down to $M_{t\bar{t}} \sim \unit{300}{\GeV}$. This is due to the relatively large decay width of the top quark. On the other hand, we have checked that the integrated cross section in the \unit{$[300,380]$}{\GeV} bin is insensitive to $\Gamma_t$. It is also clear that in and below the threshold region, the LP and NLP distributions are rather similar, showing the good convergence of the power expansion in $\beta$. Above the threshold, the difference between the LP and NLP results are mainly induced by the $\mathcal{O}(\alpha_s)$ corrections including the NLO hard functions.

\begin{figure}[t!]
\centering
\includegraphics[width=0.49\textwidth]{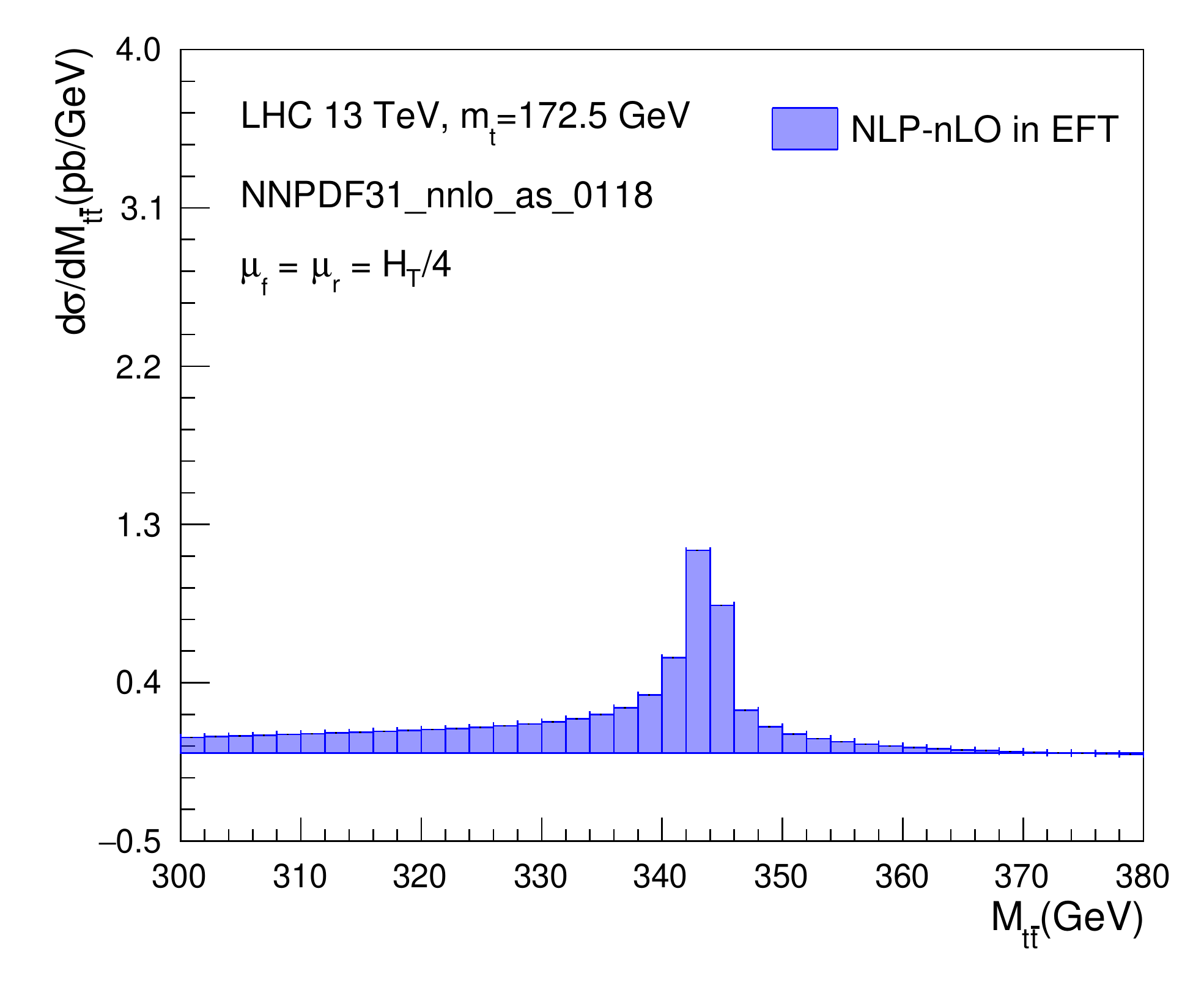}
\includegraphics[width=0.49\textwidth]{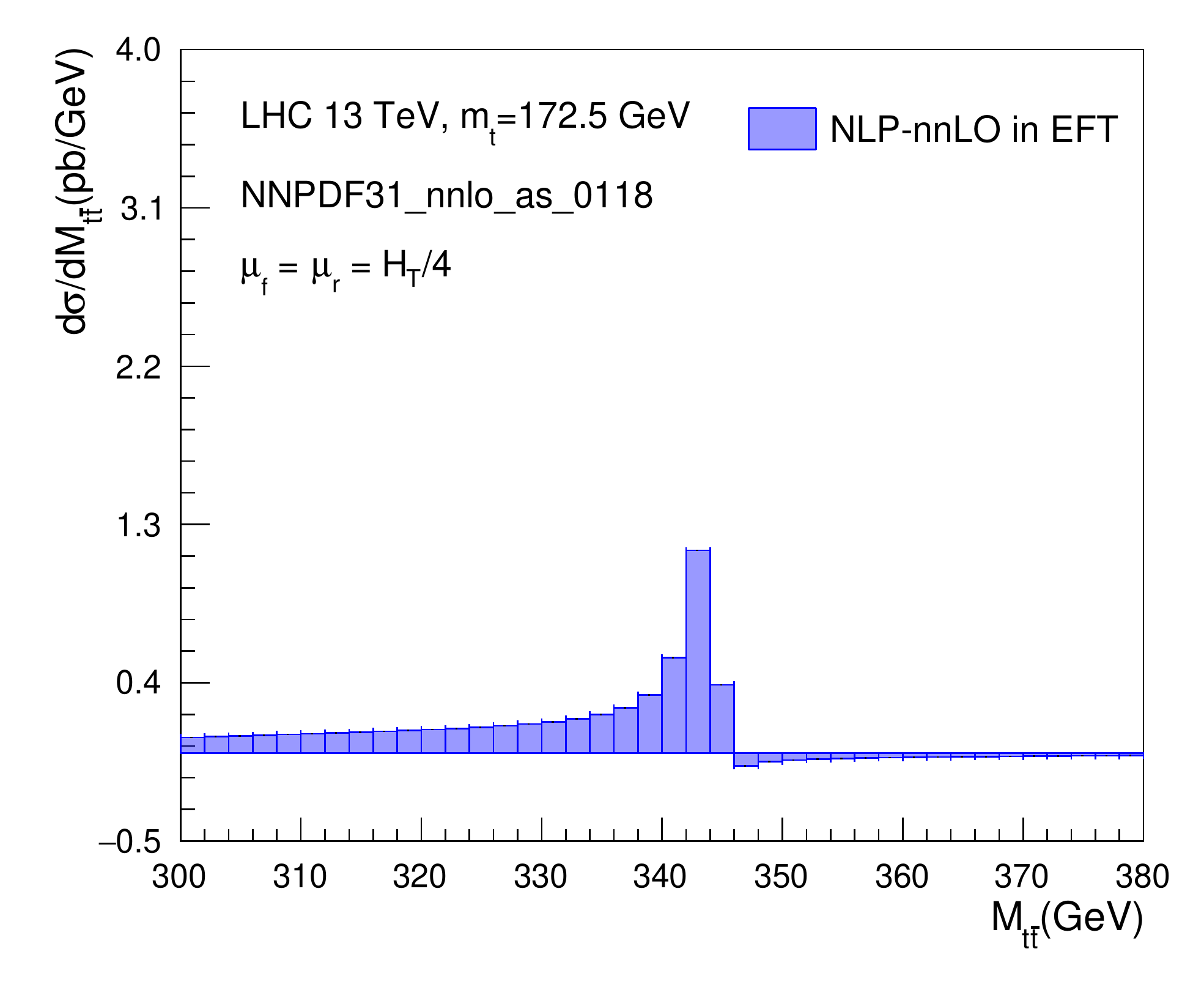}
\caption{Differences between NLP and nLO (left), and between NLP and nnLO (right). These represent the corrections induced by resummation upon the NLO and NNLO results.}
\label{fig:NLP_minus_nnLO}
\end{figure}

It is already evident from Fig.~\ref{fig:NLP_expansion} that the resummation effects are only important in and below the threshold region. As $M_{t\bar{t}}$ increases, the nLO and nnLO curves quickly approach the NLP one, meaning that the NLP corrections defined by Eq.~\eqref{eq:NLPcorrection} become small with respect to the fixed-order results when $M_{t\bar{t}}$ is far above the threshold. To see this more clearly, we directly plot the correction terms $d\sigma^{\text{NLP}}-d\sigma^{\text{nLO}}$ and $d\sigma^{\text{NLP}}-d\sigma^{\text{nnLO}}$ of Eq.~\eqref{eq:NLPcorrection} in Fig.~\ref{fig:NLP_minus_nnLO}. These quantify the corrections induced by resummation upon the NLO and NNLO results. The plots make it clear that the resummation effects concentrate in the region near and below the threshold, or more precisely, where $M_{t\bar{t}} < \unit{350}{\GeV}$. In this region $\beta<0.17$ and pNRQCD is perfectly applicable. On the other hand, for $M_{t\bar{t}} > \unit{350}{\GeV}$, the corrections are almost negligible. As a result, the NLO+NLP and NNLO+NLP predictions are dominated by the fixed-order terms away from the threshold. This demonstrates that our resummation has not been applied to regions where subleading corrections in $\beta$ might be important, and makes our predictions more robust. Later on, we will sometimes show predictions for a broader range of $M_{t\bar{t}}$, where resummation is switched off beyond $\unit{380}{\GeV}$. From Fig.~\ref{fig:NLP_minus_nnLO}, it should be clear that the results are insensitive to the the exact switch-off point, as long as it is larger than $\sim \unit{360}{\GeV}$.

\begin{figure}[t!]
\centering
\includegraphics[width=0.49\textwidth]{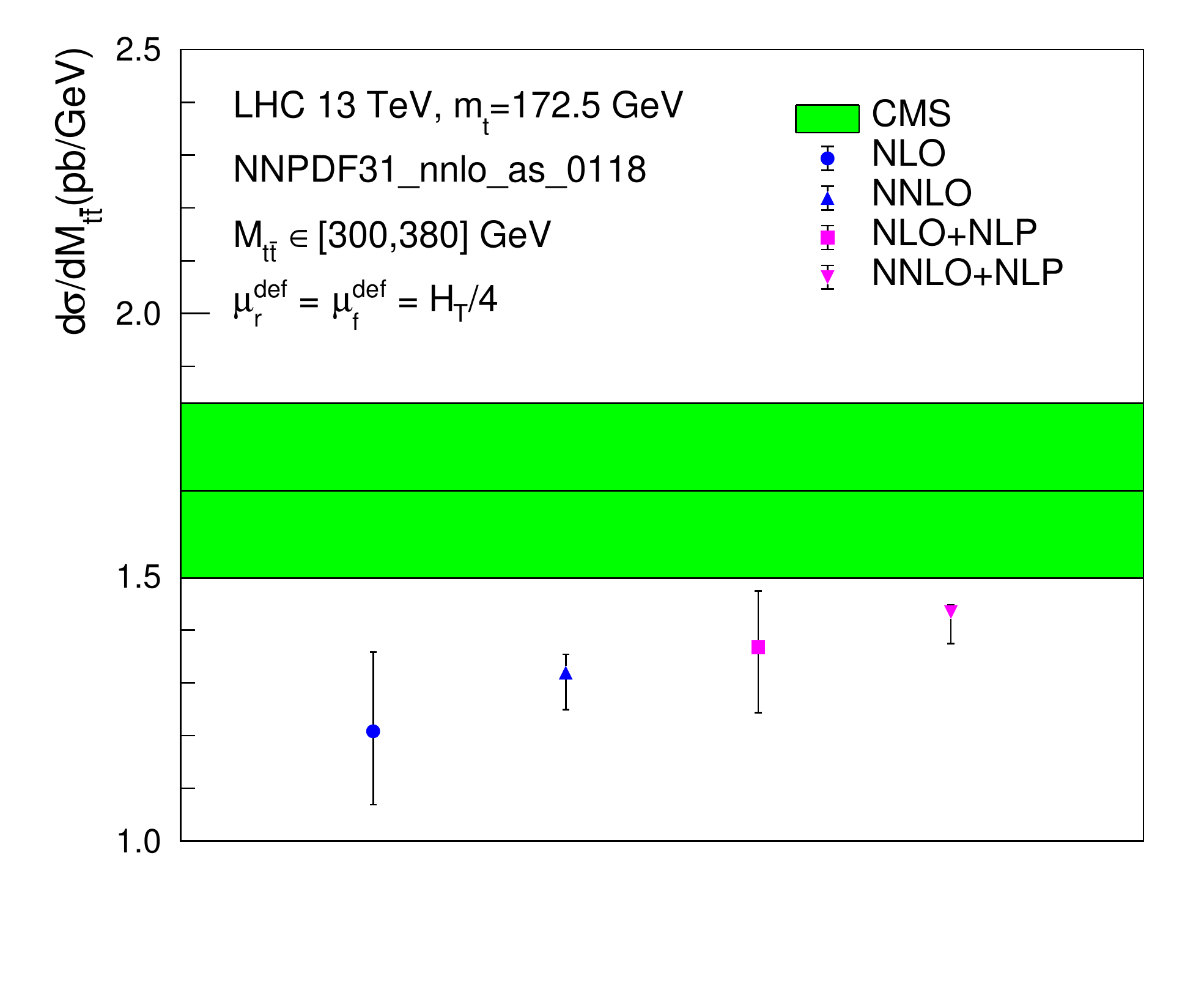}
\includegraphics[width=0.49\textwidth]{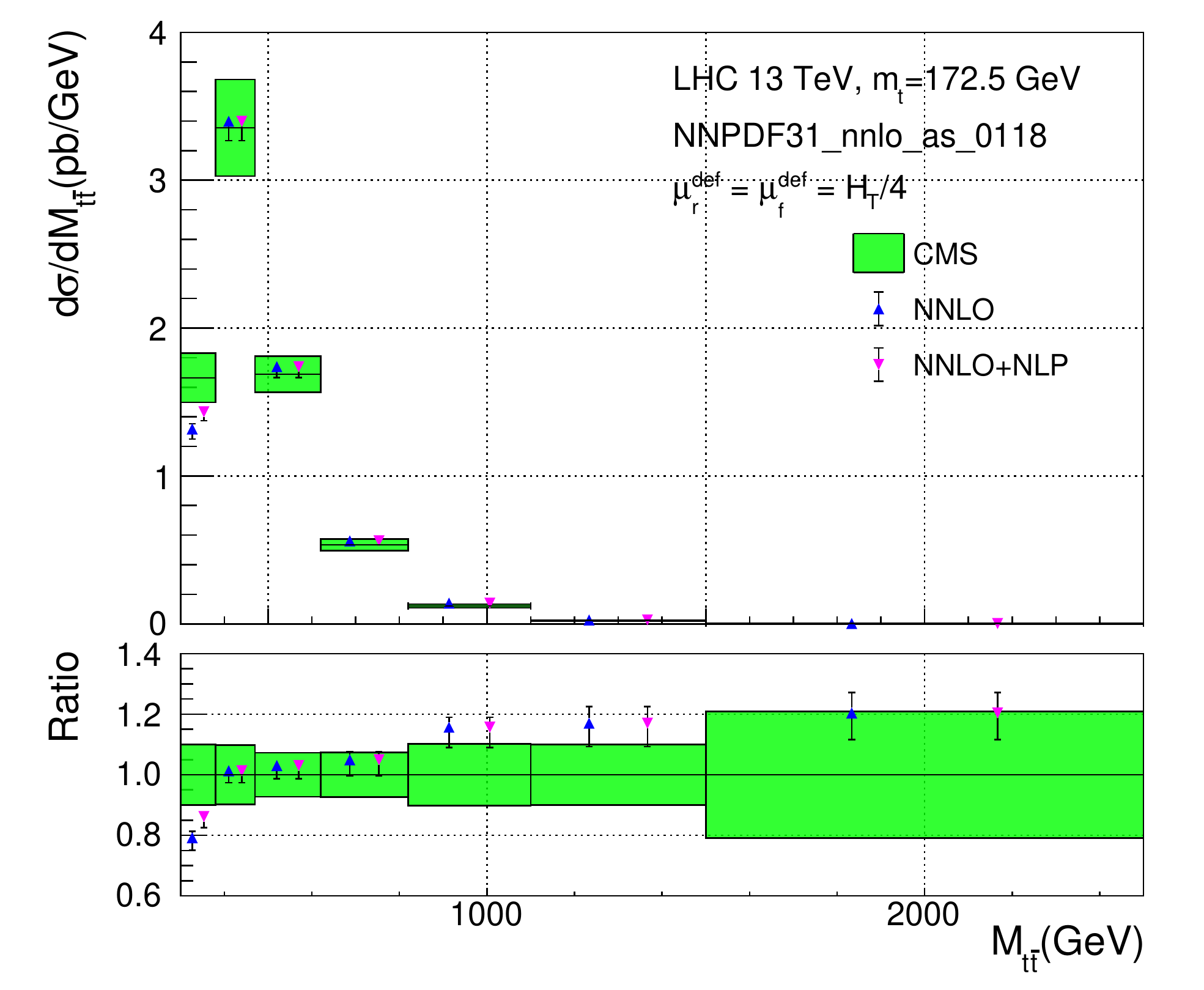}
\caption{The NLO+NLP and NNLO+NLP predictions for the absolute $M_{t\bar{t}}$ distribution against the CMS data in the di-lepton channel \cite{Sirunyan:2018ucr}. Fixed-order results are shown for comparison. The left plot shows the first bin $M_{t\bar{t}} \in \unit{$[300,380]$}{\GeV}$, while the right plot shows the full $M_{t\bar{t}}$ range.}
\label{fig:matched}
\end{figure}


We are now ready to present the matched results combining the resummation and fixed-order calculations, namely, the NLO+NLP and NNLO+NLP predictions. We show the results for the absolute differential cross sections in Fig.~\ref{fig:matched}, where the NLO and NNLO results are also given for comparison. The uncertainties estimated from scale variations are shown as the vertical bars. At central scales $\mu_r = \mu_f = H_T/4$, resummation effects increase the cross section in the first bin by 13\% with respect to NLO, and by 9\% with respect to NNLO. It should be noted that the uncertainty bar of the NNLO result does not overlap with that of the NNLO+NLP one. This shows that scale variations alone cannot faithfully account for the uncertainties of fixed-order calculations in this situation, due to the fact that the Coulomb resummation is genuinely non-perturbative. After adding the resummation effects, the NLO+NLP and NNLO+NLP predictions become more consistent with the CMS data than the fixed-order ones. This has significant impacts on the top quark mass determination, as we will discuss in the next subsection.

\begin{figure}[t!]
\centering
\includegraphics[width=0.7\textwidth]{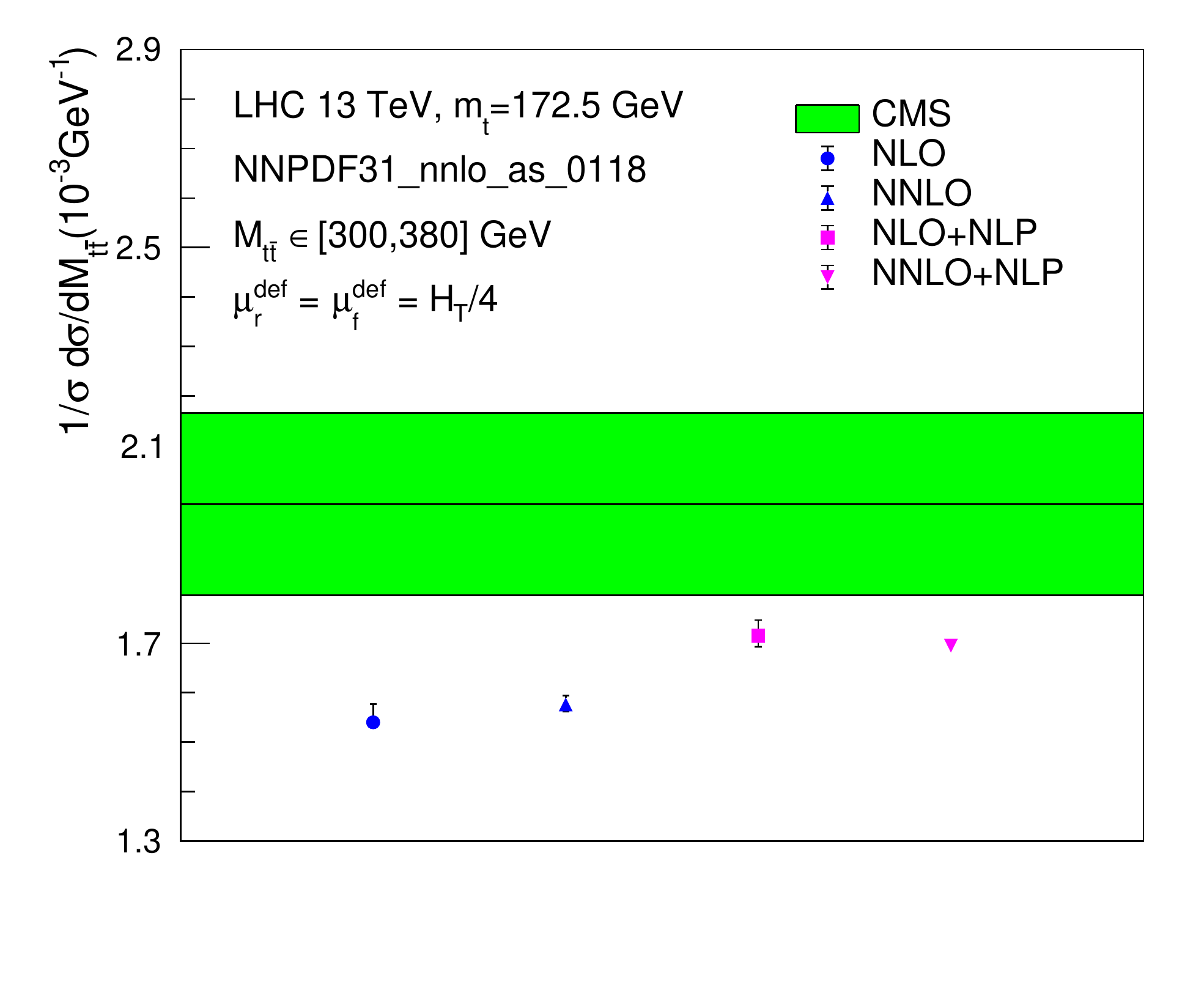}
\vspace{-6ex}
\caption{The NLO+NLP and NNLO+NLP predictions for the normalized $M_{t\bar{t}}$ distribution in the first bin $M_{t\bar{t}} \in \unit{$[300,380]$}{\GeV}$, against the CMS data in the di-lepton channel \cite{Sirunyan:2018ucr}. Fixed-order results are shown for comparison.}
\label{fig:matched_normalized}
\end{figure}

The experimental collaborations often quote the normalized differential cross sections $(d\sigma/dM_{t\bar{t}})/\sigma$ in addition to the absolute ones, where $\sigma$ is the total cross section. Normalization of the distribution has the benefit that part of the systematic uncertainties drops out when taking the ratio. On the theoretical side, normalized differential cross sections often exhibit smaller scale uncertainties as well. In Fig.~\ref{fig:matched_normalized}, we show the NLO, NNLO, NLO+NLP and NNLO+NLP predictions for the normalized differential cross section in the first bin $M_{t\bar{t}} \in \unit{$[300,380]$}{\GeV}$, in comparison with the CMS data \cite{Sirunyan:2018ucr}. We see that indeed, the scale uncertainties of all predictions are significantly reduced compared to those of the absolute differential cross sections of Fig.~\ref{fig:matched}. We also find that the NLO and NNLO results are rather close to each other. This shows that the NNLO correction to the normalized distribution is not very large. On the other hand, the resummation still shows big impact in this case: about 11\% increase from NLO to NLO+NLP, and about 8\% increase from NNLO to NNLO+NLP. This demonstrates that our conclusions in the last paragraph drawn from the absolute distribution remain unchanged when considering the normalized differential cross sections.

\begin{figure}[t!]
\centering
\includegraphics[width=0.7\textwidth]{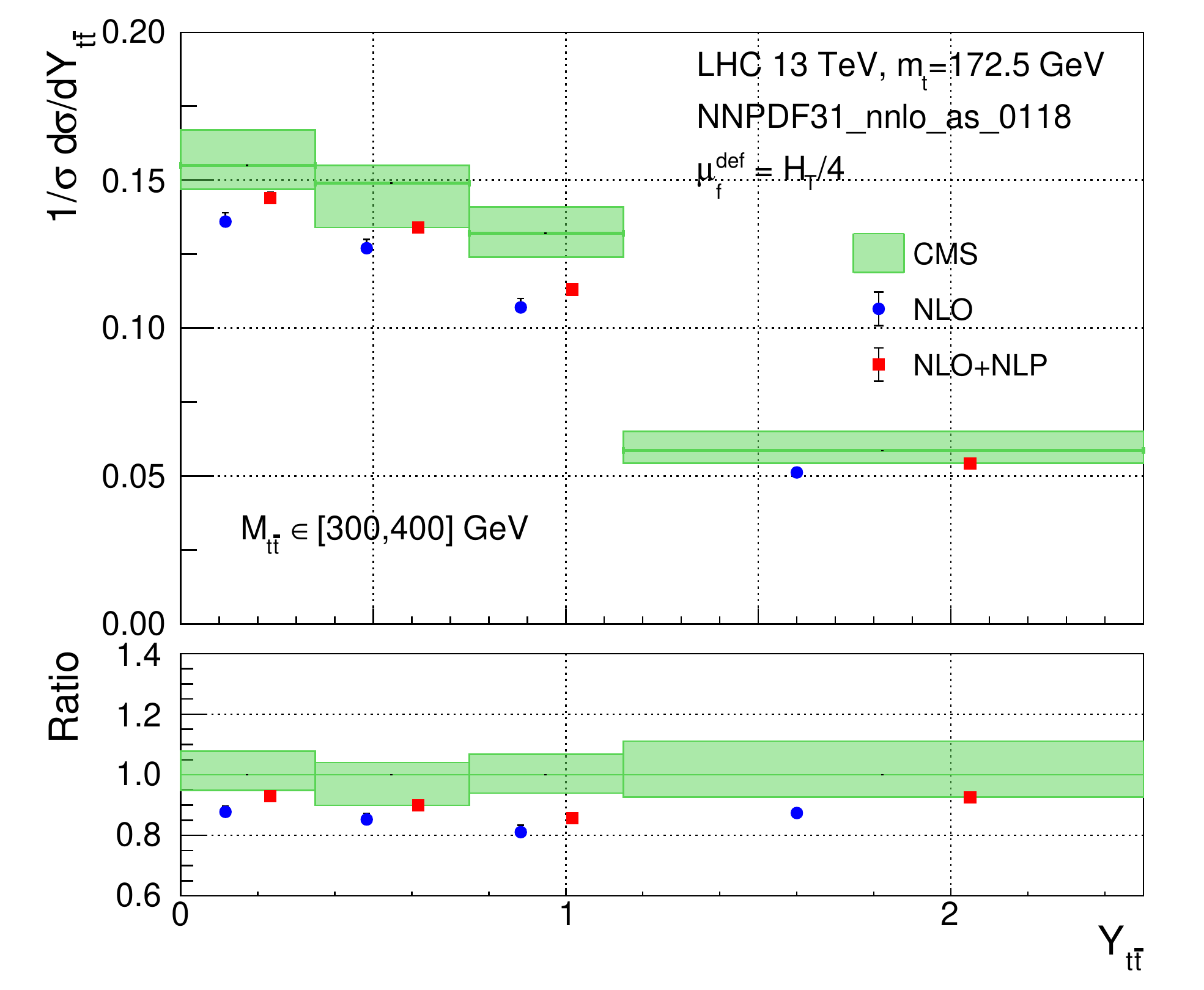}
\caption{Normalized double differential distribution with respect to $M_{t\bar{t}}$ and and the rapidity $Y_{t\bar{t}}$ of the top quark pair in the threshold region. This plot corresponds to the first bin (\unit{$[300,400]$}{\GeV}) in $M_{t\bar{t}}$ and four bins in $Y_{t\bar{t}}$. The NLO and NLO+NLP results are compared to the CMS data \cite{Sirunyan:2019zvx}.}
\label{fig:double_dist}
\end{figure}

So far we have only discussed the single differential cross section with respect to $M_{t\bar{t}}$. Thanks to the full kinematic dependence of the hard functions, our framework is flexible enough to be applied to double or triple differential cross sections, which were measured and employed to fit the top quark mass in, e.g., Ref.~\cite{Sirunyan:2019zvx}. To illustrate the idea, we have calculated the double differential cross sections with respect to $M_{t\bar{t}}$ and the rapidity $Y_{t\bar{t}}$ of the top quark pair in the laboratory frame. This can be performed using the formula
\begin{equation}
\frac{d^2\sigma}{dM_{t\bar{t}}dY_{t\bar{t}}} = \sum_{i,j} \int^1_{\tau} \frac{dz}{z} \, \frac{\tau}{z} \int d\Theta \, \frac{d\hat{\sigma}_{ij}(z,\mu_f)}{dM_{t\bar{t}} \, d\Theta} \, f_{i/h_1}(\sqrt{\tau/z} \, e^{Y_{t\bar{t}}-Y},\mu_f) \, f_{j/h_2}(\sqrt{\tau/z} \, e^{Y-Y_{t\bar{t}}}, \mu_f) \, ,
\end{equation}
where the partonic differential cross sections can be obtained using Eq.~\eqref{eq:fac2} as before. We show the normalized double differential cross sections in the threshold region in Fig.~\ref{fig:double_dist}, compared with the CMS data from \cite{Sirunyan:2019zvx}. The plot corresponds to the first bin in $M_{t\bar{t}}$, namely, $M_{t\bar{t}} \in \unit{$[300,400]$}{\GeV}$, and contains four bins in $Y_{t\bar{t}}$. Again, the resummation effects enhance the differential cross sections by about 7\% with respect to the NLO, making the theoretical predictions better consistent with experimental data. The increase here is not as big as that observed in Fig.~\ref{fig:matched_normalized}, mainly due to the larger size of the first $M_{t\bar{t}}$ bin which covers a broader range above the threshold.

\subsection{Influence on the top quark mass determination}
\label{sec:mt}

In this subsection, we discuss the influence of our resummed result on the determination of $m_t$ from kinematic distributions. Although we cannot repeat the experimental analyses in, e.g., Ref.~\cite{Sirunyan:2019zvx}, it is instructive to roughly estimate the impact of including the resummation effects in the fitting procedure.

To determine the top quark mass from kinematic distributions, one collects a set of observables $\{O_i\}$ which are theoretically functions of $m_t$, but can be experimentally measured without referring to a particular $m_t$ value. They can be the total cross section as well as single, double and triple differential cross sections in each bin. For each observable $O_i$, one has a theoretical prediction $O_i^{\text{TH}}(m_t)$ and an experimental measurement $O_i^{\text{EXP}}$. The top quark mass can then be determined by varying $m_t$ in the theoretical results and requiring a best fit between the set $\{O_i^{\text{TH}}(m_t)\}$ and the set $\{O_i^{\text{EXP}}\}$.\footnote{This can be done in any mass renormalization scheme. We will only discuss the pole mass here.} It can be understood that in such a procedure, the observables most sensitive to $m_t$ are the main driving force to decide the outcome. These include, in particular, the $M_{t\bar{t}}$ distribution near threshold and related double/triple differential cross sections.

\begin{figure}[t!]
\centering
\includegraphics[width=0.49\textwidth]{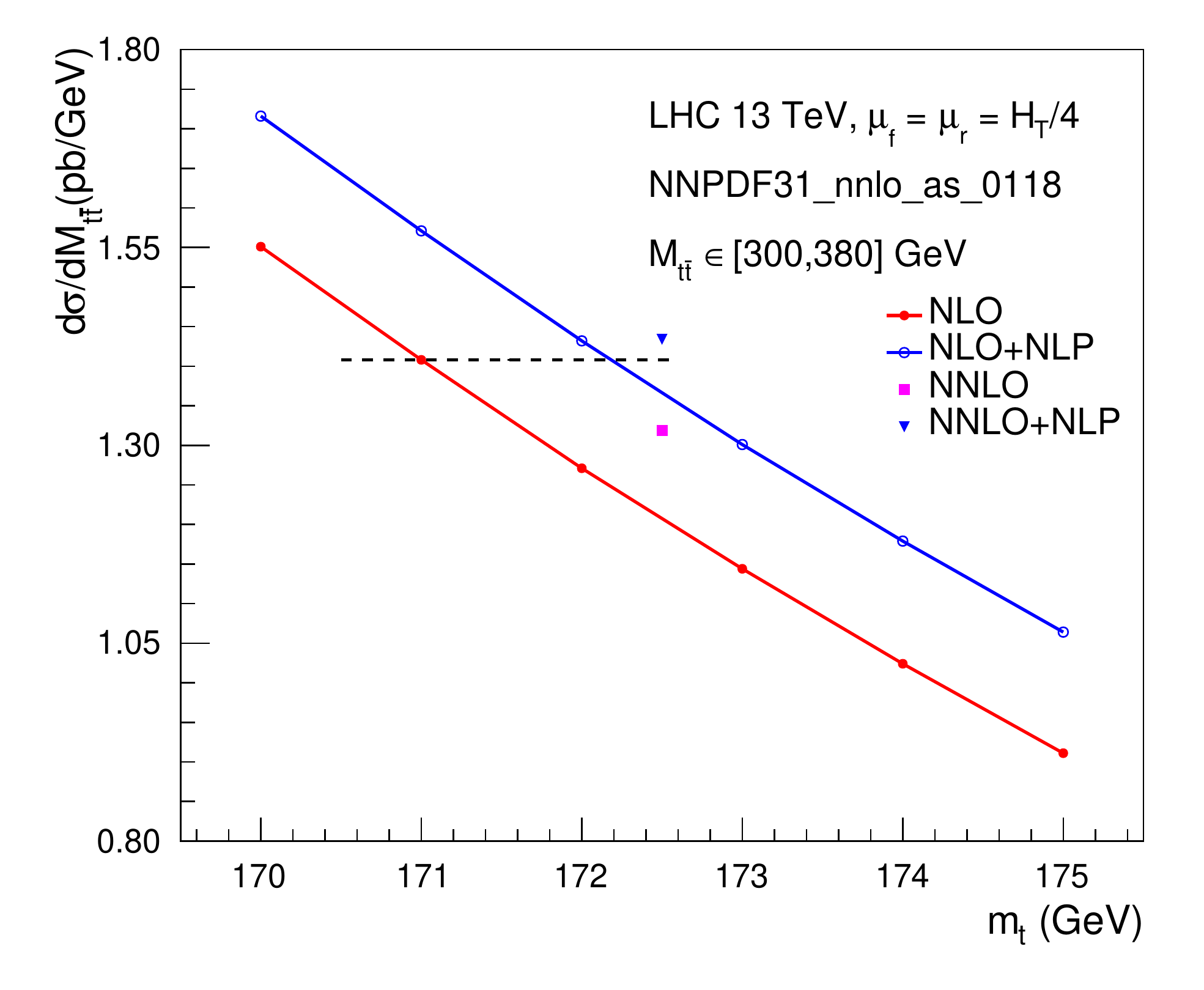}
\includegraphics[width=0.49\textwidth]{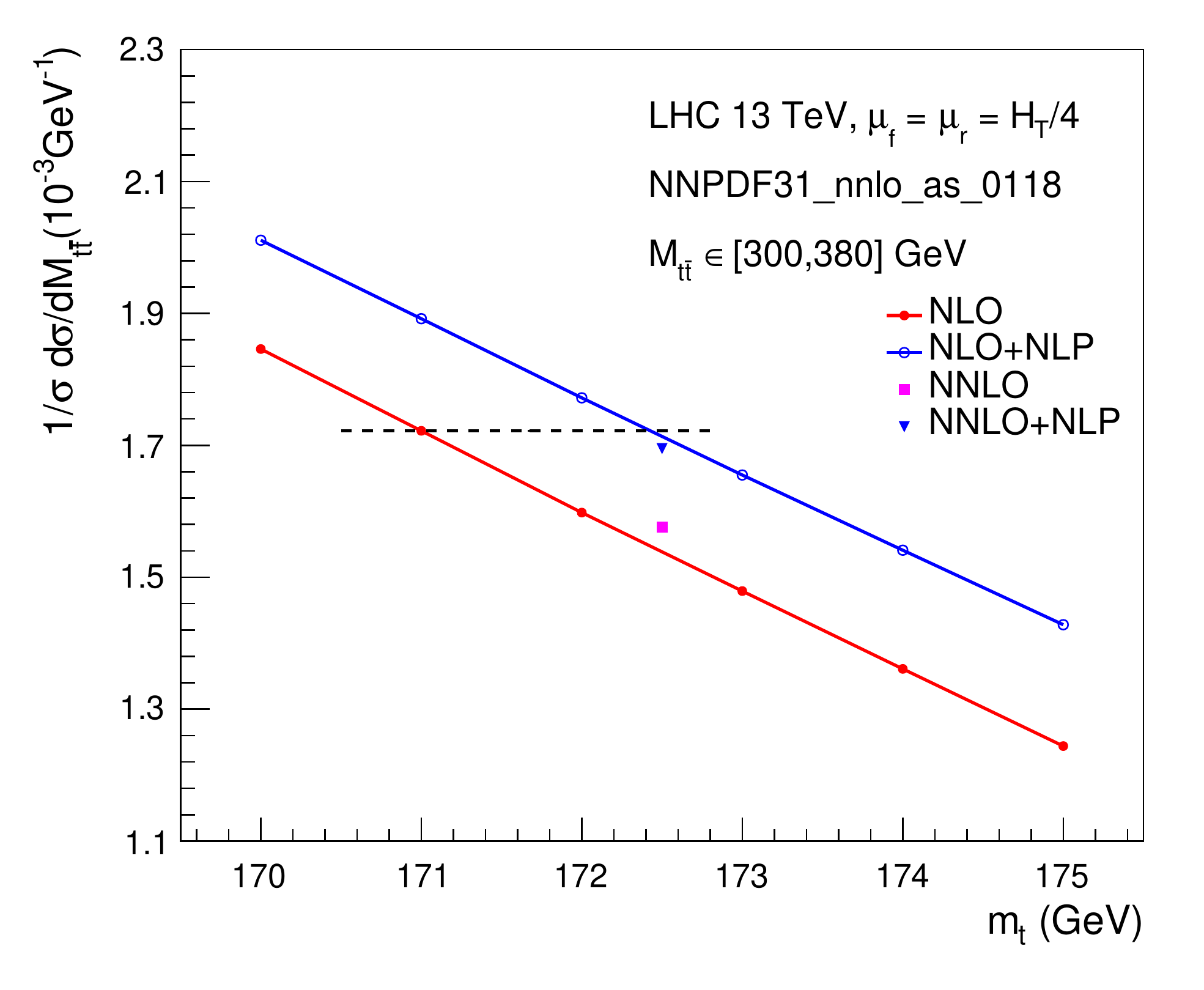}
\caption{Top-quark-mass dependence of the absolute (left) and normalized (right) $M_{t\bar{t}}$ differential cross sections in the threshold region. Only central values of the NLO and NLO+NLP results are shown here. The NNLO and NNLO+NLP predictions at $m_t=\unit{172.5}{\GeV}$ are given for reference.}
\label{fig:top_mass_fit}
\end{figure}

From the above description, it is clear that the outcome of the procedure strongly depends on the theoretical predictions entering the fit. Especially, the theoretical inputs for the $m_t$-sensitive observables are of crucial importance. For illustration, we calculate the averaged $M_{t\bar{t}}$ differential cross sections in the range \unit{$[300,380]$}{\GeV} using different top quark masses. The results are shown as functions of $m_t$ in Fig.~\ref{fig:top_mass_fit} for the absolute distribution (left plot) and the normalized distribution (right plot). As expected, we observe a strong (and nearly linear) dependence of the differential cross sections on $m_t$, and a large horizontal gap between the NLO and the NLO+NLP curves.

Ref.~\cite{Sirunyan:2019zvx} has used the NLO predictions for the normalized differential cross sections to fit the top quark mass, with the outcome $m_t \approx \unit{171}{\GeV}$. From the horizontal dashed line in Fig.~\ref{fig:top_mass_fit}, one can see that the NLO result with $m_t = \unit{171}{\GeV}$ is roughly the same as the NLO+NLP result with $m_t \approx \unit{172.4}{\GeV}$. This \unit{1.4}{\GeV} shift caused by the threshold effects is much more significant than that estimated in \cite{Sirunyan:2019zvx}. Given that the normalized NLO+NLP and NNLO+NLP results are rather close to each other, we expect a similar shift in the outcome of the fit if one uses the NNLO+NLP result as the theoretical input. We have also check that similar conclusions can be draw if the first bin is chosen as \unit{$[300,400]$}{\GeV}. Therefore, we see that the impact of the resummation effects on the top mass fit is rather concrete: the result of the fit should be much closer to the world average if one takes into account the precision theoretical predictions for the threshold region.

\subsection{Results at the \unit{8}{\TeV} LHC}
\label{sec:8tev}

The ATLAS and CMS collaborations have also performed measurements of the $M_{t\bar{t}}$ distribution at the center-of-mass energy $\sqrt{s}=\unit{8}{\TeV}$ \cite{Khachatryan:2015oqa, Aad:2015mbv}. No significant inconsistency between theory and data was spotted in those measurements, which is at first sight confusing. In this subsection, we show that the reason is simply due to the different choices of bins in the \unit{8}{\TeV} measurements than the \unit{13}{\TeV} ones.

\begin{figure}[t!]
\centering
\includegraphics[width=0.49\textwidth]{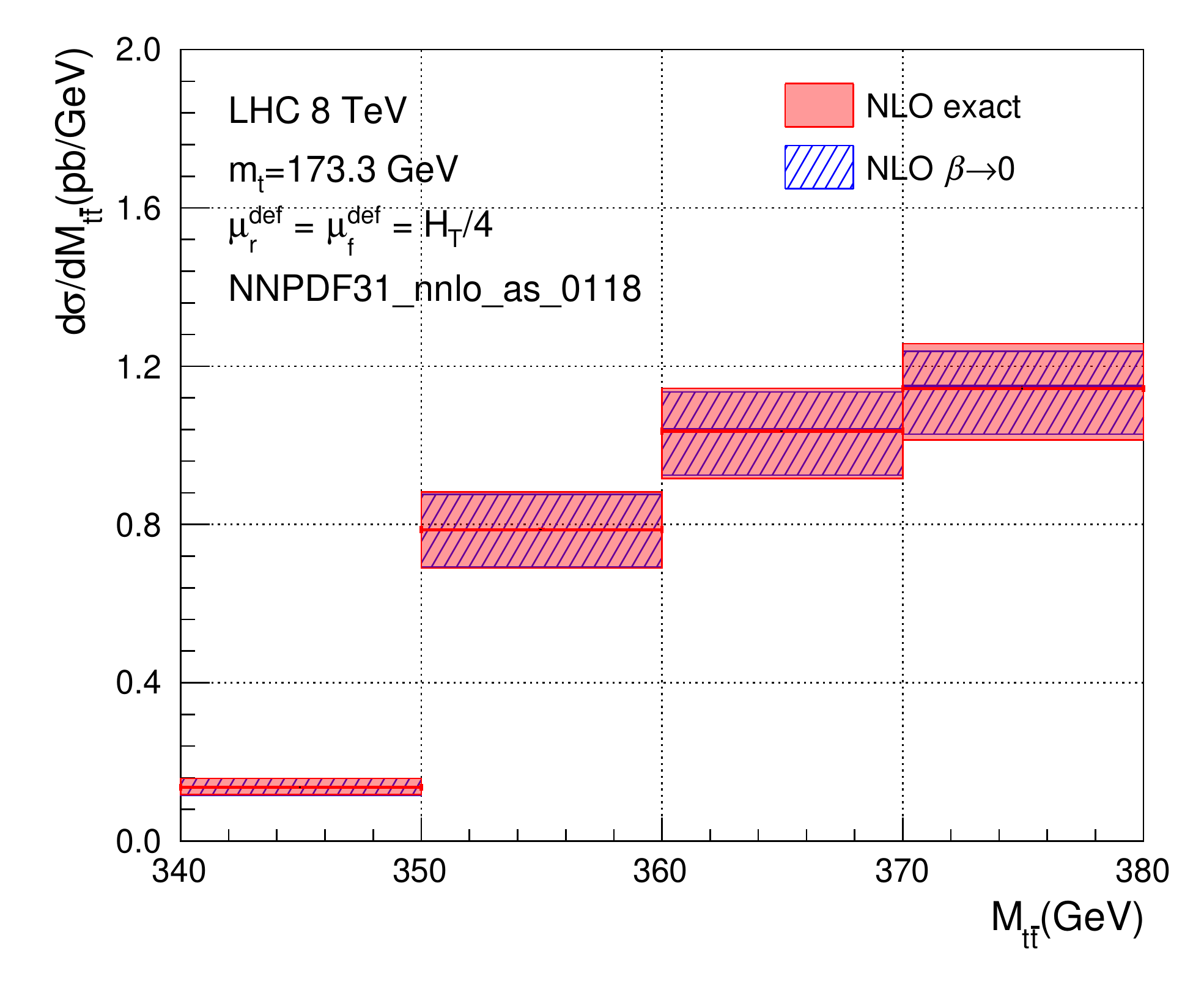}
\includegraphics[width=0.49\textwidth]{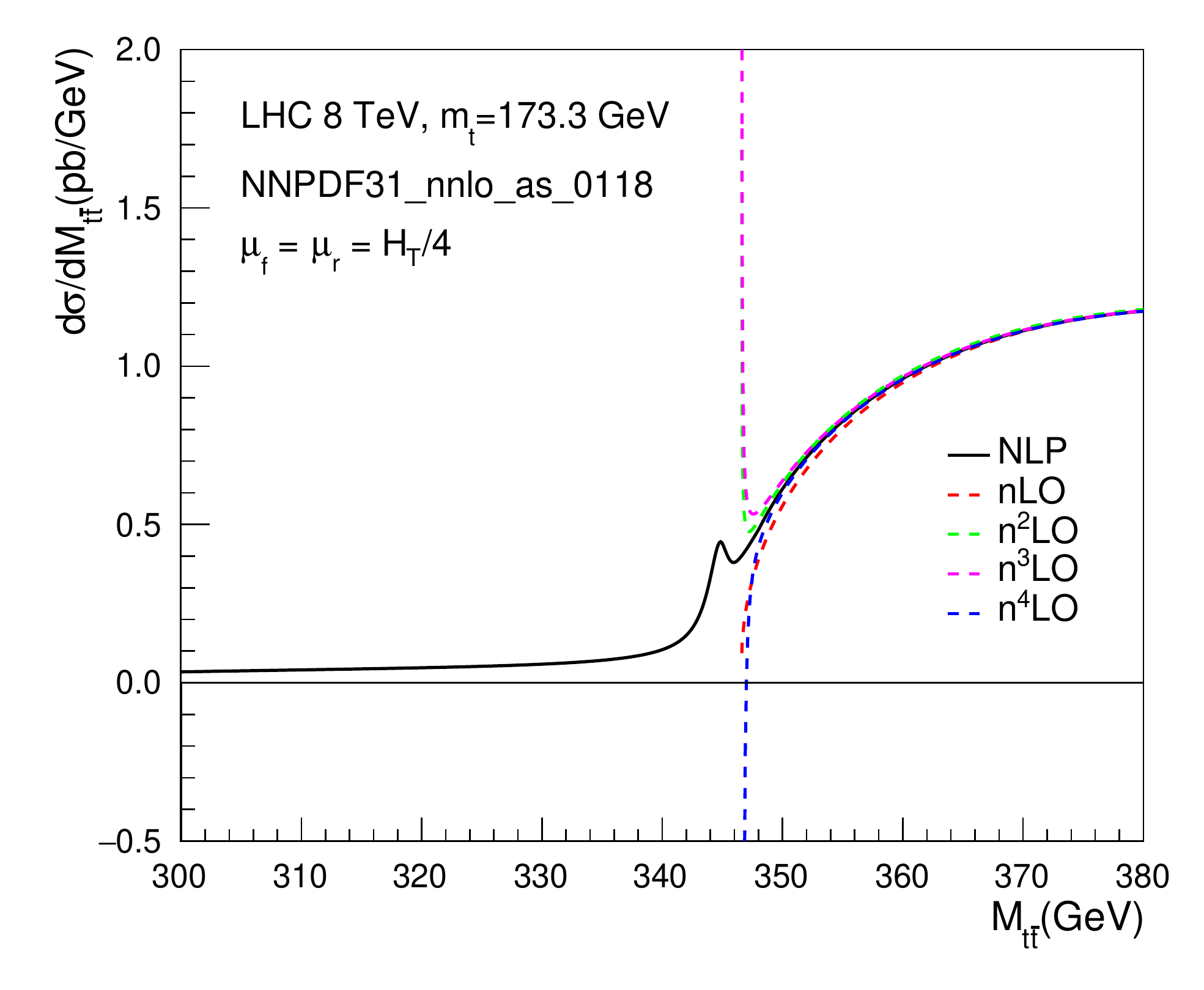}
\caption{Left: Comparison between the exact NLO result and the approximate one in the $\beta \to 0$ limit at the \unit{8}{\TeV} LHC; Right: The NLP resummed result against its fixed-order expansions at the \unit{8}{\TeV} LHC.}
\label{fig:NLP_expansion_8TeV}
\end{figure}

To begin with, we repeat the exercises we've done for the $\unit{13}{\TeV}$ LHC. In the left plot of Fig.~\ref{fig:NLP_expansion_8TeV} we compare the exact NLO distribution and the approximate one in the $\beta \to 0$ limit, while in the right plot we compare the NLP resummed distribution against its fixed-order expansions. As expected, we observe similar behaviors as the $\unit{13}{\TeV}$ case: 1) The approximate result agrees with the exact one rather well up to $M_{t\bar{t}} \sim \unit{380}{\GeV}$; 2) The resummed result regularizes the divergence near threshold, and tends to coincide with fixed-order results far above the threshold. One can then conclude that our resummation framework is reliable also for this case.

\begin{figure}[t!]
\centering
\includegraphics[width=0.49\textwidth]{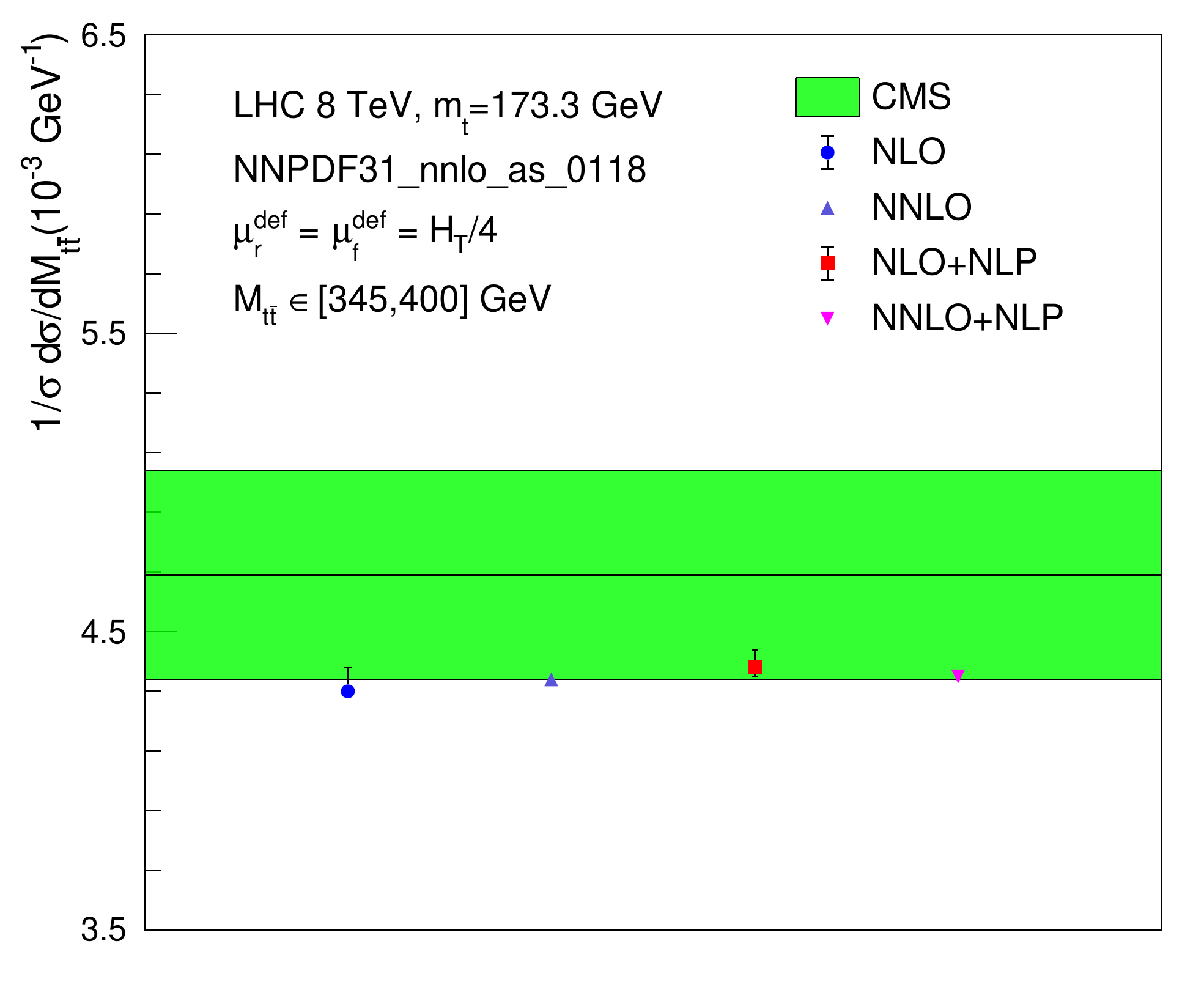}
\includegraphics[width=0.49\textwidth]{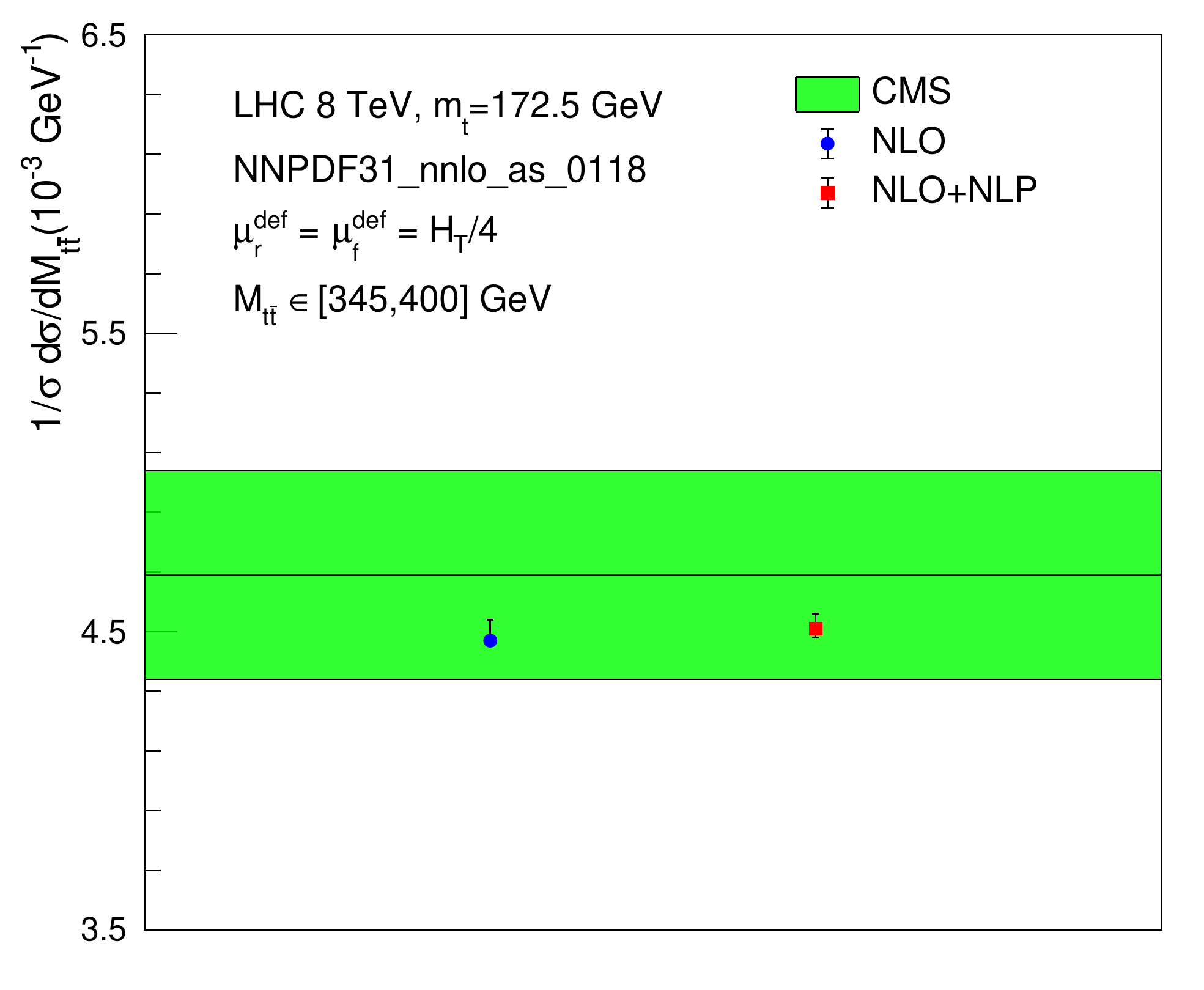}
\caption{Normalized differential cross sections at NLO, NNLO, NLO+NLP and NNLO+NLP for the range $\unit{345}{\GeV} \leq M_{t\bar{t}} \leq \unit{400}{\GeV}$ at the \unit{8}{\TeV} LHC, in comparison with the experimental data in the lepton+jets channel from the CMS collaboration \cite{Khachatryan:2015oqa}. The left plot uses $m_t=\unit{173.3}{\GeV}$, while the right one uses $m_t=\unit{172.5}{\GeV}$. For the NNLO and NNLO+NLP results, only the central values are shown.}
\label{fig:matched_results_8TeV}
\end{figure}

We now apply the resummation to the first bin of the experimental result in the lepton+jets channel from the CMS collaboration \cite{Khachatryan:2015oqa}, which is $\unit{345}{\GeV} \leq M_{t\bar{t}} \leq \unit{400}{\GeV}$. Note that the lower edge has been chosen as \unit{345}{\GeV} instead of $\unit{300}{\GeV}$ used in the \unit{13}{\TeV} measurements. We already know from Fig.~\ref{fig:NLP_minus_nnLO} that the resummation effects concentrate in the region slightly below the $2m_t$ threshold. Therefore, it can be expected that the numeric impact of resummation should not be significant for this choice of bin. Indeed, we show in Fig.~\ref{fig:matched_results_8TeV} the NLO, NLO+NLP, NNLO and NNLO+NLP predictions for the normalized differential cross sections in this bin. It can be seen that all calculations give similar numeric results, and agree with the experimental data remarkably well.

\begin{figure}[t!]
\centering
\includegraphics[width=0.7\textwidth]{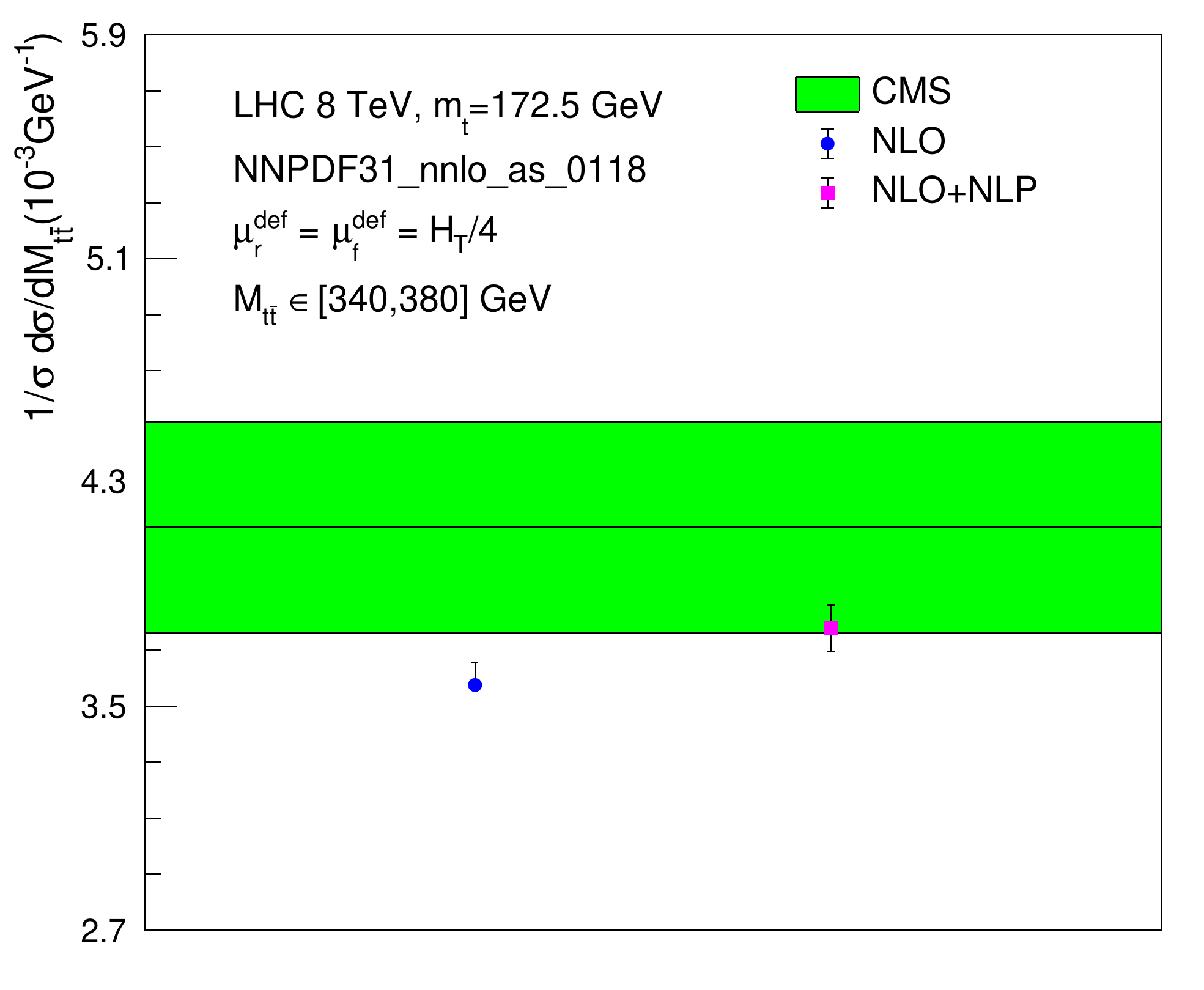}
\caption{Normalized differential cross sections at NLO and NLO+NLP for the range $\unit{340}{\GeV} \leq M_{t\bar{t}} \leq \unit{380}{\GeV}$ at the \unit{8}{\TeV} LHC, in comparison with the experimental data in the di-lepton channel from the CMS collaboration \cite{Khachatryan:2015oqa}.}
\label{fig:matched_results_8TeV_340_380}
\end{figure}

On the other hand, if the experimental data extends to lower values of $M_{t\bar{t}}$, things will be a bit different and the results will show some sensitivity to the threshold effects. Indeed, in the same CMS paper \cite{Khachatryan:2015oqa} results in the di-lepton channel are also presented. Here for the $M_{t\bar{t}}$ distribution, the first bin is chosen as \unit{$[340,380]$}{\GeV} which contains a region slightly below the threshold. We show the NLO and NLO+NLP predictions for such a bin choice in Fig.~\ref{fig:matched_results_8TeV_340_380}. We do observe a slight deficit of the NLO result compared to the experimental measurement. And a small correction from the resummation is also evident.

\begin{figure}[t!]
\centering
\includegraphics[width=0.49\textwidth]{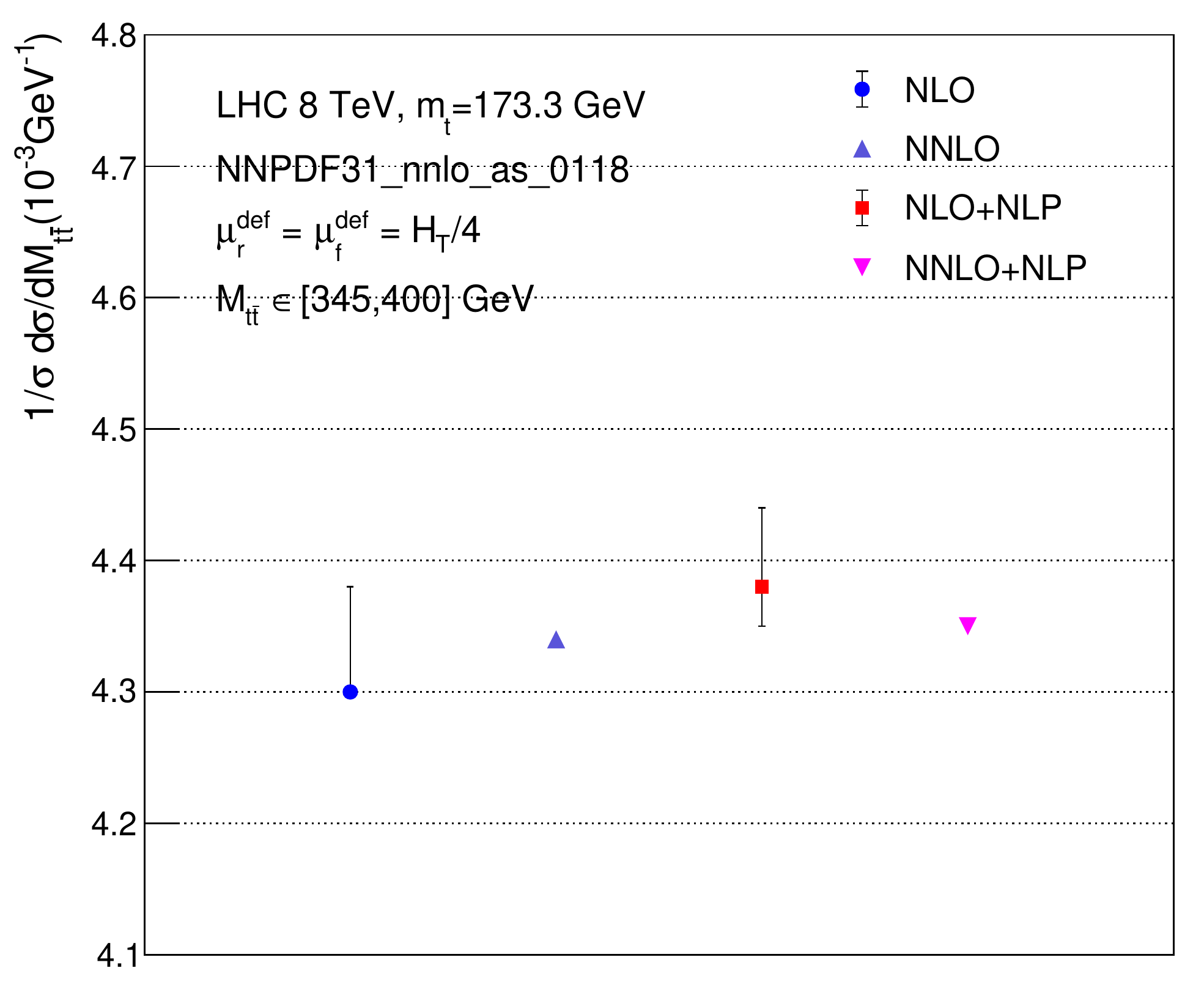}
\includegraphics[width=0.49\textwidth]{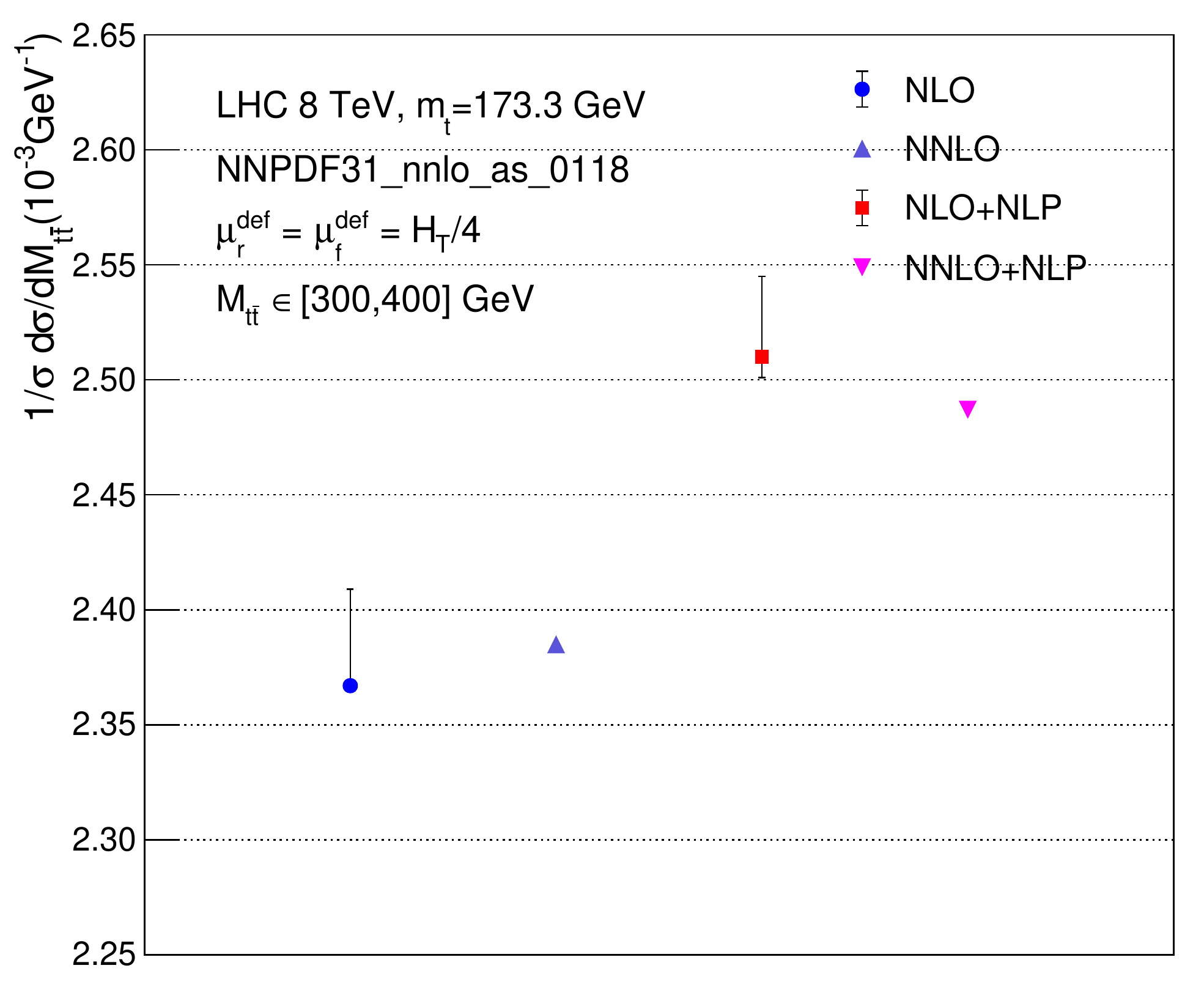}
\caption{Normalized differential cross sections in the first bin with two choices of the lower edge. Left: $\unit{345}{\GeV} \leq M_{t\bar{t}} \leq \unit{400}{\GeV}$; Right: $\unit{300}{\GeV} \leq M_{t\bar{t}} \leq \unit{400}{\GeV}$. For the NNLO and NNLO+NLP results, only the central values are shown.}
\label{fig:normalized_345_vs_300_8TeV}
\end{figure}

Had the experimental data extended further downwards, the sensitivity to the resummation effects would be more obvious. In Fig.~\ref{fig:normalized_345_vs_300_8TeV} we compare two choices of the lower edge of the first bin in the $M_{t\bar{t}}$ distribution, while keeping the upper edge at \unit{400}{\GeV}. The left plot uses the same bin choice as the experimental data in the lepton+jets channel \cite{Khachatryan:2015oqa}, and is in fact an enlarged version of Fig.~\ref{fig:matched_results_8TeV}. We see that all 4 results are similar here.
In the right plot, we extend the bin down to $\unit{300}{\GeV}$. One immediately finds that resummation has a big impact on the normalized differential cross sections in this case. We suggest that it is possible to experimentally verify the difference if one reanalyze the data in an extended range of the invariant mass.

\section{Summary}
\label{sec:summary}

To summarize, we have investigated single and double differential cross sections for $t\bar{t}$ production involving the pair invariant mass $M_{t\bar{t}}$, particularly in the threshold region $M_{t\bar{t}} \sim 2m_t$ or $\beta \sim 0$. Theoretical predictions for these observables are rather sensitive to the value of $m_t$, such that they can be used to extract the top quark mass from experimental data. The existing experimental studies at the \unit{13}{\TeV} LHC have employed the fixed-order calculations which did not take into account Coulomb effects of the form $1/\beta$ and $\ln\beta$ at and below the threshold. In this paper, we have performed a comprehensive study of these effects. Using the framework of effective field theories, we have derived a resummation formula which allows for dynamic renormalization and factorization scales. Such scale choices are often adopted in current theoretical calculations, including fixed-order ones and those with all-order resummation of soft gluon effects. As an important ingredient of our resummation formula, we have analytically calculated the hard functions up to the next-to-leading order. This enables us to perform the resummation of the Coulomb effects to all orders in $\alpha_s$ at the next-to-leading power. We further combine our resummed results with the NLO and NNLO calculations through a matching procedure. Our final predictions therefore reach the precision of NLO+NLP or NNLO+NLP.

Our resummation formula is similar to those in the literature, but differs in several important aspects. We have incorporated the leading-order coefficients with the exact dependence on $\beta$. As a result, the fixed-order expansion of our resummation formula reproduces the exact LO differential cross section, and to a good approximation the NLO one in the phase-space region of interest. Our resummation formula allows for dynamic renormalization and factorization scales, which are necessary for the combination with the existing NNLO results and for extending the prediction to a broader range of $M_{t\bar{t}}$. In our formalism, we do not consider the soft limit $z = M_{t\bar{t}}/\hat{s} \to 1$ upon the small-$\beta$ limit, since we have found that the double limit does not provide a reasonable approximation to the exact result in the threshold region. All the above make our predictions concrete and reliable. In particular, we have extensively checked that we have not introduced spurious corrections in phase-space regions where the small-$\beta$ approximation might break down. Last but not least, the full kinematic information contained in our resummation formula also enables us to study double differential cross sections, which were not available in previous studies.

In our phenomenological studies, we have concentrated on single and double differential cross sections which were employed by experimental groups to extract the top quark pole mass. We find that for the range $M_{t\bar{t}} \in \unit{$[300,380]$}{\GeV}$ at the \unit{13}{\TeV} LHC, the resummation effects increase the cross sections by about 13\% with respect to NLO, and by about 9\% with respect to NNLO. The combined NLO+NLP and NNLO+NLP results show better consistency with the experimental data. The resummation effects have a strong impact on the top quark mass determination from the $M_{t\bar{t}}$ distribution, and can change the result by about $\unit{1.4}{\GeV}$, which is much larger than the estimated uncertainties in previous experimental studies. The shifted top quark mass is much more consistent with the current world average measured using other methods. We have also investigated the double differential distribution in terms of $M_{t\bar{t}}$ and the rapidity $Y_{t\bar{t}}$ of the $t\bar{t}$ pair, and drawn similar conclusions. We therefore conclude that future experimental studies should include the Coulomb effects at and below the threshold in order to consistently extract the top quark mass.

We have also performed numeric studies for the \unit{8}{\TeV} LHC. Due to the fact that the experimental result does not cover the main portion of the phase-space below the threshold, the resummation effects do not show a big impact if using the same choice of bins. However, we have demonstrated that if one reanalyze the experimental data with an extended first bin, the threshold effects should be visible in the normalized differential cross sections.

Our NNLO+NLP result can be further combined with the NNLO+NNLL$'$ result of \cite{Ahrens:2010zv, Ferroglia:2012ku, Pecjak:2016nee, Pecjak:2018lif, Czakon:2018nun} to achieve the best prediction in the whole phase-space region. Inclusion of electroweak effects can also be done similar as \cite{Czakon:2019txp}.
Our formalism can be applied to more kinds of double and even triple differential cross sections in the future. It can be extended to study the associated production of $t\bar{t}$ with an extra jet, which is also employed in the top quark mass determination. With suitable modifications, it can be applied to $t\bar{t} + Z$ or $t\bar{t} + H$ production as well. We leave these considerations for future investigations.

\appendix

\section{Ingredients for the NLO hard functions}
\label{sec:integrals}

In this Appendix we give the relevant ingredients entering the calculation of the NLO hard functions. The hard functions correspond to the squared-amplitudes evaluated at the threshold point $M_{t\bar{t}} = 2m_t$. As a result, all the loop integrals need to be evaluated at that point. Note that this is not equivalent to first evaluating the integrals generically (whose results are well-known in the literature), and then taking the limit $M_{t\bar{t}} \to 2m_t$, due to the $1/\beta$ and $\ln(\beta)$ divergent behaviors of certain integrals.\footnote{These divergences are absent when directly evaluating the integrals at the threshold point. They reappear in the Coulomb functions in the factorization formula Eq.~\eqref{eq:fac2}.} For this reason, we explicitly list the results in the following:
\begin{align}
\label{eq:topolgy1boundary}
I_1(\epsilon) &= M^{-2+2\epsilon}_{t\bar{t}} \int [dq] \, \frac{1}{q^2 - m_t^2}  \nonumber
\\
&=\frac{1}{4\epsilon} + \frac{1}{4} \big( 1 + 2 \ln(2) \big) + \mathcal{O}(\epsilon) \, , \nonumber
\\
I_2(\epsilon) &=  M^{2\epsilon}_{t\bar{t}} \int [dq] \, \frac{1}{(q-p_1)^2 \, (q+p_2)^2 } \nonumber
\\
&= \frac{1}{\epsilon} + i\pi + 2 + \epsilon \left( 4 + 2 i\pi - \frac{7\pi^2}{12} \right) + \mathcal{O}(\epsilon^2) \, , \nonumber
\\
I_3(\epsilon) &= M^{2\epsilon}_{t\bar{t}} \int [dq] \, \frac{1}{q^2 \, \big[ (q-p_1+p_t)^2-m_t^2 \big]} \nonumber
\\
&= \frac{1}{\epsilon} + 2 + \epsilon \left( 4 - \frac{\pi ^2}{12} \right) + \mathcal{O}(\epsilon^2) \, , \nonumber
\\
I_4(\epsilon) &= M^{-2+2\epsilon}_{t\bar{t}} \int [dq] \, \frac{\big( q^2 \big)^2}{(q-p_1)^2 \, (q+p_2)^2 \, \big[ (q-p_1+p_t)^2-m_t^2 \big]}  \nonumber
\\
&= -\frac{3}{4\epsilon} - \frac{1}{12} \big( 8 i\pi + 19 + 2 \ln(2) \big) + \mathcal{O}(\epsilon) \, , \nonumber
\\
I_5(\epsilon) &= M^{4+2\epsilon}_{t\bar{t}} \int [dq] \, \frac{1}{q^2 \, (q-p_1)^2 \, (q+p_2)^2 \, \big[ (q-p_1+p_t)^2-m_t^2 \big]} \nonumber
\\
&= -\frac{4}{\epsilon^2} - \frac{2i\pi}{\epsilon} + \frac{4\pi^2}{3} + \mathcal{O}(\epsilon) \, , \nonumber
\\
I_6(\epsilon) &= M^{2+2\epsilon}_{t\bar{t}} \int [dq] \, \frac{1}{(q-p_t)^2 \, (q^2-m_t^2) \, \big[ (q-p_1)^2-m_t^2 \big]} \nonumber
\\
&= -\frac{\pi^2}{2} + \mathcal{O}(\epsilon) \, , \nonumber
\\
I_7(\epsilon) &= M^{4+2\epsilon}_{t\bar{t}} \int [dq] \, \frac{1}{(q-p_t)^2 \, (q+p_2-p_t)^2 \, (q^2-m_t^2) \, \big[ (q-p_1)^2-m_t^2 \big]} \nonumber
\\
&= \frac{4}{\epsilon^2} - \frac{7\pi^2}{3} + \mathcal{O}(\epsilon) \, , \nonumber
\\
I_8(\epsilon) &= M^{2+2\epsilon}_{t\bar{t}}  \int [dq] \, \frac{1}{(q^2-m_t^2) \, \big[ (q-p_1)^2-m_t^2 \big] \, \big[ (q-p_1-p_2)^2-m_t^2 \big]} \nonumber
\\
&= -\frac{\pi^2}{2}  + \mathcal{O}(\epsilon) \, , \nonumber
\\
I_9(\epsilon) &= M^{4+2\epsilon}_{t\bar{t}} \int [dq] \, \frac{1}{(q-p_t)^2 \, (q^2-m_t^2) \, \big[ (q-p_1)^2-m_t^2 \big] \, \big[ (q-p_1-p_2)^2-m_t^2 \big]} \nonumber
\\
&= \frac{4}{\epsilon} - 8 + \mathcal{O}(\epsilon) \, ,
\end{align}
where we have suppressed the $+i\varepsilon$ prescription in all propagators, and
\begin{align}
[dq] \equiv \frac{(4\pi)^{2-\epsilon}}{ie^{-\epsilon\gamma_E}} \frac{d^dq}{(2\pi)^d} \, .
\end{align}

Up to the NLO, the various renormalization constants are given by
\begin{align}
Z_q &= 1 + \mathcal{O}(\alpha_s^2) \, , \nonumber
\\
Z_m &= 1 - \frac{\alpha_s^{(N_f)}}{4\pi} C_F \left( \frac{3}{\epsilon} + 3L_M + 6\ln(2) +4\right) + \mathcal{O}(\alpha_s^2) \, , \nonumber
\\
Z_Q &= 1 - \frac{\alpha_s^{(N_f)}}{4\pi} C_F \left( \frac{3}{\epsilon} + 3L_M + 6\ln(2) +4\right) + \mathcal{O}(\alpha_s^2) \, , \nonumber
\\
Z_g &= 1 - \frac{\alpha_s^{(N_f)}}{4\pi} T_FN_h \left( \frac{4}{3\epsilon} + \frac{4}{3}L_M +\frac{8}{3}\ln(2) \right) + \mathcal{O}(\alpha_s^2) \, , \nonumber
\\
Z_{\alpha_s} &= 1 - \frac{\alpha_s^{(N_f)}}{4\pi}\frac{\beta_0}{\epsilon} + \mathcal{O}(\alpha_s^2) \, ,
\end{align}
where $\beta_0=(11C_A-4T_FN_f)/3$, and we have used $M_{t\bar{t}} = 2m_t$ to convert all logarithms of $m_t$ to $L_M$. The above expressions are written in terms of the strong coupling constant $\alpha_s$ with $N_f = N_l + N_h$ active flavors, where $N_l$ and $N_h$ are the number of light and heavy quark flavors, respectively. In our case we have $N_l=5$ and $N_h=1$. In practice, it is more convenient to work with $\alpha_s$ with $N_l$ active flavors, i.e., with the top quark field integrated out. This decoupling can be perform with the relation
\begin{align}
\alpha_s^{(N_f)} &=  \zeta_{\alpha_s}\alpha_s^{(N_l)} \, ,
\end{align}
where
\begin{align}
\zeta_{\alpha_s} &=  1 + \frac{\alpha_s^{(N_l)}}{4\pi}T_F N_h\frac{2}{3} \big[ L_M + 2\ln(2) \big] + \mathcal{O}(\alpha_s^2) \, .
\end{align}

\section{Implementation details of the $z$-soft limit}
\label{sec:soft_limit}

In this Appendix we discuss the implementation details of the $z$-soft limit. First we give the hard functions in the limit $z \to 1$, which were used to produce the results shown in the right plot of Fig.~\ref{fig:zsoft}. These can be obtained from the full hard functions in Eq.~\eqref{eq:HnloQTY} by keeping only the singular terms and dropping the regular terms in the $z$-soft limit.
Note that the abbreviation $d(Q_T,z)$, defined in Eq.~\eqref{eq:abbrev}, is itself a singular term, and in the other terms only the plus-distributions and delta-functions of $1-z$ are singular. Therefore, in the $z$-soft limit, the LO hard functions remain the same and the NLO ones read
\begin{align}
H_{q\bar{q},8,\text{soft}}^{(1)} &=\bigg( 2\beta_0 \, L_M + \frac{292}{9} + 8\ln(2) - \frac{20N_l}{9} - \frac{11\pi^2}{9} \bigg) \, H_{q\bar{q},8}^{(0)} + \frac{4(7+9y^2)}{27} d(Q_T,z) \nonumber
\\
&+ \bigg[ \frac{32}{3} \left( \frac{1}{1-z} \ln\frac{1-z}{2} \right)_+ - L_f P_{qq,\text{soft}}^{(0)}(z) \bigg] h_{q\bar{q},8} \, \delta(Q_T^2) \, \delta^Y_{\text{max},0} \, ,  \nonumber
\\
H_{gg,1,\text{soft}}^{(1)} &=\bigg( 2\beta_0 \, L_M - \frac{44}{3} + \frac{7\pi^2}{3} \bigg) \, H_{gg,1}^{(0)}  + \frac{1}{2}\, d(Q_T,z) \nonumber
\\
&+\bigg[ 24\left( \frac{1}{1-z} \ln\frac{1-z}{2} \right)_+ -  L_f P_{gg,\text{soft}}^{(0)}(z) \bigg]\, h_{gg,1} \,  \delta(Q_T^2) \, \delta^Y_{\text{max},0} \, ,  \nonumber
\\
H_{gg,8,\text{soft}}^{(1)}&= \bigg(2\beta_0 \, L_M  + \frac{28}{3} + \frac{5\pi^2}{6} \bigg) \, H_{gg,8}^{(0)}  + \frac{5(3+y^2)}{16} \, d(Q_T,z) \nonumber
\\
&+ \bigg[ 24 \left( \frac{1}{1-z} \ln\frac{1-z}{2} \right)_+ - L_f P_{gg,\text{soft}}^{(0)}(z) \bigg] \, h_{gg,8} \,  \delta(Q_T^2) \, \delta^Y_{\text{max},0}  \, ,
\end{align}
where the splitting functions in the $z$-soft limit are given by
\begin{align}
P_{qq,\text{soft}}^{(0)}(z) &= 2C_F \bigg[ \left(\frac{2}{1-z}\right)_+ + \frac{3}{2} \delta(1-z) \bigg]\, ,  \nonumber
\\
P_{gg,\text{soft}}^{(0)}(z) &= 4C_A \left(\frac{1}{1-z}\right)_+ + \beta_0 \delta(1-z) \, .
\end{align}

The differential cross sections in the double limit $\beta \to 0$ and $z \to 1$ can be readily obtained by inserting the above hard functions into the NLP kernel $K^{\text{NLP}}_{ij,\alpha}$ of Eqs.~\eqref{eq:K_nlp} and \eqref{eq:K_i_expansion}. However, we note that there exist alternative ways of implementing the calculation, which are equivalent at leading power in $1-z$. For example, it is legitimate to multiply $H_{ij,\alpha,\text{soft}}$ by an overall factor of $z$, since $z = 1 + \mathcal{O}(1-z)$. One is also free to use $-\ln z$ instead of $1-z$ in the hard functions, noting that $-\ln z = 1 - z + \mathcal{O}((1-z)^2)$, and
\begin{align}
\left( \frac{1}{1-z} \right)_+ &= \left( \frac{1}{-\ln z} \right)_+ - \gamma_E \, \delta(1-z) + \mathcal{O}((1-z)^0) \, , \nonumber
\\
\left[ \frac{\ln(1-z)}{1-z} \right]_+ &= \left[ \frac{\ln(-\ln z)}{-\ln z} \right]_+ + \left( \frac{\gamma_E^2}{2} + \frac{\pi^2}{12} \right) \delta(1-z) + \mathcal{O}((1-z)^0) \, ,
\label{eq:plusrelation}
\end{align}
and so on. One might wonder why one would like to employ such distributions of $-\ln z$. In fact, this is a natural consequence of the Mellin transform defined by
\begin{equation}
\mathcal{M}[f(z)](N) \equiv \tilde{f}(N) \equiv \int_0^1 dz \, z^{N-1} \, f(z) \, ,
\end{equation}
whose inverse is given by
\begin{equation}
\mathcal{M}^{-1}[\tilde{f}(N)](z) = f(z) = \frac{1}{2\pi i} \int_{-i\infty}^{i\infty} dN \, z^{-N} \, \tilde{f}(N) \, .
\end{equation}

The Mellin transform is often employed to study the $z$-soft limit, where the $z \to 1$ limit corresponds to $N \to \infty$. In particular, the Mellin transform of the plus-distributions are given by
\begin{align}
\left( \frac{1}{1-z} \right)_+ &\xrightarrow{\mathcal{M}} - ( \psi(N) + \gamma_E ) = - (\ln N + \gamma_E) + \mathcal{O}(1/N) \, , \nonumber
\\
\left[ \frac{\ln(1-z)}{1-z} \right]_+ &\xrightarrow{\mathcal{M}} \frac{1}{2} ( \psi(N) + \gamma_E )^2 + \frac{\pi^2}{12} - \frac{1}{2} \psi'(N) = \frac{1}{2} ( \ln N + \gamma_E )^2 + \frac{\pi^2}{12} + \mathcal{O}(1/N) \, ,
\end{align}
and so on. The usual exercise is then dropping the $\mathcal{O}(1/N)$ terms on the right side, performing the resummation, and carrying out the inverse Mellin transform back to the momentum space. It should be noted that the inverse transforms of powers of $\ln N$ are given by, e.g.,
\begin{align}
\ln N &\xrightarrow{\mathcal{M}^{-1}} - \left( \frac{1}{-\ln z} \right)_+ \, , \nonumber
\\
\ln^2 N &\xrightarrow{\mathcal{M}^{-1}} 2 \left[ \frac{\ln(-\ln z)}{-\ln z} \right]_+ + 2 \gamma_E \left( \frac{1}{-\ln z} \right)_+ \, ,
\end{align}
where the distributions of $-\ln z$ appear.

\begin{figure}[t!]
\centering
\includegraphics[width=0.6\textwidth]{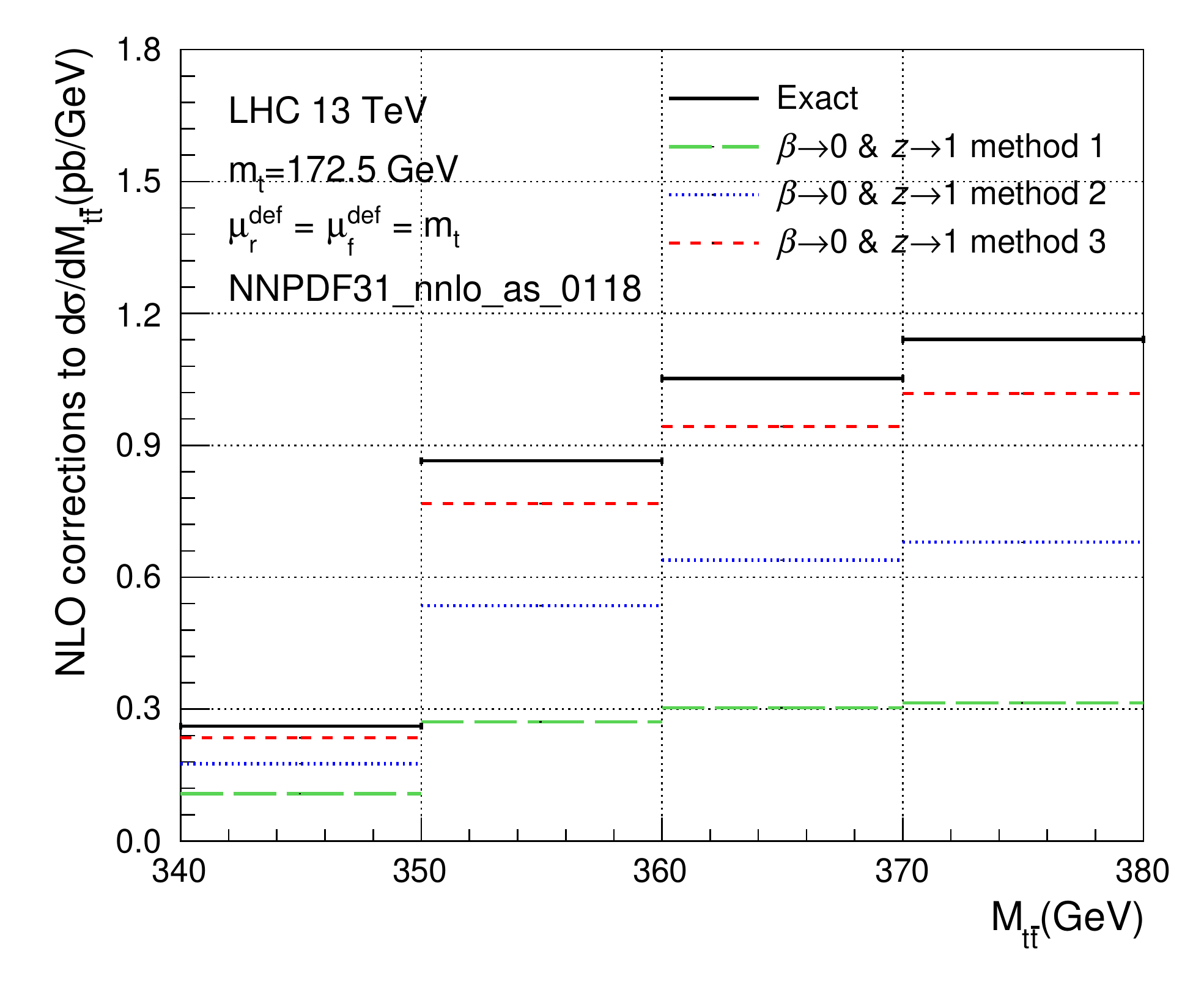}
\caption{\label{fig:zsoft_alternatives}NLO corrections using alternative methods of numerically implementing the $z$-soft limit, compared against the exact ones. See the text for details.}
\end{figure}

Following the above discussions, we then define 3 alternative methods of numerically implementing the $z$-soft limit at NLO, differing in the form of the hard functions entering Eq.~\eqref{eq:K_i_expansion}:
\begin{align}
&\text{\texttt{method 1}: } H_{ij,\alpha,\text{soft}}(z) \, , \nonumber
\\
&\text{\texttt{method 2}: } z \, H_{ij,\alpha,\text{soft}}(z) \, , \nonumber
\\
&\text{\texttt{method 3}: } z \, H_{ij,\alpha,\text{soft}}(1-z \to -\ln z) \, ,
\end{align}
where it is understood that the replacement for \texttt{method 3} should be performed through the relations in Eq.~\eqref{eq:plusrelation}. We emphasize again that the differences among the 3 methods are power-suppressed by $1-z$ (or $1/N$ in Mellin space). Therefore, if the results are indeed only sensitive to the $z$-soft region, one expect that the differences should be numerically small. We show the NLO corrections calculated using the 3 methods in Fig.~\ref{fig:zsoft_alternatives}, compared to the exact ones, for the range $M_{t\bar{t}} \in \unit{$[340, 380]$}{\GeV}$. We observe huge differences among the 3 results, which signal that the $z$-soft limit is actually not valid in this region, since the power-corrections in $1-z$ are not under-control. Nevertheless, we note that the results of \texttt{method 3} (which is closest to the implementation of Ref.~\cite{Kiyo:2008bv}) happen to be close to the exact ones at NLO. There is however no guarantee that the same is true for higher order contributions.

\section{The integrated coefficient and hard functions}

If the renormalization scale $\mu_r$ and the factorization scale $\mu_f$ only depend on $m_t$ and $M_{t\bar{t}}$, one may perform the 4-fold integration over $\theta_t$, $\phi_t$, $Q_T^2$ and $Y$ in Eq.~\eqref{eq:sigma_nlp} analytically. We define the integrated coefficient functions as
\begin{equation}
\tilde{c}_{ij,\alpha} \equiv \int_{-1}^1 d\cos\theta_t \int_0^{2\pi} \frac{d\phi_t}{2\pi} \, c_{ij,\alpha}(\cos\theta_t) \, .
\end{equation}
The results are
\begin{equation}
\begin{split}
\tilde{c}_{q\bar{q},8} &= \frac{2+\rho}{3} \, ,
\\
\tilde{c}_{gg,1} &= \bigg( 1+\rho-\frac{\rho^2}{2} \bigg) \frac{1}{\beta} \ln\frac{1+\beta}{1-\beta} - 1 - \rho \, ,
\\
\tilde{c}_{gg,8} &= \frac{16}{5} \bigg[ (2+2\rho-\rho^2) \frac{1}{\beta} \ln\frac{1+\beta}{1-\beta} - 2 - 2\rho \bigg] - \frac{3}{5} \bigg[ 6(1+\rho-\rho^2) \frac{1}{\beta} \ln\frac{1+\beta}{1-\beta} - 2 - \rho \bigg] \, ,
\end{split}
\end{equation}
where $\rho = 1 - \beta^2$.
We define the integrated hard functions by
\begin{equation}
\mathcal{H}_{ij,\alpha}(z,M_{t\bar{t}},\mu_r,\mu_f) \equiv \int_0^{Q_{T,\text{max}}^2} dQ_T^2 \int_{-Y_{\text{max}}}^{Y_{\text{max}}} dY \, H_{ij,\alpha}(z,M_{t\bar{t}},Q_T,Y,\mu_r,\mu_f) \, .
\end{equation}
The LO results are
\begin{equation}
\begin{split}
\mathcal{H}^{(0)}_{q\bar{q},1} &= 0 \, ,
\\
\mathcal{H}^{(0)}_{q\bar{q},8} &= h_{q\bar{q},8} \, \delta(1-z) = \frac{C_A C_F}{9} \, \delta(1-z) \, ,
\\
\mathcal{H}^{(0)}_{gg,1} &= h_{gg,1} \, \delta(1-z) = \frac{C_F}{32} \, \delta(1-z) \, ,
\\
\mathcal{H}^{(0)}_{gg,8} &= h_{gg,8} \, \delta(1-z) = \frac{(C_A^2-4)C_F}{64} \, \delta(1-z) \, .
\end{split}
\end{equation}
The NLO results for the integrated hard functions were already obtained in \cite{Petrelli:1997ge}. There was a small error in the result which has been pointed out in \cite{Hagiwara:2008df}. It is straightforward to re-derive the results by integrating Eq.~\eqref{eq:HnloQTY}. For completeness, we list the expressions in the following:
\begin{align}
\mathcal{H}_{q\bar{q},1}^{(1)} &= \frac{16}{81} z^2 \, (1-z) \, , \nonumber
\\
\mathcal{H}^{(1)}_{q\bar{q},8} &= 2\beta_0 \, h_{q\bar{q},8} \, \delta(1-z) \, L_M - 2z \, h_{q\bar{q},8} \, P^{(0)}_{qq}(z) \, L_f + \frac{256}{27} \bigg[ \frac{\ln(1-z)}{1-z} \bigg]_+ - \frac{16}{3} \frac{1}{(1-z)_+} \nonumber
\\
&+ \frac{1168-80N_l-44\pi^2+288\ln(2)}{81} \, \delta(1-z) - \frac{128(2+z+z^2)}{27} \, \ln(1-z) \nonumber
\\
&+ \bigg[ \frac{64(2+z+z^2)}{27} - \frac{128}{27(1-z)} \bigg] \ln(z) + \frac{8(54+36z+23z^2-5z^3)}{81} \, , \nonumber
\\
\mathcal{H}^{(1)}_{gg,1} &= 2\beta_0 \, h_{gg,1} \, \delta(1-z) \, L_M - 2z \, h_{gg,1} \, P^{(0)}_{gg}(z) \, L_f + 2 \bigg[ \frac{\ln(1-z)}{1-z} \bigg]_+ \nonumber
\\
&+ \bigg( \frac{7\pi^2}{72} - \frac{11}{18} \bigg) \delta(1-z) - 2z(2-z+z^2)\ln(1-z)
- \frac{7}{18(1-z)} \bigg[ 1 + \frac{\ln(z)}{1-z} \bigg]
\nonumber
\\
&+ \bigg[ \frac{-17+19z-27z^2+9z^3}{9} - \frac{1}{12(1-z)} + \frac{85+210z+129z^2}{36(1+z)^3} \bigg] \ln(z) \nonumber
\\
&+ \frac{20+12z-53z^2+33z^3}{36} - \frac{21+23z}{18(1+z)^2} \, , \nonumber
\\
\mathcal{H}^{(1)}_{gg,8} &= 2\beta_0 \, h_{gg,8} \, \delta(1-z) \, L_M - 2z \, h_{gg,8} \, P^{(0)}_{gg}(z) \, L_f + 5 \bigg[ \frac{\ln(1-z)}{1-z} \bigg]_+ - \frac{5}{4} \frac{1}{(1-z)_+} \nonumber
\\
&+ \bigg( \frac{35}{36} + \frac{25\pi^2}{288} \bigg) \delta(1-z) - 5z(2-z+z^2)\ln(1-z)
+ \frac{58}{3(1-z)} \bigg[ 1 + \frac{\ln(z)}{1-z} \bigg]
\nonumber
\\
&+ \bigg[ \frac{77+23z-45z^2+15z^3}{6} - \frac{29}{(1-z)} - \frac{19+84z+3z^2}{6(1+z)^3} \bigg] \ln(z) \nonumber
\\
&+ \frac{-338+24z-143z^2+115z^3}{24} + \frac{15-47z}{6(1+z)^2} \, , \nonumber
\\
\mathcal{H}^{(1)}_{qg,1} &= -z \, h_{gg,1} \, P^{(0)}_{gq}(z) \, \big[ L_f - 2\ln(1-z) \big] - \frac{4(1-z)}{9} - \frac{z^2}{9} \ln(z) \, , \nonumber
\\
\mathcal{H}^{(1)}_{qg,8} &= -z \, \big[ h_{gg,8} \, P^{(0)}_{gq}(z) + h_{q\bar{q},8} \, P^{(0)}_{qg}(z) \big] \, \big[ L_f - 2\ln(1-z) \big] \nonumber
\\
&+ \frac{8+7z+13z^2-24z^3}{9} + \frac{z(28+47z-16z^2)}{18} \ln(z) \, .
\end{align}

\section{Results with alternative scale choices}

\begin{figure}[t!]
\centering
\includegraphics[width=0.7\textwidth]{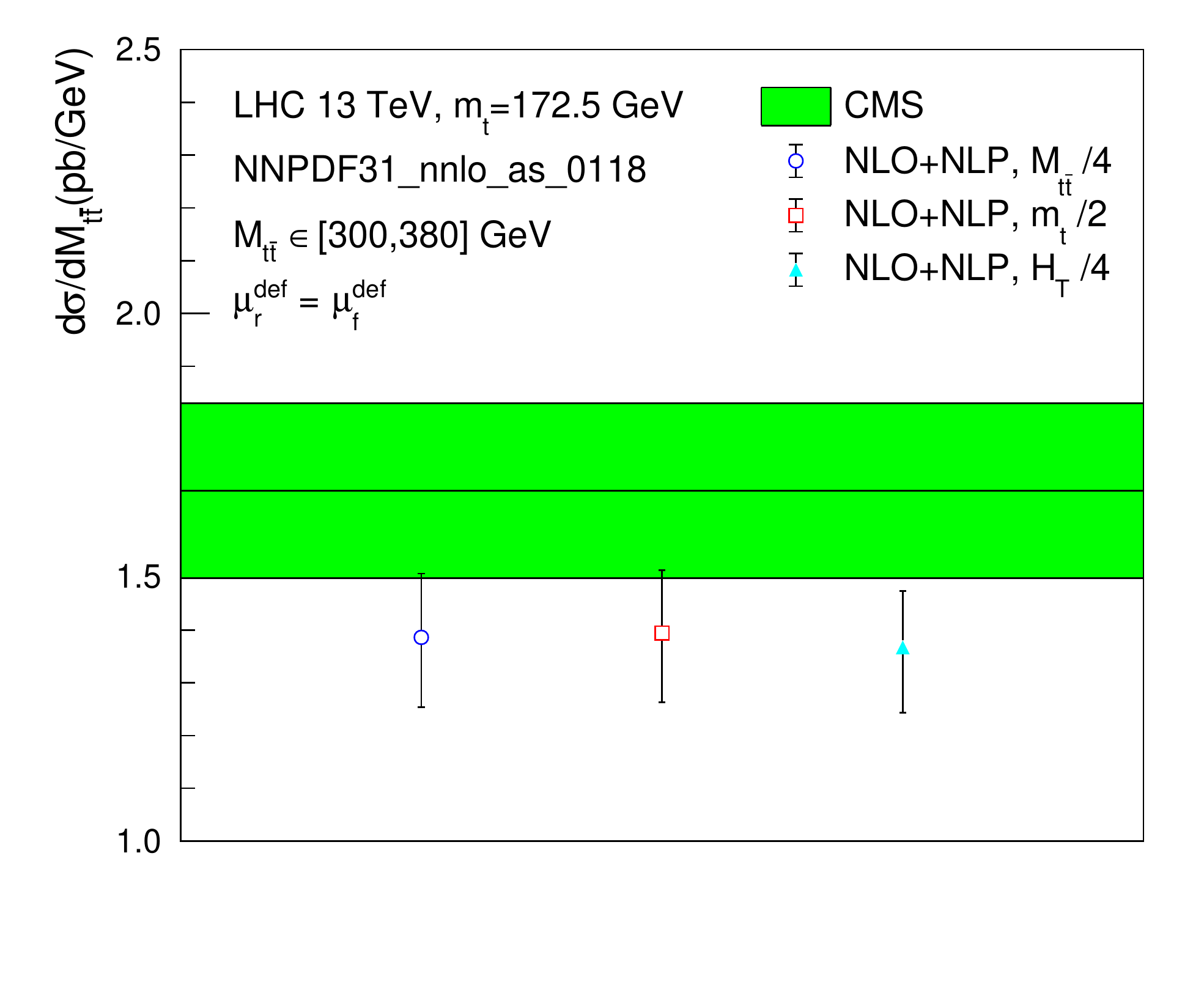}
\vspace{-6ex}
\caption{NLO+NLP results with alternative scale choices.}
\label{fig:scale_choices}
\end{figure}

So far we have used the dynamical scale choice $\mu_r \sim \mu_f \sim H_T/4$ exclusively, which is necessary to match with the NNLO results of \cite{Czakon:2016dgf, Catani:2019hip} and to make predictions for a wide range of $M_{t\bar{t}}$ values up to a few TeV. If we restrict ourselves to the threshold region, other scales choices are also reasonable. For example, one may let the scales correlated with $M_{t\bar{t}}$ or $2m_t$, which are expected to be numerically similar to $H_T$ at low invariant masses. For completeness, we take a sketchy glance at the NLO+NLP results with two different choices of default scales:  $M_{t\bar{t}}/4$ and $m_t/2$. We show the outcome in Fig.~\ref{fig:scale_choices}. As expected, these two results are in good agreement with the one using $H_T/4$ as the default scale.

\section{Possible contributions at NNLP}

While it is beyond the scope of the current paper, it is interesting to discuss possible contributions at the next-to-next-to-leading power (NNLP). At this order, there can be double insertions of the NLP Lagrangians and effective operators, as well as single insertions of the NNLP ones. One of the complications here is that crosstalk among different sectors through sub-leading ultrasoft interactions is activated, which cannot be removed by the decoupling transformations of Eq.~\eqref{eq:decoupling}. As an example, we consider the double insertion of the NLP pNRQCD Lagrangian term $\mathcal{L}_{\text{pNRQCD}}^{1a}$, which contains ultrasoft interactions. In particular, the double insertion of the first term in Eq.~\eqref{eq:L1apNRQCD} induces a new contribution to the potential function with the matrix element
\begin{align}
\braket{0 | \mathrm{T} \big[ \chi^\dagger(0) \psi(0) \, \psi^{\dag}(x_1) \vec{x}_1 \psi(x_1) \, \psi^{\dag}(x_2) \vec{x}_2 \psi(x_2) \big] | t_{a_3}\bar{t}_{a_4}} \braket{t_{a_1}\bar{t}_{a_2} | \psi^\dagger(0) \chi(0) | 0} \, ,
\end{align}
and a new contribution to the soft function with the matrix element
\begin{align}
\braket{ 0| \mathrm{T} \big[ O_s^\dagger(0) \, (S^\dagger_v \vec{E}_{\text{us}} S_v) (x_1^0,\vec{0}) \, (S^\dagger_v \vec{E}_{\text{us}} S_v) (x_2^0,\vec{0}) \big] | X_{s} } \braket{ X_{s}| O_s(0) |0 } \, ,
\end{align}
where $O_s$ is a product of ultrasoft Wilson lines. These two functions are convoluted together in momentum space due to their common dependence on the coordinates $x_1$ and $x_2$.
As a result, the ultrasoft integrals are no longer scaleless and may have a non-zero contribution. Note that similar contributions have also been discussed in the context of heavy quarkonium fragmentation \cite{Nayak:2005rt, Nayak:2005rw}. It remains unknown whether this kind of corrections persist when considering the full NNLP contributions, which is an interesting question for future investigations.

\acknowledgments
L.~L.~Yang would like to thank M.~Czakon, A.~Mitov and Ben~D.~Pecjak for useful discussions and collaborations on related subjects, and to thank A.~Hoang for very stimulating discussions.
This work was supported in part by the National Natural Science Foundation of China under Grant No. 11975030 and 11635001, and the China Postdoctoral Science Foundation under Grant No. 2017M610685.
W.-L.~Ju was also partly funded by the Royal Society through an Enhancement Award (grant agreement RGF\textbackslash{}EA\textbackslash{}181033).
The research of X.~Wang was supported in part by the Cluster of Excellence PRISMA$^+$ (project ID 39083149).
The research of X.~Xu was supported in part by the Swiss National Science Foundation (SNF) under Grant No. 200020\_182038.

\end{document}